\documentclass[preprint,12pt]{aastex}
\newcommand{\wmap}{\textit{WMAP}}

\shorttitle{\WMAP\ 7-year calibration sources}
\shortauthors{Weiland et al.}
\slugcomment{version accepted to ApJS}

\begin{document}
\title{Seven-Year Wilkinson Microwave Anisotropy Probe 
  (\wmap\altaffilmark{1})
  Observations:  Planets and Celestial Calibration Sources}
\author{%
{J. L. Weiland} \altaffilmark{2},
{N. Odegard} \altaffilmark{2},
{R. S. Hill}    \altaffilmark{2},
{E. Wollack} \altaffilmark{3},
{G. Hinshaw} \altaffilmark{3},
{M. R. Greason} \altaffilmark{2},
{N. Jarosik} \altaffilmark{4},
{L. Page}    \altaffilmark{4},
{C. L. Bennett} \altaffilmark{5},
{J. Dunkley} \altaffilmark{6},
{B. Gold}    \altaffilmark{5},
{M. Halpern} \altaffilmark{7},
{A. Kogut}   \altaffilmark{3},
{E. Komatsu} \altaffilmark{8}, 
{D. Larson}  \altaffilmark{5},
{M. Limon}   \altaffilmark{9}, 
{S. S. Meyer}   \altaffilmark{10}, 
{M. R. Nolta}   \altaffilmark{11},
{K. M. Smith}  \altaffilmark{12}, 
{D. N. Spergel} \altaffilmark{12,13},
{G. S. Tucker}  \altaffilmark{14},
and {E. L. Wright}  \altaffilmark{15}
}

\altaffiltext{1}{\wmap\ is the result of a partnership between Princeton
                 University and NASA's Goddard Space Flight Center. Scientific
                 guidance is provided by the \wmap\ Science Team.}
\altaffiltext{2}{{ADNET Systems, Inc., %
                    7515 Mission Dr., Suite A100 Lanham, MD 20706, USA}; {jweiland@sesda2.com}}
\altaffiltext{3}{{Code 665, NASA/Goddard Space Flight Center, %
                    Greenbelt, MD 20771, USA}}
\altaffiltext{4}{{Department of Physics, Jadwin Hall, %
                    Princeton University, Princeton, NJ 08544-0708, USA}}
\altaffiltext{5}{{Department of Physics and Astronomy, %
                    The Johns Hopkins University, 3400 N. Charles St., %
                    Baltimore, MD  21218-2686, USA}}
\altaffiltext{6}{{Astrophysics, University of Oxford, %
                    Keble Road, Oxford, OX1 3RH, UK}}
\altaffiltext{7}{{Department of Physics and Astronomy, University of %
                    British Columbia, Vancouver, BC V6T 1Z1, Canada}}
\altaffiltext{8}{{Department of Astronomy, University of Texas,   %
                    2511 Speedway, RLM 15.306, Austin, TX 78712, USA}}
\altaffiltext{9}{{Columbia Astrophysics Laboratory, %
                    550 W. 120th St., Mail Code 5247, New York, NY  10027-6902, USA}}
\altaffiltext{10}{{Departments of Astrophysics and Physics, KICP and EFI, %
                    University of Chicago, Chicago, IL 60637, USA}}
\altaffiltext{11}{{Canadian Institute for Theoretical Astrophysics, %
                    60 St. George St, University of Toronto, %
                    Toronto, ON M5S 3H8, Canada}}
\altaffiltext{12}{{Department of Astrophysical Sciences, %
                    Peyton Hall, Princeton University, Princeton, NJ 08544-1001, USA}}
\altaffiltext{13}{{Princeton Center for Theoretical Physics, %
                    Princeton University, Princeton, NJ 08544, USA}}
\altaffiltext{14}{{Department of Physics, Brown University, %
                    182 Hope St., Providence, RI 02912-1843, USA}}
\altaffiltext{15}{{UCLA Physics and Astronomy, PO Box 951547, %
                    Los Angeles, CA 90095-1547, USA}}
\begin{abstract}
We present \wmap\ seven-year observations of bright sources which are often used as calibrators
at microwave frequencies. Ten objects are studied in five frequency bands (23 - 94~GHz):
the outer planets
(Mars, Jupiter, Saturn, Uranus and Neptune) and five fixed celestial sources
(Cas~A, Tau~A, Cyg~A, 3C274 and 3C58). 

The seven-year analysis of Jupiter provides temperatures which are within
$1\sigma$ of the previously published \wmap\ five-year values, with slightly tighter 
constraints on variability 
with orbital phase ($0.2\% \pm 0.4\%$), and limits (but no detections) on linear polarization.
Observed temperatures for both Mars and Saturn
vary significantly with viewing geometry.   
Scaling factors are provided which, when multiplied by the Wright Mars thermal model 
predictions
at 350~\micron, reproduce \wmap\ seasonally averaged observations of Mars within $\sim2\%$.
An empirical model is described which fits brightness variations of Saturn due to geometrical
effects and can be used to predict the \wmap\ observations to within 3\%.
Seven-year mean temperatures for Uranus and Neptune are also tabulated.  Uncertainties
in Uranus temperatures are 3\%-4\% in the 41, 61 and 94~GHz bands; the smallest uncertainty for Neptune
is ~8\% for the 94~GHz band.
Intriguingly, the spectrum of Uranus appears to show a dip at $\sim30$~GHz of unidentified origin,
although the feature is not of high statistical significance.

Flux densities for the five selected fixed celestial sources are derived from the 
seven-year \wmap\ sky maps,
and are tabulated for Stokes I, Q and U, along with
polarization fraction and position angle.  Fractional
uncertainties for the Stokes I fluxes are typically 1\% to 3\%. Source variability over 
the seven-year baseline is also estimated. Significant secular decrease is seen for Cas~A and Tau~A:
our results are consistent
with a frequency independent decrease of about 0.53\% per year for Cas~A and
0.22\% per year for Tau~A.  We present
\wmap\ polarization data with uncertainties of a few percent for Tau~A.

Where appropriate, \wmap\ results are compared against previous findings in the literature.
With an absolute calibration uncertainty of 0.2\%, 
\wmap\ data are a valuable asset for calibration work.

\end{abstract}
\section{Introduction}

The primary goal of the Wilkinson Microwave Anisotropy Probe (\wmap) mission \citep{bennett/etal:2003} 
is to study the cosmic microwave background (CMB) anisotropy,
an undertaking that requires detailed knowledge of the instrumental angular response and noise,
low systematic errors, and absolute calibration at sub-percent levels.  These attributes
make \wmap\ data of natural interest for ancillary studies of sky-based calibration objects.
This paper focuses on \wmap\  observations of selected planets 
and fixed celestial objects which have been commonly used as calibrators in the
microwave, either by ground- or space-based instruments.  We provide 
fundamental measurements for these objects which will be of use
not only for calibration purposes, but also other studies, such as planetary 
radiative transfer.

A brief discussion of \wmap\ planetary radiometric data was made by 
\citet{hill/etal:2009}, with an earlier discussion of Jupiter temperatures
by \citet{page/etal:2003}. We extend these presentations to include 
\wmap\ observations of Mars, Jupiter, Saturn, Uranus
and Neptune over a seven-year interval (2001 - 2008), in five microwave 
passbands ranging from $\sim$23 to 94~GHz.
We also select five bright celestial sources for more detailed study:
Cas~A, Cyg~A, Tau~A, 3C58 and 3C274.  
The \wmap\ project routinely compiles and distributes a point source catalog,
which tabulates mean flux values in those same five passbands for bright
sources off the Galactic plane \citep{wright/etal:2009, hinshaw/etal:2007, 
bennett/etal:2003c}. 
However, four of the five objects discussed here (Cas~A, Tau~A, Cyg~A, 3C 58)
have low Galactic latitudes and are not included in the catalog.
An earlier analysis of three-year \wmap\ Tau~A observations by \citet{page/etal:2007}
included both intensity and linear polarization measurements.

This paper is one of a suite that details the analysis of \wmap\ seven-year data. 
An overview and description of data processing methods is provided by 
\citet{jarosik/etal:prep}. \citet{gold/etal:prep} 
discuss Galactic foregrounds and techniques for removing them; an updated
point source catalog is included. Analysis of CMB power spectra, best fit models
and cosmological interpretations are presented by \citet{larson/etal:prep} and 
\citet{komatsu/etal:prep}. \citet{bennett/etal:prep} assess 
potential CMB ``anomalies'' relative to the best-fit six parameter $\Lambda$CDM model.

We organize this paper as follows: a short discussion of the \wmap\ instrument, plus 
data strengths and limitations for calibration source work, is provided in 
Section~\ref{sec:wmapintro}.  
The paper then divides into two major sections: analysis and results
for planets (Section~\ref{sec:planets}), and then Section~\ref{sec:celsrc}
discusses analysis methods and results for the five bright celestial sources.
A closing summary and conclusions follow in Section~\ref{sec:summary}.

\section{\wmap\ Overview\label{sec:wmapintro}}

Since its arrival at the Earth-Sun L2 point in 2001 August, \wmap\  
has produced a well-calibrated, multi-wavelength microwave survey of the 
entire sky with high sampling redundancy. 
Detailed descriptions of the mission profile and hardware can be found in
\citet{bennett/etal:2003}, \citet{jarosik/etal:2003} and \citet{page/etal:2003}.
To assist with 
clarifying the terminology and analysis methods presented in this paper, we describe
the key features of the mission.

\wmap\ observes the sky differentially: signal from the sky is simultaneously
sampled in two different directions $\sim140\arcdeg$ apart and guided by 10 feed horn pairs
into 10 receivers, referred to as differencing assemblies (DAs).  
The receivers cover five
frequency bands (K, Ka, Q, V and W) with approximate center frequencies of $\sim$23, 33,
41, 61 and 94~GHz, respectively.  There is one DA each for K and
Ka bands (K1, Ka1), two DAs each for Q and V bands (Q1, Q2, V1, V2) and four DAs
(W1, W2, W3, W4) for W band.  The frequency passbands are broad ($\sim20\%$ fractional bandwidth), 
so that the exact frequency associated with an observation depends on the
source spectrum. An orthomode transducer at the throat of each feed
separates the sky signal into two orthogonal polarizations, allowing 
for reconstruction of both intensity and linear polarization signals.
Angular resolution is moderate, ranging from roughly $0.2\arcdeg$ FWHM in W band
to $\sim0.9\arcdeg$ FWHM in K band.

As a survey mission, \wmap\ is not commanded to point at specific targets,
but executes a pole-to-pole sky-scanning strategy whose position on the ecliptic advances $\sim1\arcdeg$ 
each day. To facilitate high redundancy in sky coverage, the spacecraft 
rotates about a spin axis, which also precesses.  
These combined motions produce a $\sim45\arcdeg$ wide scan swath which looks like a ``belt'' 
projected onto the sky, passing through the ecliptic poles. The scan swath returns to its 
original starting position on the ecliptic every year, after completing a full-sky survey
with multiple observations per sky pixel.  

A primary strength of \wmap\ data is the calibration accuracy, with an
absolute calibration error of $0.2\%$ \citep{jarosik/etal:prep}.
The absolute calibration is tied to knowledge of the CMB monopole and
the spacecraft orbital velocity, and thus is independent of any reliance
on previously established measurements of the sources under study.
When coupled with {\wmap}'s temporal and spectral coverage, the accurate flux
measurements can provide a useful measure of intrinsic or geometrically induced 
source variability.

There are some limiting factors, however.  
Temporal coverage of an object is constrained by the sky-scanning method.
For a fixed source on the ecliptic, the \wmap\ scan strategy 
provides for two $\sim45$-day windows a year in which to observe
that source.  These ``observing seasons'' are spaced roughly six months apart.
While this geometry is modified somewhat for moving objects, only Mars
is near enough that the orbital motion can extend a single viewing window 
from 45 days to as much as 60 to 90 days.  
Since \wmap\ does not dwell on an object, 
but simply scans as it passes through the viewing swath, the number of observations
for an individual object is also limited, which affects the signal-to-noise.
Finally, \wmap's moderate spatial resolution,  which allows us to
characterize the planets as calibrators, also places
detailed knowledge of structures such as Saturn's rings beyond instrument capabilities.
\section{Planets\label{sec:planets}}
There are five planets with sufficient signal-to-noise for analysis.
Listed in order of decreasing signal-to-noise, these are
Jupiter, Saturn, Mars, Uranus and Neptune.  Since \wmap\  points away from
the Sun, Earth and Moon in its L2 orbit, it cannot observe Mercury or Venus.
\subsection{Data Processing and Analysis Methods\label{sec:methods}}
Planets are excluded from the
standard map-based products used for \wmap\ studies of the CMB and
Galactic foregrounds. Observations of the planets
must be extracted from the calibrated time-ordered data (TOD), and
then undergo quality assessment and background removal
before they can be analyzed. 
The extraction process gathers
all observations from the TOD taken within several degrees of each planet
and associates ancillary information such as position
in the focal plane, attitude, planet viewing aspect and data quality flags with each
observation.  Observations whose sky positions lie within a $7\arcdeg$ radius of other 
planets or which are
of suspect quality are immediately rejected.  In addition, observations
are excluded from use if either beam of a DA feed-horn pair samples sky coordinates
that fall within the boundaries of a spatial processing mask \citep{jarosik/etal:2007}.
The purpose of this ``Galactic masking'' is to mitigate
background removal error by excluding observations located near strong-signal, high-gradient
regions of the Galaxy.  For those observations which survive these quality
control steps, Galactic and CMB background signals are subtracted.  The background
signal subtracted from each observation is computed from the DA-appropriate
seven-year mean map (from which planetary data were excluded), using the value from the $\sim6\arcmin$ wide
pixel which is closest in spatial coordinates to those
of the time-ordered observation.
Table~\ref{tab:planetseasons}  lists the beginning and ending
dates of the observing seasons for the five planets, i.e., when they are directly 
in the \wmap\ viewing swath.
The ``sky proximity'' column in the Table indicate probable reasons for an
unusually high rejection of observations within a season.

As noted in \citet{page/etal:2003}, planets are ``point sources'' for \wmap\, and thus the
brightness temperature of a planet $T_{p}$ can be computed
via the relation
\begin{equation}
  T_{p} \Omega_{p} =  T^m \Omega_{B},
  \label{eq:tempeqn}
\end{equation}
where $T^m$ is the peak response observed by \wmap,  $\Omega_B$ is the main-beam solid angle,
and $\Omega_{p}$ is the solid angle subtended by the planet.  Changing distance between
the planet and \wmap, plus changes in viewing aspect of an oblate planet, will cause  
$\Omega_{p}$ 
(and also $T^m$) to vary with time.  We account for both geometrical effects and compute the
brightness temperature relative to a fiducial solid angle $\Omega_{p}^{\mathrm{ref}}$ via the use
of time-dependent geometrical scaling factors $f_d$ and $f_A$. The factors $f_d$ and $f_A$ may 
be thought of as ``distance correction'' and ``disk oblateness
correction'' factors:

\begin{eqnarray}
  T_{p}  =  T^m \Omega_{B}/ \Omega_{p}^{\mathrm{ref}} \times {f_d / f_A } \ \\
  f_A = A_{\mathrm{disk}}^{\mathrm{proj}}/A_{\mathrm{ref}}\ \\
  f_d = d_{p}^2/d_{\mathrm{ref}}^2,
  \label{eq:tempeqn2}
\end{eqnarray}
where $A_{\mathrm{ref}}$ and $d_{\mathrm{ref}}$ are a fixed fiducial disk area and distance, respectively.
$A_{\mathrm{disk}}^{\mathrm{proj}}$ is the projected area of the planetary disk, and $d_{p}$ the distance of the
planet from \wmap, at the time of each observation.
The projected area of the oblate disk is computed from
\begin{equation}
  A_{\mathrm{disk}}^{\mathrm{proj}} =  \pi R_{\mathrm{pole}}^{\mathrm{proj}} R_{\mathrm{eq}},
  \label{eq:diskarea}
\end{equation}
where  $R_{\mathrm{eq}}$ is the planet's equatorial radius and 
$R_{\mathrm{pole}}^{\mathrm{proj}}$ is the projected polar radius given by
\begin{equation}
   R_{\mathrm{pole}}^{\mathrm{proj}} =  R_{\mathrm{pole}} [ 1 - \sin^2(D_{W}) (1 -
   (R_{\mathrm{pole}}/R_{\mathrm{eq}})^2)]^{1/2},
  \label{eq:projrad}
\end{equation}
where $D_{W}$ is the planetary latitude of \wmap\ (``sub-\wmap\ latitude'').
The fiducial disk area $A_{\mathrm{ref}}$ is simply $A_{\mathrm{disk}}^{\mathrm{proj}}$ evaluated at $D_W$ of $0\arcdeg$.
Planetary radii and north pole directions used to compute $D_{W}$ are adopted from \citet{seidelmann/etal:2007};
the polar and equatorial radii are listed in Table~\ref{tab:radii}.
Because \wmap\ is at L2, $D_{W}$ is not very dissimilar to that of the sub-Earth
latitude, $D_{e}$.

\begin{figure}
  \begin{center}
    \includegraphics[height=7.0in]{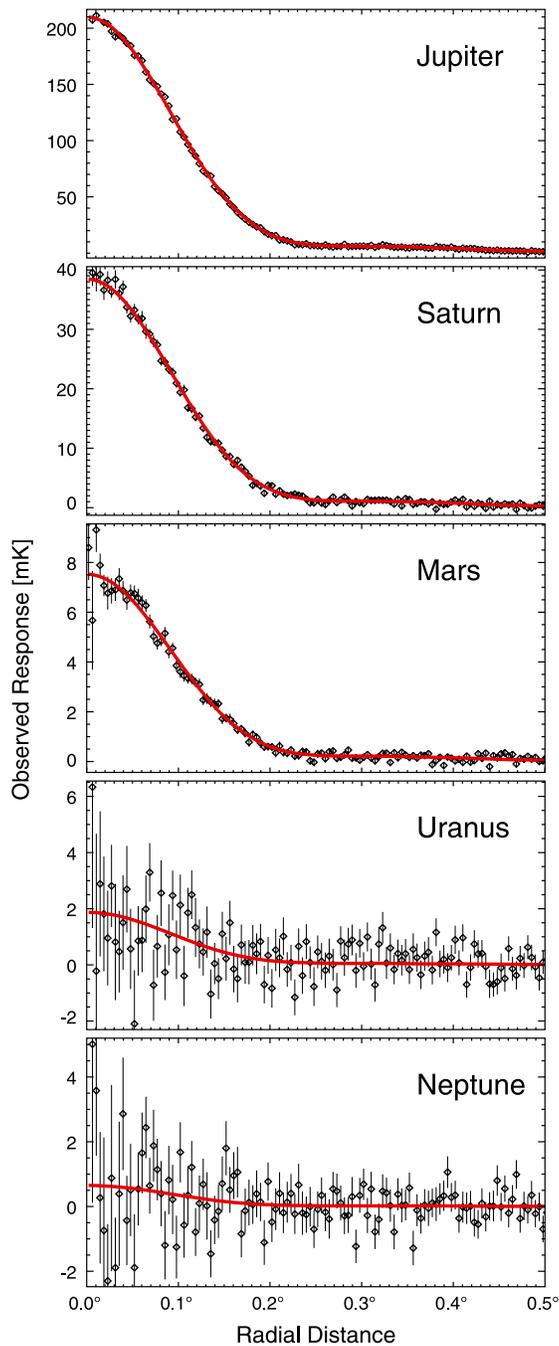}
    \caption{ Typical single-season radial profiles of Jupiter, Saturn, Mars, Uranus and Neptune 
    for the W1 differencing assembly. Black points are data, red line is the fit
    of the smooth beam profile template from which the temperatures are computed.  The signal-to-noise
    ratio is best for Jupiter and worst for Neptune.
    \label{fig:radialprofiles}}
  \end{center}
\end{figure}

As the brightest ``point source'' visible to \wmap, Jupiter is used to characterize the
main beam response of the instrument.  This process and analysis are described
in detail in \citet{hill/etal:2009}, \citet{jarosik/etal:prep}, and
\citet{page/etal:2003}.  The end result is a
set of seven-year, azimuthally-symmetrized, one-dimensional radial beam profiles per DA.
These radial beam profiles are then used to derive window functions used for CMB power spectrum 
deconvolution. Unlike the rest of the planetary temperatures in this section, the 
seven-year mean Jupiter temperatures are derived from the $l=0$ value of the unnormalized 
window functions (which
measures $T^m \Omega_{B}$), coupled
with the fiducial solid angle for Jupiter \citep{hill/etal:2009}.  
For all other planetary analysis, brightness temperatures are derived using a template-fitting
technique, which takes advantage of the high-quality beam information obtained from 
seven years of Jupiter observations.  After the processing steps described above,
radial response profiles are produced for each planet, per DA, per season,
by assigning observations to 0.25 arcmin bins increasing in angular distance from 
the planet's center.
Each bin is assigned a mean value and an error based on instrument noise appropriate to the number
of observations within the bin. 
The known radial beam response, based on seven-year Jupiter data, is used as a fitting template
to the seasonal radial profile for each DA.  The fit returns $T^m$ for the planet. The brightness 
temperature is then derived from $T^m$, the main-beam solid angle \citep{jarosik/etal:prep} and the 
reference solid angle for the planet as per the above equations.
Errors include the formal error in the template
coefficient and the propagated error in the beam solid angle.
Figure~\ref{fig:radialprofiles} illustrates sample W1 radial profiles for a single observing season
for each of the planets. The beam template is overplotted in red.  As well as illustrating the technique,
the figure is also a useful comparator of signal quality for each planet at W-band.  
Beam dilution increases with decreasing \wmap\ frequency:  $\Omega_{B}$ is roughly  24.6, 14.4, 8.98,
4.20 and 2.10 $\times 10^{-5}$ sr for the K - W bands respectively \citep{jarosik/etal:prep}.

Tabulated brightness temperatures are appropriate for the Rayleigh-Jeans (RJ) approximation. 
Brightness temperatures provided in the Tables do not include the CMB contribution
blocked by the planet but included in the background.  As noted in the Tables,
absolute brightness temperature is obtained by adding 2.2, 2.0, 1.9, 1.5 and 1.1 K in bands
K, Ka, Q, V and W respectively \citep{page/etal:2003b}, since 
in the RJ approximation the CMB temperatures are frequency dependent.
\wmap\ data points shown in comparison to measurements in the literature are
converted to absolute brightness.

\subsection{Jupiter\label{sec:jupsec}}

Jupiter is a bright, accessible source whose whole-disk observations 
have long been used for calibration in the infrared through radio.  
\citet{hill/etal:2009} recommend the use
of \wmap\ radiometry of Jupiter as the preferred method of transferring the \wmap\ dipole calibration
to another microwave instrument.  For some instruments, Jupiter is unfortunately too 
bright, hence the interest in the use of fainter sources, and a set of observations which
tie several common sources together on the same calibration scale.  We continue to recommend
Jupiter as the primary comparison calibrator, however.

Seven-year mean brightness temperatures for Jupiter, computed as described in Section~\ref{sec:methods}, 
are presented in Table~\ref{tab:juptemp}.
The seven-year values differ from those of the five-year analysis of \citet{hill/etal:2009} by 
$1\sigma$ or less.    
The largest changes, which hover near $1\sigma$, are for K band and some of the W band 
DAs.  A detailed explanation of these changes is given in \citet{jarosik/etal:prep}.
Briefly, the change in K band results from improved background subtraction, and the W band
changes are driven by small revisions to the beam solid angles, based on modeling.   
The seven-year temperatures for W band show a decreased scatter between the individual
DAs compared to five-year, implying an improved calibration. The Jupiter temperature
uncertainties listed in Table~\ref{tab:juptemp} are  
derived from the quadrature sum of the solid angle error and absolute calibration error of $0.2\%$
\citep{jarosik/etal:prep}, as they were in the five-year analysis.
  
\begin{figure}
  \begin{center}
    \includegraphics[height=4.5in]{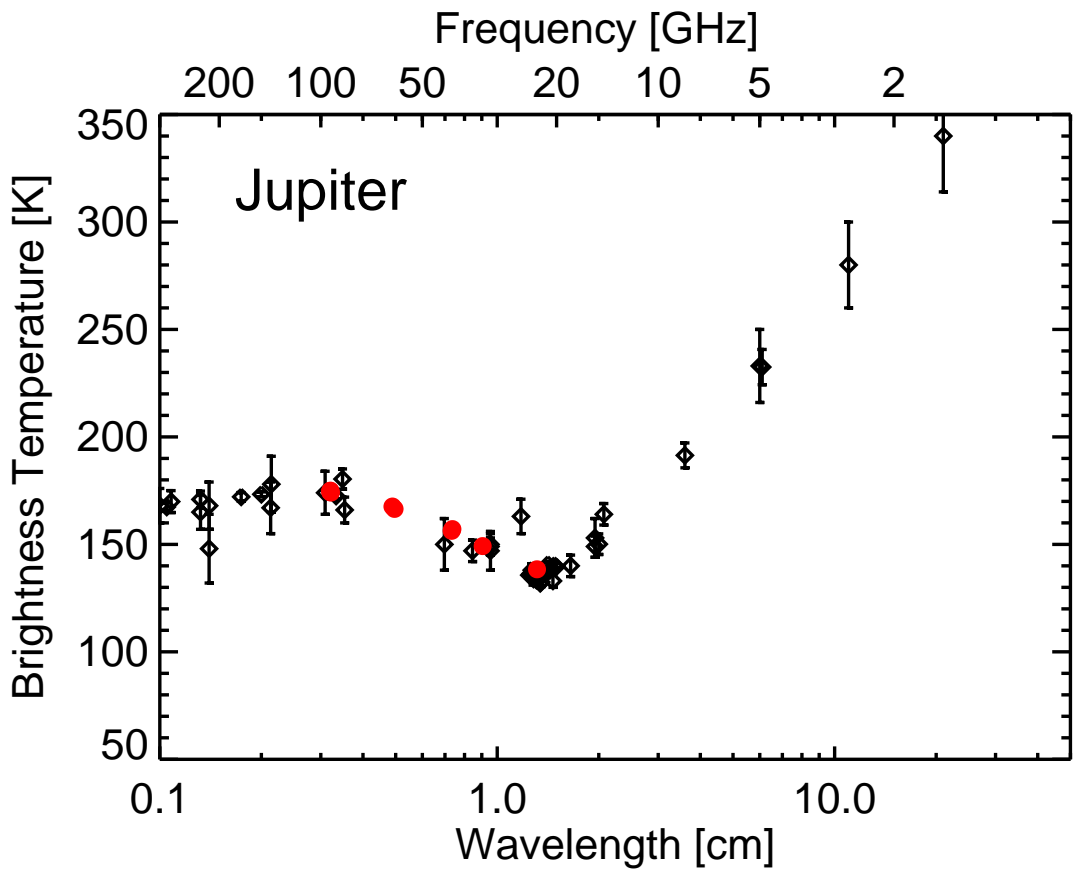}
    \caption{ Microwave spectrum of Jupiter. Black diamonds
      are selected observations compiled from
      the literature (see the text).  
      The seven-year mean \wmap\ temperatures, corrected to absolute brightness
      and including error bars, are shown in red. 
      Strong absorption by {NH}$_{3}$ inversion band is
      centered near 1.3 cm.  Minor contributions from synchrotron
      emission have not been removed from the \wmap\ data (see Section~\ref{sec:jupsec}).
      \label{fig:jupspec}}
  \end{center}
\end{figure}

Thermal emission from Jupiter's atmosphere is the primary signal source at wavelengths 
spanned by \wmap.  As such, the observations are helpful for providing constraints on
atmospheric models for Jupiter, which in turn could be used for calibration purposes.
However, as noted by e.g., \citet{gibson/welch/depater:2005}, there
is a fine point to consider: small contributions to the total observed intensity from a synchrotron
component up to about 40~GHz are expected. This non-thermal ``decimetric emission'' arises from charged 
particles trapped close to Jupiter by its strong magnetic field, and dominates the 
spectrum at frequencies less than about 3~GHz. 
Estimates in the literature place the synchrotron contribution at 23 GHz 
near 1\% of the total intensity  \citep{depater/dunn:2003, gibson/welch/depater:2005}, with 
a roughly $T \sim \nu^{-2.4}$ frequency dependence \citep{klein/gulkis:1978}.  However, the synchrotron emission is
known to be variable on a wide range of time scales \citep{santos-costa/bolton/sault:2009,
depater/dunn:2003, depater/etal:2003, miyoshi/etal:2000}, and accurate removal is difficult.
The \wmap\ seven-year mean Jupiter brightness temperatures are graphically compared
against those in the literature in Figure~\ref{fig:jupspec}.  Ground-based
observational values are taken from compilations of \citet{joiner/steffes:1991} 
and \citet{klein/gulkis:1978},
along with individual observations from \citet{depater/kenderdine/dickel:1982,depater/etal:2001}, 
\citet{greve/etal:1994} and \citet{goldin/etal:1997}.
Estimates of the synchrotron emission have been removed from the  
observations in the figure by many, but not all, of the authors, and not 
necessarily to the same level.  
The \wmap\ observations are plotted ``as observed''.  

As an estimate of the stability of Jupiter's emission with orbital phase over the seven-year 
interval, we extend the 
five-year analysis of 
\citet{hill/etal:2009}, in which temperatures for individual Jupiter observing seasons are computed and
compared to the multi-year mean values. Individual seasonal temperatures are derived using the
same beam-profile fitting technique which is used for the other planets (Section~\ref{sec:methods}).  
We compute $\Delta T/T$
as the mean deviation of all DAs from their seven-year mean values, and include a $1\sigma$ standard
deviation as a measure of coherency.  These results are listed in Table~\ref{tab:juptime}.  
Using the \wmap\ five-year data, \citet{hill/etal:2009} 
quote an upper limit of $0.3\% \pm 0.5\%$ based on the largest deviation and scatter seen in the first 
10 observing seasons, which occurred in season four.
A careful examination of the entries in Table~\ref{tab:juptime} with the largest deviations and/or scatter 
(observing seasons 4, 11, 12)  shows that these particular instances have more to do
with template fitting and background subtraction quality than with intrinsic source variability.
The two seasons with high scatter (4, 12) are those in which the number of available observations 
for fitting has been significantly reduced because of foreground masking (``\% masked'' in the Table),
which increases the statistical error in the temperature determination.  Season 11, with the
highest $\Delta T/T$, is one in which Jupiter is nearly aligned with the Galactic center, but outside the processing
mask.  This is a region in which background subtraction is subject to greater error because of
higher temperature gradients across the pixels in the map used for background removal.
If one ignores these three entries, then an upper limit of $0.2\% \pm 0.4\%$ is more appropriate.
We also searched for indications that K band might be more intrinsically variable than W band, because
of the small synchrotron contribution.  There is no indication of such a frequency dependence in 
the data, but the upper envelope here is about 0.5\%.

The \wmap\ Jupiter observations were analyzed to determine if any linear polarization signal could be
detected. A special version of the beam mapping software was run for Jupiter, in
which the angle between the polarization axis of each radiometer and the
planet's magnetic dipole was computed for each observation.  For each differencing assembly, all observations 
within a tight radius of beam center were accumulated, corrected to peak beam response, 
background subtracted, loss imbalance corrected, separated by radiometer, and then 
used to solve for Stokes parameters I, Q, U and their errors.  For a more complete description of the 
relevant equations, see \citet{hinshaw/etal:2003} and \citet{jarosik/etal:2007}. 
Normalized Stokes parameters Q/I and U/I are reported in 
Table~\ref{tab:juppoln}, along with their propagated errors and estimated linear polarization fraction,
$p_{\mathrm{lin}} = \sqrt{(Q/I)^2 + (U/I)^2}$.  No attempt has been made to correct for noise bias in either
$p_{\mathrm{lin}}$ or its reported error, which is simply the propagated error.  The estimated signal-to-noise is low, 
with no significant detections in any differencing assembly.  
Linear polarization of the synchrotron emission at 15 and 22~GHz has been detected by \citet{depater/dunn:2003}
using the Very Large Array (VLA).  They were unable to obtain a synchrotron polarization fraction at 22~GHz,
but estimated 20\% polarization based on the 15~GHz observations, which implies roughly a 0.2\% polarization 
fraction at K band using a combined thermal and non-thermal intensity.   
The \wmap\ results are consistent with this estimate, but do not provide any further constraint.
\subsection{Mars}

Mars' value as a calibrator depends upon the accuracy to which its whole-disk
brightness variations (both apparent and intrinsic) can be modeled.  
Much of the perceived variability from earth-orbit is attributable to geometrical factors: 
the apparent phase as seen by the observer and 
changes in solar irradiance as Mars' distance from the Sun varies in its
elliptical orbit \citep{wright:1976}. There are a number of additional complexities, 
however.  The day/night cycle imparts a significant rotational modulation to the
observed temperature as the surface heats and cools. In the infrared, variable 
atmospheric effects including molecular line absorption and dust storms can be of importance.
Additionally, the Martian surface is itself far from homogenous, with polar ice-caps of changing 
extent and variations in albedo and thermal inertia on relatively small scales 
\citep{putzig/mellon:2007, christensenpr/etal:2001}.

Planetary science studies notwithstanding, the astronomical community has most consistently 
referenced two basic models when calibrating to Mars: the ``Wright model''
(\citealt{wright:1976}, \citealt{wright/odenwald:1980}, \citealt{wright:2007}) and the ``Rudy model''
(\citealt{rudy/etal:1987}, \citealt{muhleman/berge:1991}), both developed before the current level of 
detail available from orbiters such as Mars Global Surveyor (MGS).  
In the thermal emission regime ($\lambda \gtrsim 10$ \micron), 
the planet's brightness temperature can be computed to first order as re-emission 
of absorbed solar energy by a surface layer with a characteristic albedo, emissivity and 
thermal inertia \citep{depater:1990}. The Wright model uses this approach to predict thermal 
emission in the 10-350 \micron\ range, with no positional variations of parameters such 
as albedo and thermal inertia.  The Rudy model originates from VLA observations at 2 and 6~cm,
which sample further into the Martian subsurface.  Concurrent measurements of the polarized flux
are used to estimate radio reflectivities and absorption lengths as a function of 
planetary latitude, and brightness temperatures calculated via radiative transfer as a function of 
position on a
longitude/latitude grid.  Typical errors on $T_{b}$ quoted for the Wright model are $\pm 5\%$, or
about $\pm 10$K, 
and those for the Rudy model at 5\%-10\%.
We have chosen in this paper to compare the \wmap\ observations against predictions from the
Wright model, primarily because of its long history of use and readily available model code
\citep{wright:2007}, but also because the seasonal brightness temperatures quoted for \wmap\ are 
the resultant average over several weeks of data, which means we are only sampling mean 
properties at best.  We have modified the 2007
Wright model code in three ways: (1) JPL ephemeris positions are substituted for the lower
precision formulae originally used; (2) an option is added to use \wmap\ spacecraft 
positions rather than those of the Earth; and (3) computations are extended out to \wmap\ frequencies.
The first two changes are not critical to the analysis, as the 
combined modifications introduce at most at 0.6~K difference from the original results.
Typically the model predicts brightness temperatures at 3~mm which are a few Kelvins lower than those at
350~\micron, and the temperature spread between the extrema of the
5 \wmap\ frequencies is quite low, of order 0.5 K.

\begin{figure}
  \begin{center}
    \includegraphics[height=4.5in]{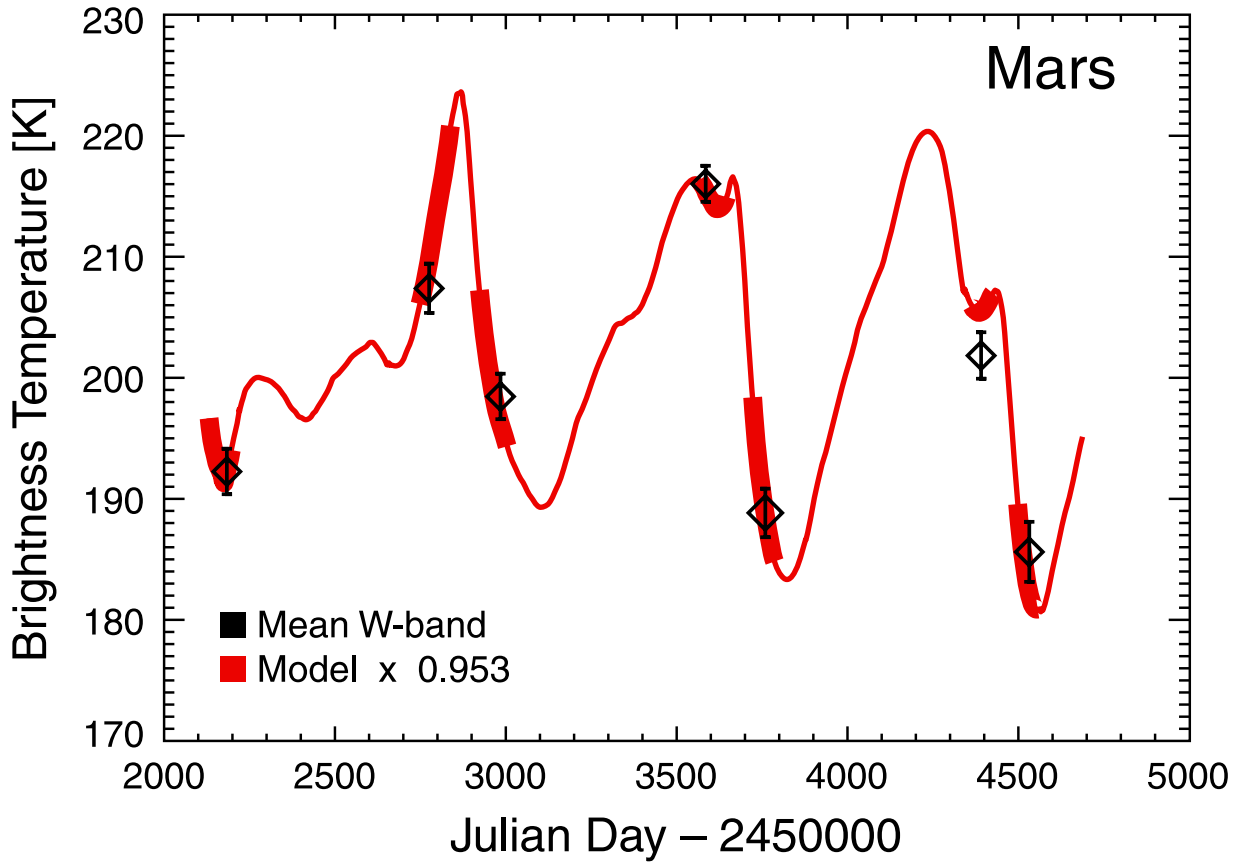}
    \caption{ Comparison of \wmap\ W-band seasonal averages (black diamonds; Table~\ref{tab:mars}) to
    the Mars model of Wright (1976, 2007).  The \wmap\ observations have been corrected  
    to absolute brightness.  Model values (red line) have been rescaled by a factor of 0.953 to bring
    them into overall agreement with the observations; thick portions of the line indicate
    observing seasons.  Data quality masking can skew the mean times of observations from the
    mean of the seasonal interval, as is evident in the second observing season.
    \label{fig:marswband}}
  \end{center}
\end{figure}

\citet{hill/etal:2009} presented the first five years of \wmap\ Mars observations 
at 94~GHz.  Here we extend the W-band observations to seven years, and
include the observations at the other four \wmap\ frequencies.  Mean brightness 
temperatures for Mars at five frequencies and seven \wmap\ observing seasons 
derived from the seven-year data are presented in Table~\ref{tab:mars}. 
After correcting to absolute brightness temperature and converting from Rayleigh-Jeans
to Planck brightness temperature \footnote{The Wright model predictions are tabulated as 
Planck brightness temperatures,
which reconstruct the model flux assuming a blackbody with no limb darkening},
the seven seasonal W-band averages
are shown in comparison to the Wright model
predictions at 3.2~mm in Figure~\ref{fig:marswband}.  This plot is similar to Figure~15 of
\citet{hill/etal:2009}, whose preliminary analysis
found that the Wright model was a good predictor
of the mean temporal variation, but the model $T_b$ seemed high
by roughly 10\%.  Our results generally agree, but we
find that multiplying the Wright model by a factor of $0.953$ fits
the seven-year data best.  If one uses the Wright model tabulated at 350 \micron, the
plot looks nearly identical, but the scaling factor is 0.941, or about a 6\% reduction.
A previous comparison by \citet{griffin/etal:1986}
indicated that the 3~mm whole-disk observations of \citet{ulich:1981} were within
7.5\% of Wright model predictions at 350 \micron.
The first \wmap\ Mars observing season occurred 
during the time of the 2001 July-October massive global dust storm which elevated
atmospheric temperatures in the mid-IR by nearly 40~K \citep{smith/etal:2002}.  
The effect of dust storms on microwave data is expected to be less of a concern,
but we did fit for the model scaling factor both with and without inclusion of
the first observing season, with no significantly different results.

\begin{figure}
  \begin{center}
    \includegraphics[height=5in]{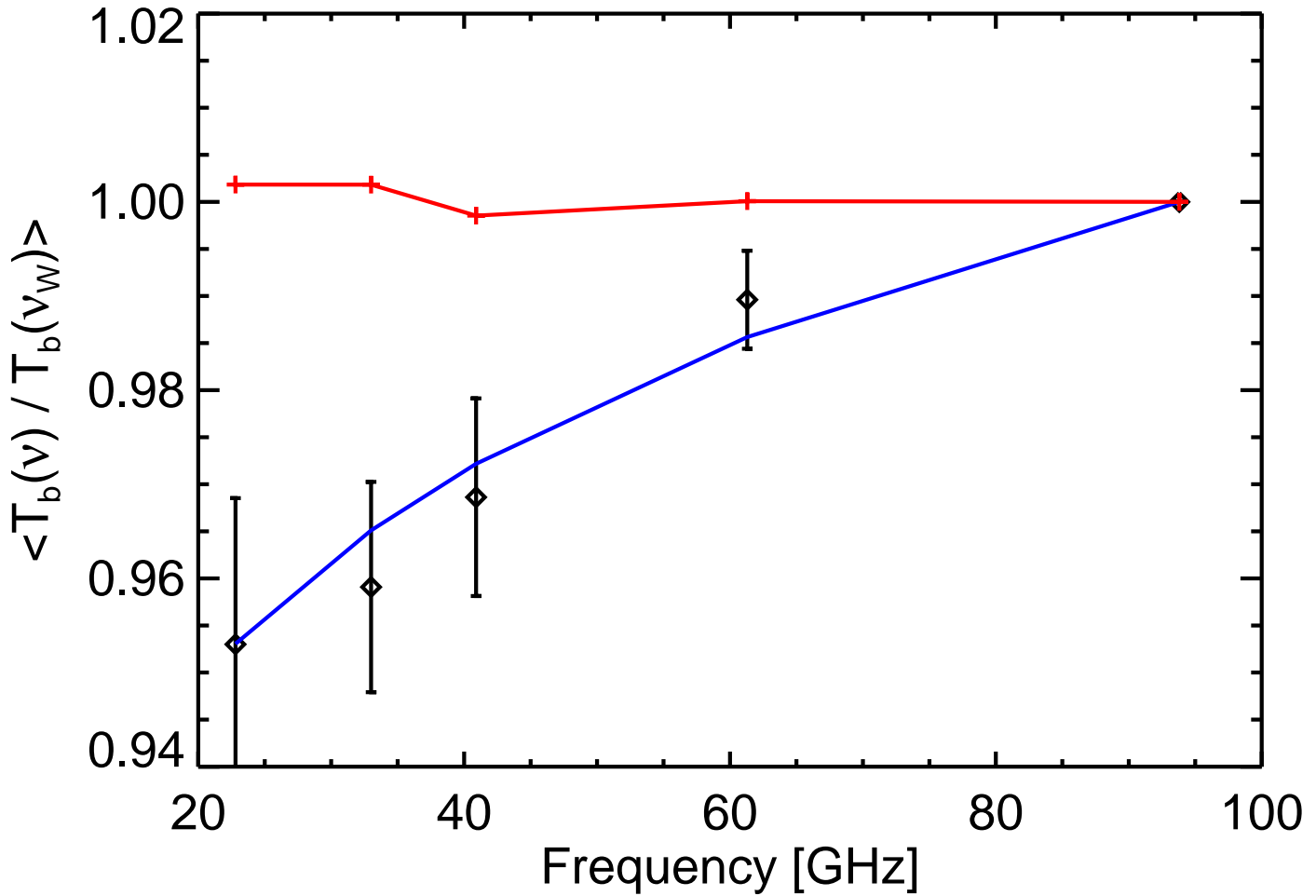}
    \caption{ Seven-year mean Mars brightness temperatures as a function of
    frequency, normalized within each season to $T_{b}$ at W band.  Black diamonds
    are the \wmap\ data; error bars are obtained from the variation over
    the seven observing seasons within each frequency band. 
    The same averaging procedure is carried out using the
    Wright model predictions for the mean time of observation (red line): 
    small time-sampling and spectral variations between the
    frequencies cause a slightly jagged appearance.  The
    \wmap\ data show a weak frequency dependence which can be approximated
    as ($\nu$(GHz)/94)$^{0.03}$ (blue line). This can be interpreted as
    a frequency-dependent emissivity.
    \label{fig:marsfreqdep}}
  \end{center}
\end{figure}

The uncertainties on the derived \wmap\ seasonal brightness temperatures increase with 
decreasing frequency, which follows as a result of decreasing signal-to-noise in the data.  
However, it is possible to extract a trend in temperature with frequency by 
finding within each season the ratio of the brightness temperature
at each frequency to that at W band,
 and then computing the
mean over all seven observing seasons as a function of frequency. These means are plotted
in Figure~\ref{fig:marsfreqdep} in black.  Errors are standard deviations of the mean.
The weak frequency dependence can be characterized roughly as $\nu^{0.03}$, indicated by the
blue line. The Wright model, processed in the same fashion, predicts almost no variation with
frequency (red line). 
Such a frequency dependence could be attributed to an emissivity change of $\sim5\% \pm 2\%$ over the
wavelength interval.  In a similar vein, a weak $5\% \pm 3\%$ decline in surface emissivity with
increasing wavelength was noted by 
\citet{burgdorf/etal:2000} in the 50 to 180 \micron\ range, based on analysis of Infrared Space Observatory
(ISO) observations of 
water vapor absorption lines.

Some consideration was given to placing the \wmap\ observations in a wider context,
with the goal of gaining further insight into the model parameters. 
A search was made for a dataset that would provide absolutely calibrated, whole-disk 
observations of Mars which 
spanned both sides of the thermal energy distribution peak and were taken over a reasonably
contiguous interval of time.  Mars data taken by the COBE/DIRBE instrument 
\citep{boggess/etal:1992, hauser/etal:1998}
for a 146 day interval (1990 April 28 - September 20) most closely
matched these requirements, although no analysis of the 
publicly available~\footnote{\texttt{http://lambda.gsfc.nasa.gov/}} data appears to have been
previously published.  There are six DIRBE wavebands which sample the thermal emission portion
of the Mars spectrum, with effective wavelengths of 12, 25, 60, 100, 140 and 240 \micron.  Of these, 
the 25~\micron\   observations
cannot be used as an independent check of the Wright model, since the model itself was
the ultimate source of the DIRBE absolute calibration in this band (S.~Moseley \& W.~Glaccum 2009, 
private communication, \citealt{hauser/etal:1998}).
In addition, while the 12, 25, 140 and 240~\micron\  detectors are well-behaved linear
devices, the 60 and 100~\micron\ detectors are known to suffer non-linear response, and Mars
is on the bright end of that response.
DIRBE is a broad-band photometer, and reports monochromatic flux values at the effective 
wavelength of each band assuming 
a constant source spectrum in $\nu I_{\nu}$ across the band.
One must apply a ``color correction factor'' to the reported fluxes if the source
spectrum is different from that assumed.  The DIRBE Explanatory Supplement provides such factors
for gray-body and power-law spectra.  However, in the case of Mars, some of the DIRBE bands 
intercept known atmospheric absorption lines.  A significant CO$_2$ absorption centered at 15~\micron\
lies well within the 12~\micron\ passband, and grazes the blue edge of the 25~\micron\  band.
Numerous minor water vapor absorptions are known to occur at wavelengths sampled by the 60 and 
100~\micron\ bands \citep{sidher/etal:2000}.  
We have used spectral data available from the MGS Thermal Emission Spectrograph (TES) public 
archives~\footnote{\texttt{http://tes.asu.edu/}}
and the ISO spectrum published by \citet{sidher/etal:2000} to form approximate 
correction factors for the DIRBE  data at these bands: only the 12~\micron\ band required significant 
correction. For the 140 and 240~\micron\ bands, color corrections
were calculated using the spectrum predicted by the Wright model.

\begin{figure}
  \begin{center}
    \includegraphics[width=6in]{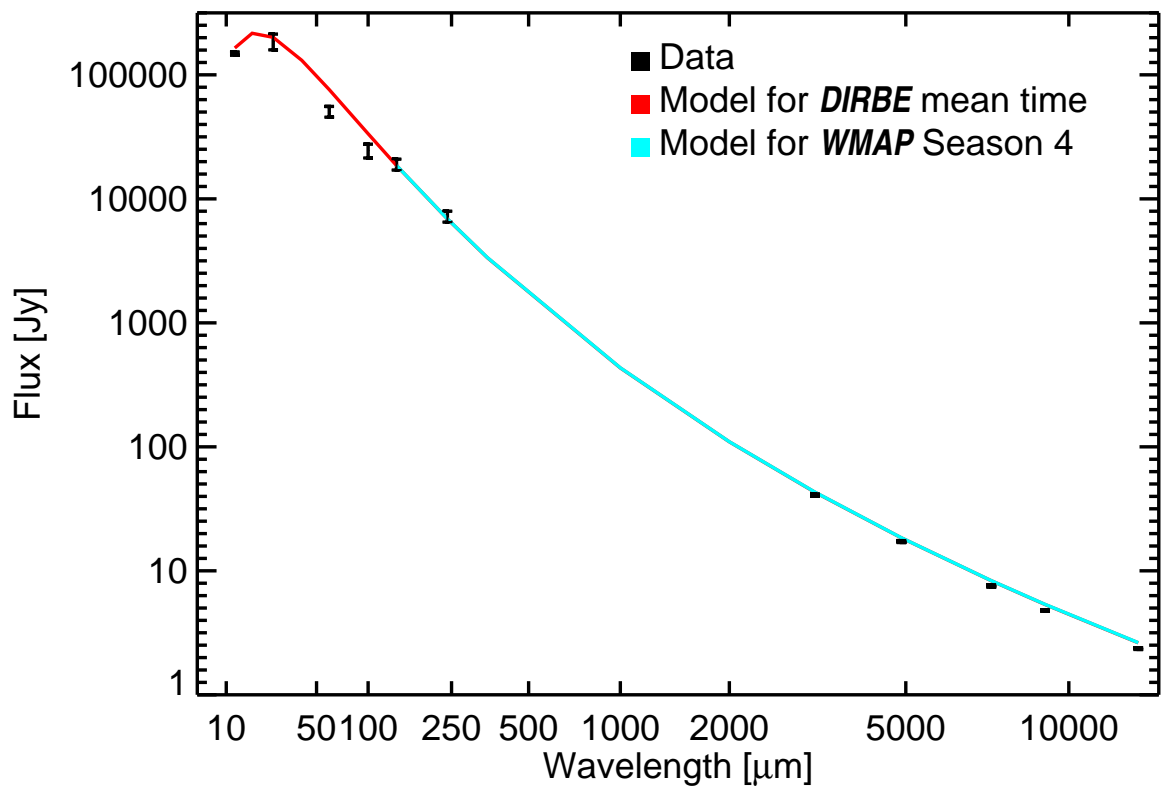}
    \caption{ Mean Mars fluxes derived from DIRBE observations in 1990 and \wmap\ observing 
    season 4 in 2005.  Although separated by several years, the geometrical observing
    configuration is similar between the two dates, and the Wright model predictions 
    (red and cyan lines) for the two time intervals for wavelengths between 140 \micron\ and 1.3~cm overlay
    each other on this scale. DIRBE measurements for wavelengths $\le 100 $ \micron\ do not provide
    good constraints for testing model accuracy.  The \wmap\ data lie a few percent
    below the model, which is significant given the error bars. 
    \label{fig:marsdirbecomp}}
  \end{center}
\end{figure}

A mean flux at each DIRBE effective wavelength was evaluated over the 146 day interval, including 
application of color corrections and scaling to a fiducial solid angle at a distance of 1.5 AU
(Table~\ref{tab:dirbemars}).
Although the epochs of \wmap\ and DIRBE observations are separated by more than 10 years,
the viewing geometry at the time of the DIRBE observations is most closely duplicated
during \wmap\ observing season four.  The \wmap\ mean brightness temperatures for this season were also 
converted to fluxes using the same fiducial solid angle, and both sets of data
plotted as a spectrum in Figure~\ref{fig:marsdirbecomp}.  The spectra predicted by the Wright model
for the mean time of the two epochs are also shown on the plot.  For clarity, the quoted
model errors of about $\pm 5\%$ are not shown on the plot.
On the logarithmic scale needed to show both sets of data together, the two model spectra overlay
each other.  Within the quoted data and model errors, the DIRBE measurements at 60 and 100~\micron\ 
do not match well with the model.  The DIRBE 12~\micron\ data deviate by a little more than 
$1\sigma$, accounting for both data errors and model accuracy.
In the case of the 60 and 100~\micron\ bands, we suspect a non-linear response to a bright object is
the root cause.  The 12~\micron\ band has a low absolute calibration error, but there is a significant
and uncertain color correction due to the CO$_2$ atmospheric absorption.
Ratios of the DIRBE fluxes to Wright model predictions are presented in Table~\ref{tab:dirbemars};
the major source of DIRBE error is that of the absolute calibration, which is also listed.
DIRBE observations at 140 and 240~\micron\ agree within the 10\% calibration errors, but cannot be
used to deduce any trend with frequency, such as seen by \wmap.  We decided to forgo adjustments to
the model at this point.

As currently implemented, the Wright model may be used to predict microwave observations
at \wmap\ frequencies by applying a ``post-processing'' multiplicative scaling factor to the published values:
$T_{\mathrm{new}} = T_{\mathrm{pub}} \times f_{\mathrm{scl}}$.  For both model and data evaluated at 
3.2~mm, we have already noted a scaling factor of 0.953. We use the mean ratios
of the remaining four frequencies to W band (Figure~\ref{fig:marsfreqdep}) as scaling factors
for these frequencies, which are in Column~2 of Table~\ref{tab:marsmodelresid}.  
The W-band observations
can be reproduced to $\sim 0.5\%$, with confidence decreasing with frequency to $\sim 2\%$
at K band.  Since the 3.2~mm model values we have computed are not published, we also provide
scaling factors to the 350~\micron\ temperatures, which are derived from the
the scaling factor of 0.941 noted earlier and multiplied through by the 3.2~mm scaling ratios.

Differences between measurements
and the Wright model at \wmap\ wavelengths are not unexpected. The Wright model 
was designed to predict infrared emission arising from the top layer of the Martian surface, 
whereas the microwave signal originates from subsurface layers of order several centimeters deep 
(\citealt{depater:1990, orton/burgdorf:2003}), with potentially different composition and compaction.
This was recognized by \citet{griffin/etal:1986} and \citet{griffin/orton:1993}, who recommended
a logarithmic interpolation between the Wright 350~\micron\ prediction and the 3.3~mm brightness
temperature versus Mars-Sun distance fit of \citet{ulich:1981}, citing agreement with the Rudy model 
within 1.2\% for a dielectric constant revised upward from 2.25 to 2.8.  On average, the \citet{ulich:1981} fit
predicts brightness temperatures roughly 5\% higher than the \wmap\ mean seasonal values at 3.2~mm.
This suggests that the lower value for the dielectric constant is preferable.  
We obtained disk-averaged Rudy model 
predictions at \wmap\ seasonal mean times of observation using a web-based, slightly revised version of the model
maintained at NRAO
\footnote{ available at \texttt{http://www.aoc.nrao.edu/$\sim$bbutler/work/mars/model/} }. 
In general, Rudy model predictions using the ``standard dielectric constant'' were slightly high compared 
to the \wmap\ observations, by 
$\sim1$\% at 3.2 and 4.9~mm and roughly 3\% at 7.3, 9.1 and 13.1~mm.

It should be noted that the \wmap\ observations provide model constraints only in the disk-averaged sense.
The post-processing scaling of the Wright model suggested here simulates a frequency-dependent emissivity,
which implies slow changes in mean subsurface properties with depth. 
In the Rudy model, which already incorporates subsurface sampling, use of
model parameters derived from the radio data produces agreement with \wmap\ 
within a few percent. 
It is possible that this agreement could be improved either through 
the use of a revised dielectric function or updated thermal inertia and albedo maps.
The thermophysical parameters used by the Rudy model
are based on $\sim$20 year old maps of thermal inertia and albedo from the 
Infrared Thermal Mapper (IRTM) aboard Viking
\citep{muhleman/berge:1991}. More recent MGS maps of these quantities obtained using the TES instrument 
are available with roughly 10 to 40 times
better spatial resolution, full planet coverage and approximately twice the signal-to-noise 
(\citealt{christensenpr:1998, christensenpr/etal:2001}).  The surface albedo has been observed to have
some time dependence:
a visual comparison between the Viking IRTM and TES albedo maps shows alterations in the spatial pattern
have occurred since the Viking mapping \citep{christensenpr/etal:2001}.

\subsection{Saturn}

As with Mars, Saturn's microwave brightness temperature varies considerably over time for
near-Earth observers,
primarily as a result of geometrical effects.  Here the dominant factors are the observer's
changing viewing angle of the rings and projected disk of the oblate planetary spheroid.  

The ring system presents the main difficulty in Saturn's use as a calibrator in the microwave,
although substantial progress has been made in understanding and modeling it.
As evidenced by images from Hubble Space Telescope (HST) and data from orbiter and flyby missions such 
as Cassini and Voyager, Saturn's ring system is complex and rich in detail.  
The brightest rings (A, B, C) are composed of objects typically less than 5~m in size 
\citep{marouf/etal:1983} and
are composed primarily of water ices mixed with a small amount of impurities (\citealt{poulet/etal:2003},
and references therein).  
Those portions of the rings between the observer's line-of-sight and the 
planet serve to attenuate emission from Saturn's disk. The rings also contribute to
the microwave signal via a mix of scattering and thermal re-emission of 
planetary radiation.  
The crossover point between the two emission mechanisms occurs
near 1~cm
\citep{dunn/molnar/fix:2002, schloerb/muhleman/berge:1980, epstein/etal:1980}, with thermal 
emission dominating shortwards and scattering
longwards. Optical depths have been measured at a variety of wavelengths:
\citet{dowling/muhleman/berge:1987}
found that the optical depth of the combined ABC rings is roughly gray over
a wide frequency range.
Voyager I radio occultation observations \citep{tyler/etal:1983} at 3.6 and 13 cm produced
radial profiles of optical depth for the A and C rings and Cassini division, but only
approximate results for portions of the B ring with $\tau > 1$.
The most complex and recent modeling of microwave data is that of 
\citet{dunn/molnar/fix:2002, dunn/etal:2005, dunn/depater/molnar:2007}
who
use a Monte Carlo radiative transfer code to simulate the scattered and thermal emission
of the rings based on VLA and other high-resolution observations.

\begin{figure}
  \begin{center}
    \includegraphics[width=7.0in]{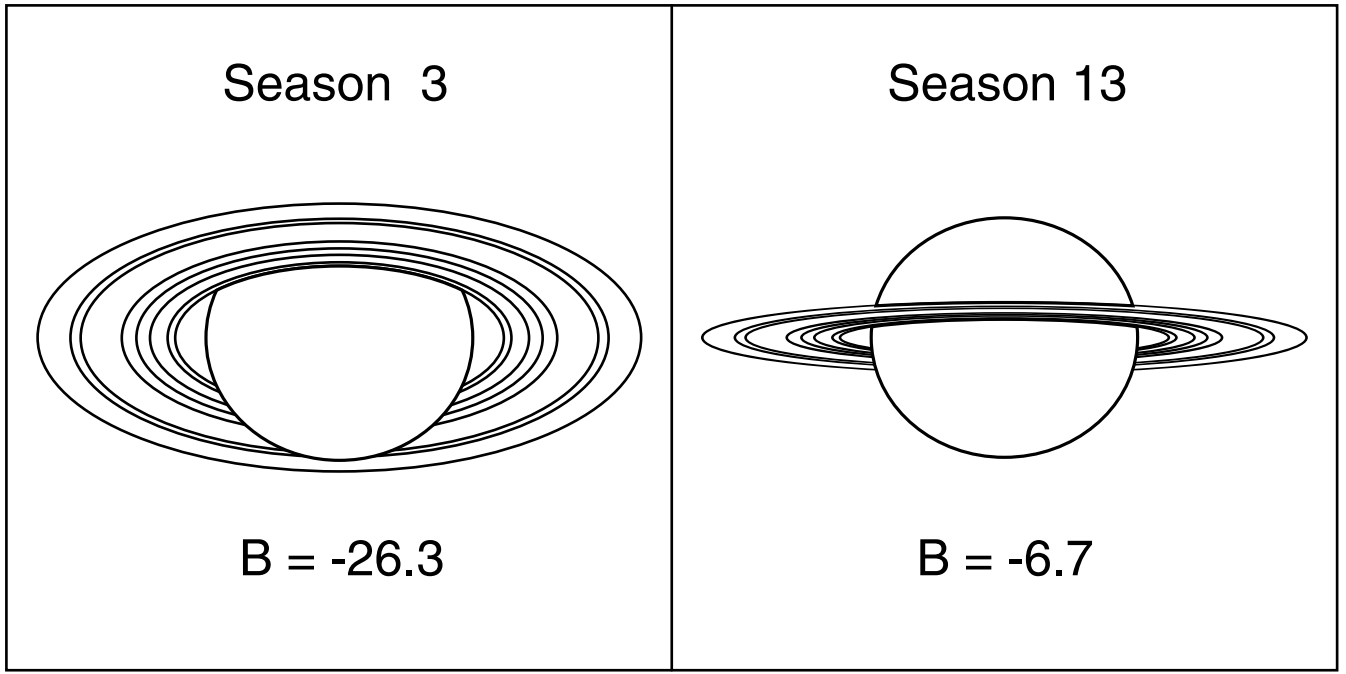}
    \caption{ Extrema of Saturn viewing geometry sampled during the 14 observing
    seasons available in the seven-year \wmap\ data;
    $B$ is the ring opening angle at mid-season.
    The contours illustrate the components used in an empirical model consisting
    of the planetary disk and seven ring regions, each of a different fixed normal optical depth.
    Only the A, B, C and Cassini division rings are represented, and the radial divisions 
    follow those used by \citet{dunn/molnar/fix:2002}.
    \label{fig:satincl}}
  \end{center}
\end{figure}

\wmap\ does not spatially resolve the
individual components of the Saturn system, nor is the spectral coverage sufficient to definitively
isolate ring scattering and thermal
spectral regimes.  However, the changes in viewing geometry with orbital phase
can be used to break
degeneracy between ring and disk components, especially as the viewing aspect of the rings
becomes more ``edge-on''. Over the currently available seven-year observing baseline, 
Saturn's rings are seen at inclinations between $-28\arcdeg$ and $-6\arcdeg$: the sign convention 
indicates that the planet's south pole
is tilting slowly away from the observer as the $0\arcdeg$ edge-on equatorial configuration is approached.  
Figure~\ref{fig:satincl} illustrates the two extrema
and Table~\ref{tab:saturn} lists mean times and ring opening angles for each observing season.

\citet{hill/etal:2009} briefly discussed \wmap\ five-year W-band observations of Saturn, which
sampled ring opening angles $B \le -17\arcdeg$. They found that the brightness temperature could be fit 
remarkably well with a simple $\sin B$ dependence.  This convenient relationship does not effectively 
characterize the behavior at lower inclinations, however, and so a more complex formulation
is needed for the seven-year data. 
We adopt a simple empirical model of Saturn's microwave emission, variants of which have
appeared in the literature for over 30 years (e.g.,  \citealt{ klein/etal:1978, epstein/etal:1980}).
The goal is to provide a predictive formula for the unresolved observed brightness
of Saturn with as low an error as possible: however,  given the \wmap\ data constraints, such a 
model will lack physical detail.  At a given frequency $\nu$, we assume a single temperature 
for the  planetary disk, $T_{\mathrm{disk}}(\nu)$, and that all rings are characterized by the same temperature
$T_{\mathrm{ring}}(\nu)$. The model variant we adopt allows for seven radially concentric ring divisions as 
defined by \citet{dunn/molnar/fix:2002}, which are based on optical depth variations observed by
Voyager \citep{tyler/etal:1983}.  
Each of the seven ring sectors has its own ring-normal optical depth
$\tau_{0,i}$, with $1 \le i \le 7$, but each $\tau_{0,i}$ is assumed to be both constant within its ring and 
frequency independent.  

Those portions of the planetary disk which are obscured by ring cusps
will have their emission attenuated by a factor 
$e^{-\tau_{0,i} |\csc B|}$,
where $B$ is the ring opening angle seen by the observer.  Thus at a given frequency
and ring opening angle $B$:

\begin{equation}
\label{eq:satmodeqn}
T(\nu,B) = T_{\mathrm{disk}}(\nu) [A_{\mathrm{ud}} + \sum_{i=1}^7 e^{-\tau_{0,i}|\csc B|} A_{\mathrm{od},i}] +
T_{\mathrm{ring}}(\nu) \sum_{i=1}^7 A_{r,i},
\end{equation}
where $A_{\mathrm{ud}}$, $A_{\mathrm{od},i}$ and $A_{r,i}$ are the projected areas of the unobscured disk, the portion
of the disk obscured by ring i, and $i^{th}$ ring, respectively.  These areas are normalized to the total 
(obscured+unobscured) disk area.
$T_{\mathrm{ring}}(\nu)$ is an observed brightness temperature and hides such physical details
as ring emissivity and the apportionment of the scattering versus thermal emission.  
The assumption
of a single mean ring temperature is a convenience rather than reality
\citep{spilker/etal:2006, grossman/muhleman/berge:1989, dunn/molnar/fix:2002}. However, the apportionment between 
individual rings is not critical to the model fit, as it only depends on the summed total ring emission.   
Potential contributions to modeling error 
from nonuniform planetary disk emission are discussed later in this section.

The \wmap\ brightness temperatures in Table~\ref{tab:saturn} are sorted by frequency band and $B$ and then
a single
simultaneous fit for the model parameters is made.  Two separate fits were tried, differentiated
from one another in the handling of the ring optical depth.  In one fit, the $\tau_{0,i}$ were
held fixed at the values chosen by \citet{dunn/molnar/fix:2002} (Table~\ref{tab:satmodel1}), and the ten disk and ring 
temperatures (for each of 10 DAs) were
solved for.  In an alternate fit, the relative ratios between the $\tau_{0,i}$ were fixed as per the
table, but a single $\tau_{0,\mathrm{max}}$ was solved for in addition to the ten free temperatures.  Both fits
returned similar results, with reduced $\chi^2$ of $\sim1.2$ for $\sim100$ degrees of freedom.
We show only the results for the fixed-$\tau$ fit.

\begin{figure}
  \begin{center}
  \includegraphics[height=6.5in]{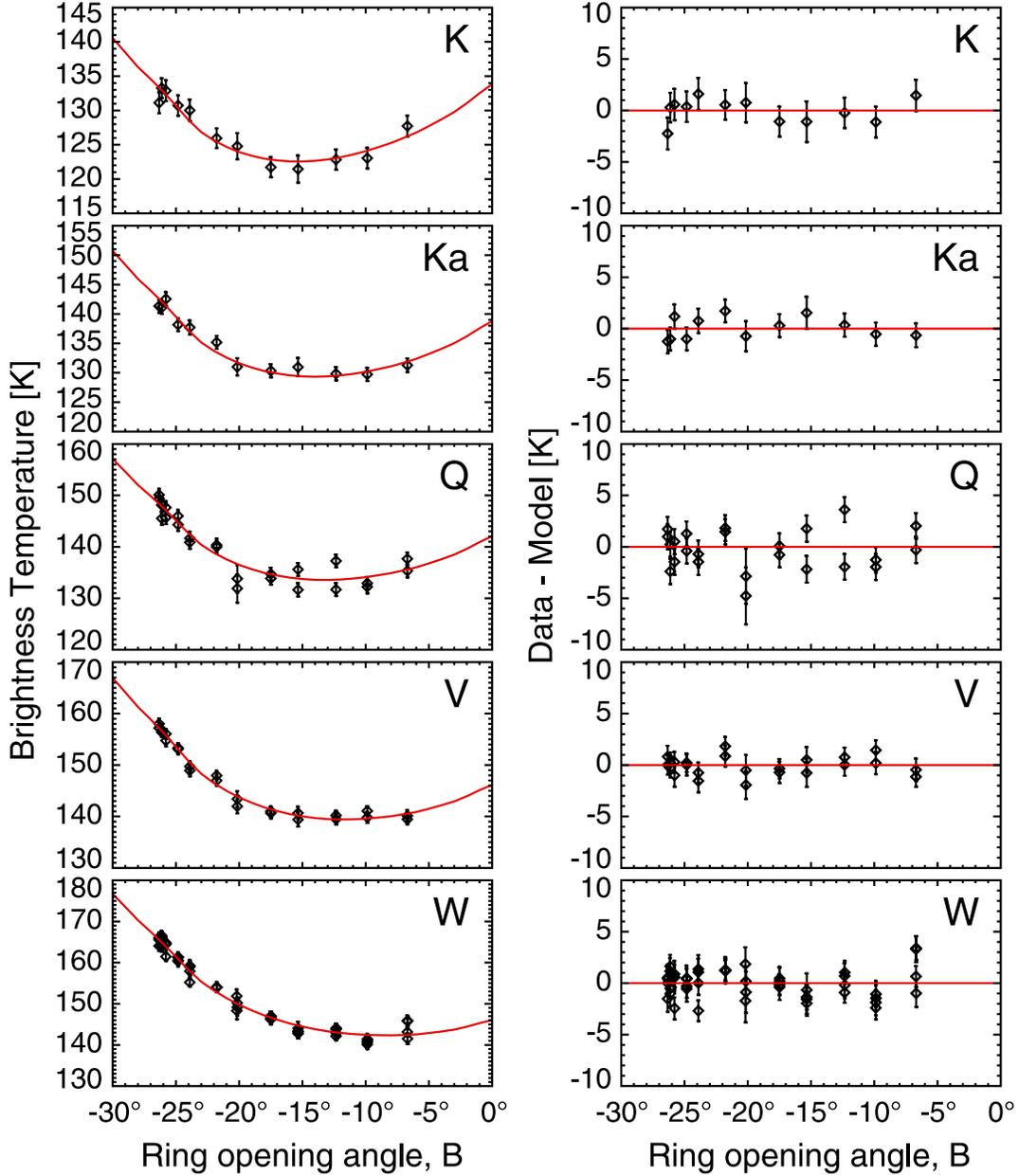}
    \caption{ Modeling results for Saturn. Left: brightness temperatures based on unresolved Saturn observations
    as a function of ring inclination $B$ are shown in black for each \wmap\ frequency band.  Where there are
    multiple differencing assemblies per frequency, multiple points are plotted at each inclination.
    An empirical model including both ring and disk components (see the text) is overplotted in red.  
    The temperature of the planetary disk predicted
    by the model occurs at $B$=0$\arcdeg$, when the rings are viewed edge-on. The model is symmetric about
    $B$=0$\arcdeg$.
    Right: residuals (data-model).  The model
    predicts the observed brightness temperature within $\approx$3\%.
    \label{fig:satmod}}
  \end{center}
\end{figure}

\begin{figure}
  \begin{center}
    \includegraphics[width=6.5in]{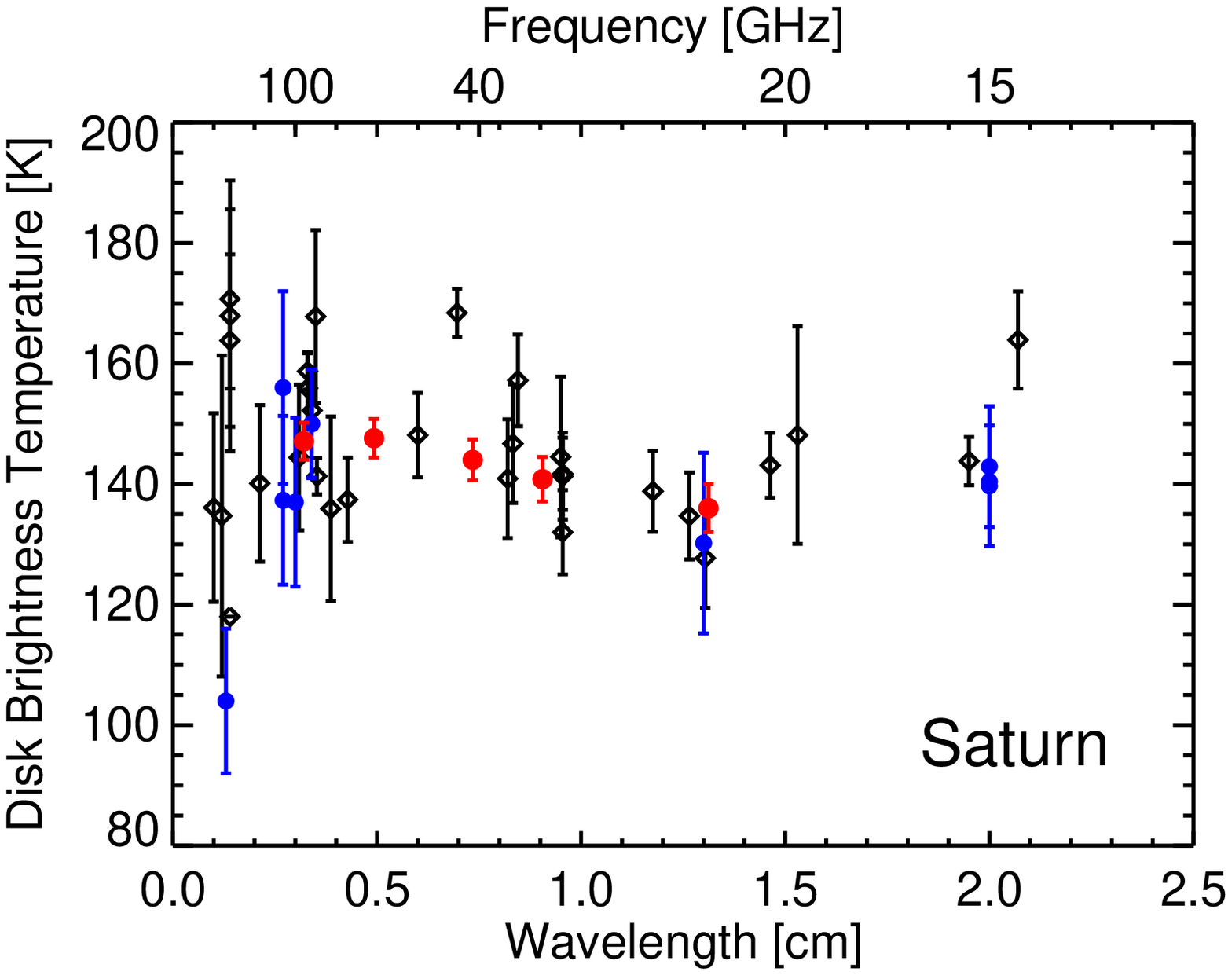}
    \caption{ Spectrum of Saturn's disk. Black diamonds are from \citet{klein/gulkis:1978}, consisting of 
    a compendium of
    unresolved observations which were corrected for a ring contribution.  Filled blue circles are 
    interferometric observations
    from \citet{grossman/muhleman/berge:1989}, \citet{depater/dickel:1991} and \citet{dunn/etal:2005}.  
    Red circles are disk temperatures derived from \wmap\ data
    using an empirical model to separate disk and ring contributions. \wmap\ disk temperatures are 
    corrected to absolute brightness.  The \wmap\
    values agree with the interferometric observations within the uncertainties,  while providing
    tighter constraints at multiple frequencies.
    Uncertainties in the \wmap-derived  disk temperatures 
    are expected to decrease when observations at ring inclinations near $0\arcdeg$ become available.
    Ammonia absorption near 24~GHz is less pronounced for Saturn than for Jupiter, making
    tighter constraints useful. 
    \label{fig:satspec}}
  \end{center}
\end{figure}

\begin{figure}
  \begin{center}
    \includegraphics[width=6.5in]{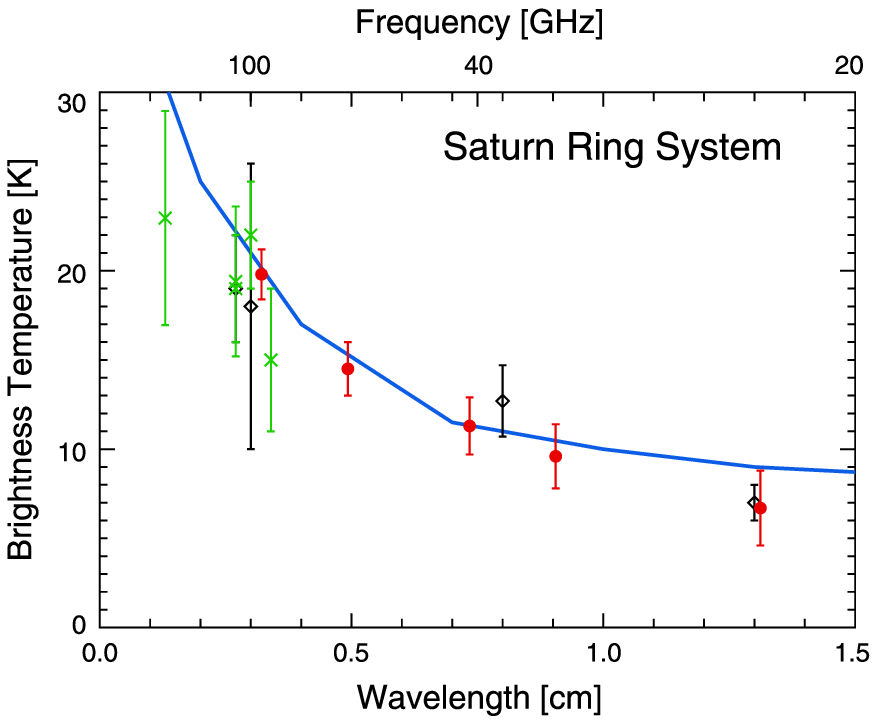}
    \caption{Brightness temperature of Saturn's combined rings. Interferometric measurements taken from 
    Table 4 of \citet{dunn/etal:2005} 
    are shown in green.  Black points are older observations from \citet{janssen/olsen:1978} and 
    \citet{schloerb/muhleman/berge:1980}.  Model predictions from Figure 7 of
    \citet{dunn/etal:2005} are plotted as the blue line. 
    The derived \wmap\ ring temperatures, shown in red, compare well with the observations of others and 
    the Dunn et al. model.
    Results from \citet{dunn/molnar/fix:2002} indicate that the rings primarily scatter rather
    than absorb CMB radiation; thus ring temperatures have not been corrected to
    absolute brightness as is necessary for the disk temperatures.
    \label{fig:satringspec}}
  \end{center}
\end{figure}

\begin{figure}
  \begin{center}
    \includegraphics[width=7.0in]{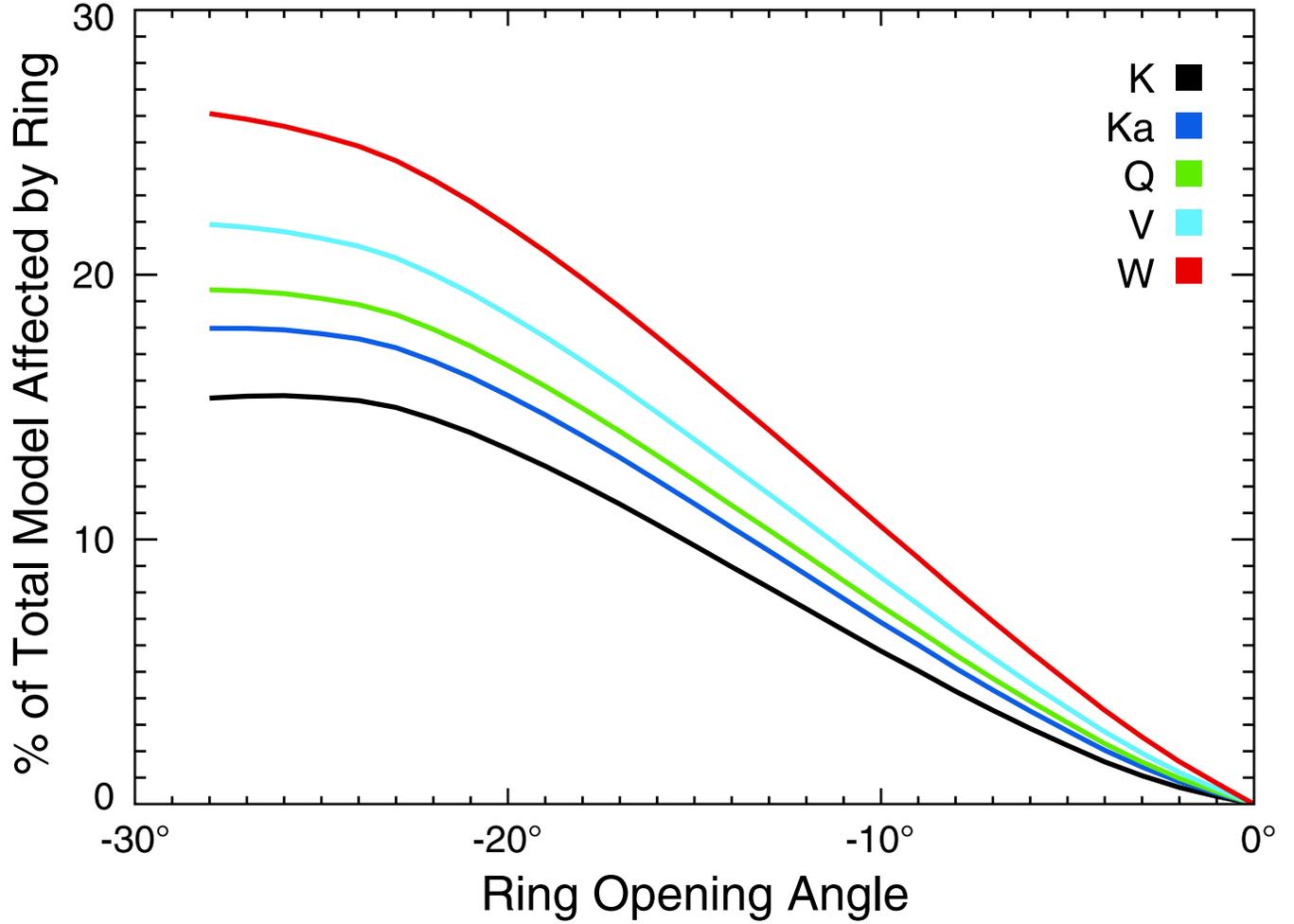}
    \caption{ Contribution of ring emission plus ring-attenuated disk emission to the
    total computed brightness temperature of Saturn, expressed as a percentage.
    Exact values are model dependent; this set is for the empirical model 
    discussed in the text, and whose parameters are given in Tables~\ref{tab:satmodel1} and
    \ref{tab:satmodel2}.
    The ring system contribution is lowest in K band, where thermal emission is
    dropping and scattering is more dominant.
    \label{fig:satringaffect}}
  \end{center}
\end{figure}

The disk and ring brightness temperatures derived from this model fit are presented in 
Table~\ref{tab:satmodel2}.  
Formal errors for the ring and disk temperatures at lower frequencies are higher because 
the ring contribution is smaller and there is a larger 
covariance between the two components.   
The model fit and residuals are shown in Figure~\ref{fig:satmod}.  The model
reproduces the observations to within $\sim3\%$, but observations closer to $B=0\arcdeg$ would better
constrain the disk temperatures and so reduce the covariance between the disk and rings.  
The parameterization chosen for the ring optical depths, together with the incomplete sampling of the 
ring system orientation,
allows for alternate models which would fit the data equally well but return somewhat 
different mean disk and ring temperatures.
This systematic is not reflected in the formal fitting errors in Table~\ref{tab:satmodel2}.
In order to assess the magnitude of this error, we explored a small number of
model variants.
The model which fit the data equally well but returned temperatures with the largest
differences compared to Table~\ref{tab:satmodel2} consisted of a combined ABC ring represented by
a single $\tau_{0}$, which was a free parameter in the fit.  Disk temperatures derived for this model 
were $\sim2.5$~K higher, and ring
temperatures $\sim1$~K lower.  As a conservative approach, we additively combine
this estimated systematic error with the formal fitting errors to produce the
``adopted error'' columns in Table~\ref{tab:satmodel2}. 
The derived disk and ring temperatures
match well with those already in the literature.  Figure~\ref{fig:satspec} shows the 
\wmap\ derived disk temperatures in context with the compendium of \citet{klein/etal:1978} plus
more recent interferometric observations.  The ring temperatures also compare well with those 
measured directly from high-resolution images of the ring ansae \citep{dunn/etal:2005,
schloerb/muhleman/berge:1980, janssen/olsen:1978}: see Figure~\ref{fig:satringspec}.
Figure~\ref{fig:satringaffect} illustrates the percentage contribution of the ring to our total model, which sums 
the ring emission and attenuation of disk radiation, as a function of frequency and ring inclination $B$.

The empirical model described here assumes Saturn's mean whole-disk temperature is
time invariant at each frequency.  
The presence
of band-like structures (which can persist for years) and latitudinal brightness gradients at the $\sim5$\% 
level have been noted by several observers at wavelengths of 2~cm and longward
\citep{grossman/muhleman/berge:1989, depater/dickel:1991, vandertak/etal:1999}.
There is less discussion in the literature of disk temperature structure and/or variability at
the frequencies observed by \wmap: there have been no reports of bands or localized
structures, but some indication of north/south latitudinal brightness differences.
Within the seven-year epoch under consideration,
\citet{dunn/etal:2005} observed a latitudinal disk brightness gradient at 3~mm, such that the
north pole was brighter than the south by $\sim5$\%.  They interpreted differences between their 1.3~mm and 3~mm data,
taken roughly 4 months apart, as possible evidence for atmospheric changes on week to month timescales.   
In order to estimate potential model bias caused by the assumption of a uniform disk temperature, Saturn seasonal 
brightness temperatures 
were simulated using geometrical models similar to those described in this section, with the 
addition of a disk latitudinal gradient as described by \citet{dunn/etal:2005}. If one makes the assumption that such 
a temperature gradient is persistent over seven years and exists at all \wmap\ frequencies, then such structure
would produce a temperature behavior with inclination which is asymmetric about $B=0$\arcdeg.
This is because the rings
would obscure hotter regions in the northern portion of the disk but colder regions in the south,  
producing a maximum deviation from symmetry of $\sim1$\% at 
the largest ring opening angles.  Since the \wmap\ seven-year observations
sample slightly less than half of the full $B$ range, this asymmetry would not be evident, and 
a model fit which assumes a constant disk temperature would use the ring model component to compensate for 
the disk temperature gradient. For a simulated dataset matching the \wmap\ seven-year observing seasons,
the model fit returns a mean ring temperature biased by less than 1~K, with an unbiased 
recovery of the mean whole-disk temperature.  This estimate is of uncertain quality, however, because
it is based on a snapshot of Saturn at one wavelength and a narrow time window.
For this reason,
the adopted errors in Table~\ref{tab:satmodel2} have not been adjusted to reflect this potential effect.

An extended, low optical depth dust ring associated with Saturn's moon Phoebe was recently
discovered using Spitzer imaging at 24 and 70~\micron\ \citep{verbiscer/skrutskie/hamilton:2009}.  With a quoted
radial extent between at least 128 and 207 Saturn radii, or roughly $1\arcdeg$ in apparent diameter, it would
be possible for \wmap\ to resolve the Phoebe ring, with W band as the best chance for doing so.
However, no signal from the ring was detected
by \wmap. Properties of the grains are not well specified, but naive estimates
assuming typical dust emission and optical depth frequency dependencies would predict temperatures 
at 94~GHz of less than 1 $\mu$K, well below achievable noise levels of about 300 $\mu$K ($1\sigma$).

\subsection{Uranus and Neptune}
With their lower brightness, lack of a dominant ring system, and small solid angle,
Uranus and Neptune are used as primary calibrators in infrared through
radio wavebands for both ground- and space-based instruments.  Spectral coverage in the 
microwave is somewhat undersampled, however, leaving room for interpretation in atmospheric 
modeling efforts.  The microwave spectra of Uranus and Neptune lack the broad
NH$_{3}$ absorption centered near 24~GHz which is characteristic of Jupiter and Saturn.
At millimeter wavelengths, collision-induced absorption by H$_{2}$ is considered the dominant
continuum opacity source. CO rotational absorptions have been observed for Neptune in the sub-millimeter
\citep{marten/etal:2005}.  At centimeter wavelengths, ammonia (which is depleted relative
to solar nitrogen levels) and hydrogen sulfide are the main opacity contributors
\citep{deboer/steffes:1996, spilker:1995, depater/romani/atreya:1991}. 
Of special note is Uranus' unique $98\arcdeg$
obliquity, which allows for a slowly changing pole-to-pole panorama of the planet as
it moves in its 84 year orbit, and may also play a role in determining atmospheric 
conditions.

For selected frequencies, observational databases spanning decades
have permitted characterization of the variability of 
these two planets.  Uranus in particular has exhibited long-term whole-disk temperature 
changes since 1966 which correlate with the viewing aspect of the south pole.  
After accounting for geometrical solid angle changes, \citet{kramer/moreno/greve:2008} reported 
a gradual temperature drop of order $10\%$ over 20 years (1985-2005) at 90~GHz. 
This is interpreted as a true integrated-disk temperature change, as the 1985 face-on contribution 
from the bright south pole progressively morphs into a 2007 view dominated by colder equatorial regions. 
Similar findings, with similar phasing albeit different amplitudes, have been reported 
in the radio \citep{klein/hofstadter:2006} and visible \citep{hammel/lockwood:2007}.  There is weaker evidence 
that some portion of variability in the light curves is attributable to 
changes deep in the atmosphere \citep{klein/hofstadter:2006}; \citet{hofstadter/butler:2003} used 
VLA 2 and 6~cm ``snapshots'' of the disk over several years to argue for opacity changes in zonal bands.
Neptune's microwave variability is less well documented, although there is ample evidence
for variability in the visible and near-IR, as summarized by \citet{hammel/lockwood:2007}.  
\citet{kramer/moreno/greve:2008} find the 90~GHz integrated disk temperature to be constant to 
within $\sim8\%$ over 20 years.

Peak \wmap\ antenna temperatures for these distant ``ice giants'' range from roughly 1~mK
at W band to $\sim$0.2 mK at K band.  The signal-to-noise for individual observations of  
these objects is low ($\sim$0.3 at W band, $\sim$0.14 at K band), resulting in large statistical errors
in single-season disk temperature determinations.  Tables~\ref{tab:uran} and \ref{tab:nept}
present the
single-season brightness temperatures computed for Uranus and Neptune.
Brightness temperatures are listed per frequency rather than per DA as a means
of boosting signal-to-noise for those frequencies with multiple DAs. 
Temperatures for Uranus observing seasons four and five exhibit somewhat larger error bars than 
other seasons. For these two seasons, there was a reduction in the number of observations available 
for analysis because
data quality checks excluded observations in close proximity to Mars sky coordinates.

Sub-\wmap\ latitudes for Uranus range between $-30\arcdeg$ and $4\arcdeg$ over our seven-year baseline.  
Linear correlations of $T_{b}$ against both sub-\wmap\ latitude and time
produced no statistically significant trend other than a flat line.  This is not surprising 
given that the tightest seasonal temperature errors are of order 10\%, which encompasses the 
entire twenty-year
range of variation seen in the 90~GHz light curves of \citet{kramer/moreno/greve:2008}.
We performed similar correlations for Neptune, with again the same null variability result.  Note,
however, that the sub-\wmap\ latitude for this planet is relatively unchanged over the epochs of
observation.  
With no discernable variability over the observing baseline, we computed seven-year
means of the disk brightness for each planet; these are listed as the last line
in Tables~\ref{tab:uran} and \ref{tab:nept} .  These seven-year means
in turn may be compared to observations in the literature and placed in context
with the microwave spectra in general. 

\begin{figure}
  \begin{center}
    \includegraphics[height=4.5in]{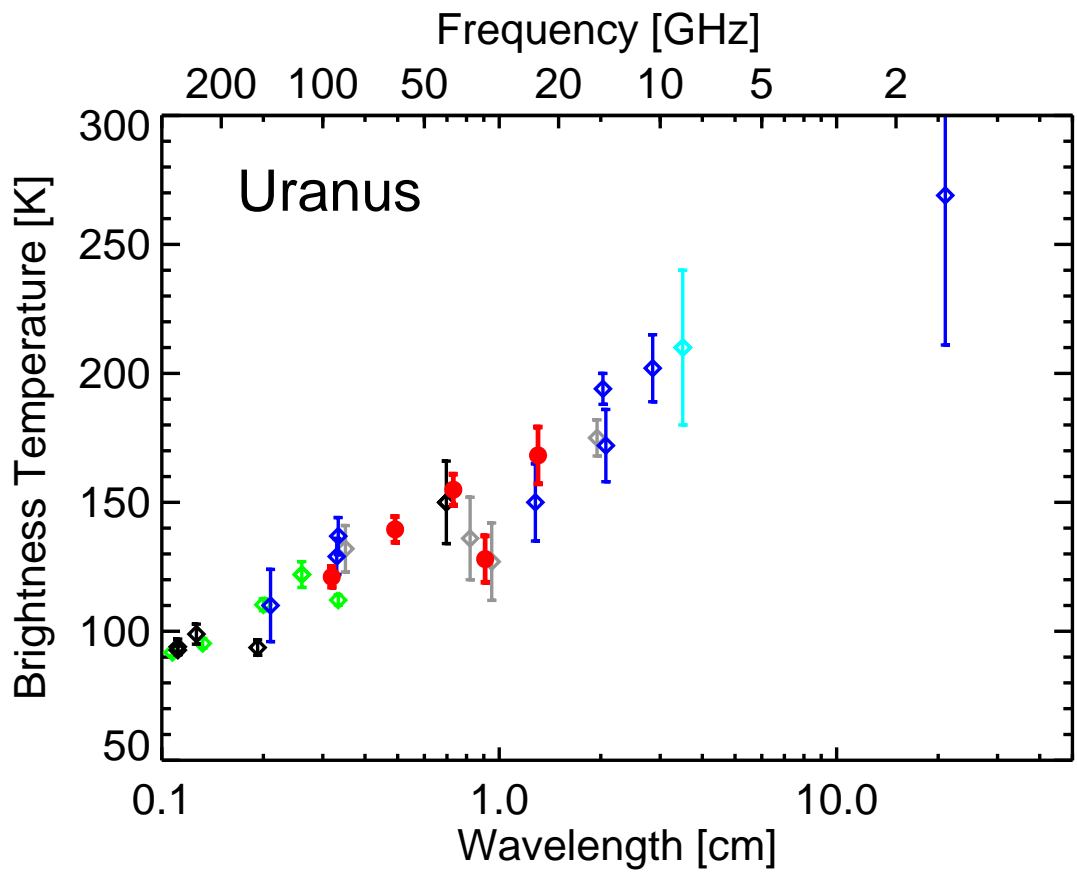}
    \caption{ Microwave spectrum of Uranus, compiled from
      selected observations in the literature.  Points
      have been color-coded into wide bins of observational
      epoch: gray for 1966-1969, blue for 1970-1979, green for 1980-1989, black
      for 1990-1999, red for \wmap\ points (2001-2008) and
      cyan for the 1966-2002 mean and peak-to-peak
      variability observed at 3.5 cm by Klein \& Hofstadter
      (2006). See the text for detailed data references.
      \wmap\ data have been corrected to absolute brightness.
      Of potential atmospheric modeling interest is the ``dip'' near
      30~GHz. 
      \label{fig:uraspec}}
  \end{center}
\end{figure}

Figure~\ref{fig:uraspec} overplots the seven-year mean \wmap\ Uranus temperatures at K, Ka, Q, V and W 
center frequencies
on a composite spectrum derived from several sources in the literature.
Points are color-coded to reflect the decade in which observations were taken.  Data
have been culled from \citet{griffin/orton:1993}, \citet{muhleman/berge:1991}, \citet{greve/etal:1994},
\citet{cunningham/etal:1981}, and \citet{gulkis/janssen/olsen:1978}.  
Also included is the mean 3.5~cm value from 
the 1966-2002 compilation of \citet{klein/hofstadter:2006}, with a single large error bar indicating
the $\sim30$~K peak-to-peak variation in the light curve.
Within the quoted errors, there is general agreement between the \wmap\ observations and
those taken at previous epochs.  The \wmap\ observations show a ``dip'' in Ka band temperature
compared to neighboring frequencies.  Although not of high statistical significance,
a feature of this nature is clearly of interest.  A literature search for corroborating observations near 
30~GHz only returned those few acquired in the late 1960s, listed in Table~1 of
\citet{gulkis/janssen/olsen:1978}.  For a variety of reasons, more recent observations would
be preferable, but the older epoch does have the advantage of sharing a similar viewing geometry 
to that of the \wmap\ data.
These older observations are plotted in gray in
the Figure, and agree well with the \wmap\ measurements, but also do not define a ``dip'' at
high statistical significance.  
Although suggestive of an interesting atmospheric opacity constraint for future study, 
the possibility of such a feature has gone unremarked in the literature.

\begin{figure}
  \begin{center}
    \includegraphics[height=4.5in]{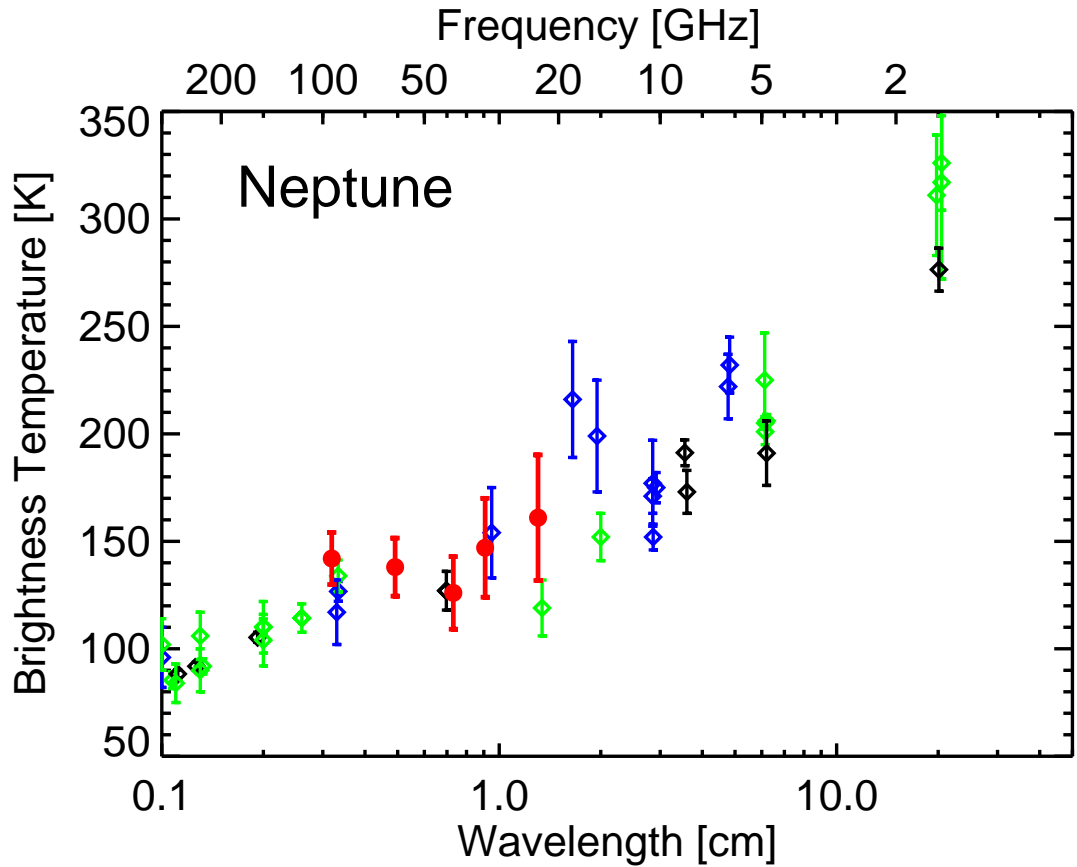}
    \caption{ Microwave spectrum of Neptune, compiled from
      selected observations in the literature (see the text).  Points
      have been color-coded into wide bins of observational
      epoch: blue for 1969-1979, green for 1980-1989, black
      for 1990-1999, and red for \wmap\ points (2001-2008).
      \wmap\ data have been corrected to absolute brightness, and
      in general agree well with previously published values.
      \label{fig:nepspec}}
  \end{center}
\end{figure}

A composite microwave spectrum for Neptune is shown in Figure~\ref{fig:nepspec}.  The color coding
again indicates observational epoch, with ground-based data taken from 
\citet{depater/richmond:1989}, \citet{depater/romani/atreya:1991}, \citet{muhleman/berge:1991}, \citet{griffin/orton:1993}, 
\citet{hofstadter:1993}, and \citet{greve/etal:1994}.
\citet{deboer/steffes:1996} provide a summary of observations in the 1990s.  Observations with
error bars larger than about 30~K are not included in the plot.
\wmap-derived temperatures for Q band agree well with those of \citet{greve/etal:1994}, while the
94~GHz observations lie on the upper envelope of data taken in the 1970s and 1980s.
\wmap\ observations occur when Neptune's south pole is most fully viewable.
The observational error on the seven-year W-band mean is $\sim8\%$, which is equivalent to the
level to which \citet{klein/hofstadter:2006} claimed 20 year stability.

Reasonable temperature error bars for the \wmap\ Uranus and Neptune observations are
only achieved through averaging the seven-year data.  Observers wishing to use
these data for calibration of observations taken outside of the 2001-2008 epoch should keep 
in mind the variability and related geometric issues described above.

\section{Celestial Sources\label{sec:celsrc}}

Non-variable, spatially isolated fixed celestial calibrators for millimeter
wavelengths are not common. At high Galactic latitude, 
bright sources are predominantly identified as some form of AGN/QSO, which are prone to
outbursts and variability on a wide range of timescales and frequencies.  Brighter sources in 
the Galactic plane tend to be HII regions (which may not be ``point-like'') or 
supernova remnants (SNRs) with
potential variability.  At moderate spatial resolution, confusion with neighboring
diffuse and compact sources in the Galactic plane can be an issue for
background subtraction.  

Five sources were chosen for study out of an original list which included
some of the brightest sources from \citet{baars/etal:1977} and  \citet{ott/etal:1994}, 
in addition to the brightest, least variable
objects from the seven-year \wmap\ source catalog \citep{gold/etal:prep}. 
The five selected sources are listed in Table~\ref{tab:calobj}.  Some of the sources which 
were initially considered but ultimately rejected, primarily because of low
background contrast, included 3C286, NGC 7027, 3C84, 3C218, 3C123, and 3C147.
There are unfortunately few suitable calibration sources in the Southern hemisphere
with a long-term history of observation.

\subsection{Data Processing and Analysis Methods\label{sec:calsrcmethods}}

Flux densities of the selected sources are measured from
the seven-year sky maps at HEALPix \footnote{\wmap\ sky map products are in HEALPix format; 
see \texttt{http://healpix.jpl.nasa.gov/}.
HEALPix divides the sky into pixels of equal area.  The number and spacing between
pixel centers depends on the chosen resolution.  Nside=512 corresponds to
a pixel area of $\sim3.995 \times 10^{-6}$ sr, with a mean spacing of 6.87 arcmin
between pixel centers.}
resolution 9 (Nside=512).  
For each frequency band, the
azimuthally symmetrized beam profile (see Section~\ref{sec:methods}) is convolved with 
a sky map pixel to produce a map-based beam template.
The sum of the pixel-convolved beam template plus a sloping planar base level is fit
to the Stokes I, Q, and U sky map data at each source position, using pixels
within 3.5 times the beam width $\sigma$ (1.5 times the FWHM) in each band.
The peak source temperature from each fit is converted from thermodynamic
temperature to Rayleigh-Jeans brightness temperature and then translated to 
a source flux density
using a conversion factor $\Gamma$ that is a weak function of the source
spectral index \citep{jarosik/etal:prep}.  
For point sources, this method of flux measurement is more accurate than
the aperture photometry method as used for example by \citet{page/etal:2003}
for Tau~A.  The \wmap\ beam profiles have extended wings \citep{hill/etal:2009},
so an aperture radius $\gtrsim 3$ times the beam FWHM should be used and the
results are more susceptible to error due to background confusion.

The uncertainty in flux density is calculated as the quadrature sum of 
(1) map measurement uncertainty, (2) the uncertainty in $\Gamma$ 
(0.5\% to 0.7\% depending on the band), and (3) the 0.2\% absolute calibration
uncertainty.  For Stokes I, map measurement uncertainty is estimated from
the rms fit residual in the fit region.  The Stokes I residuals generally
appear to be dominated by beam asymmetry effects, but also include background
confusion and noise. (The source profiles in the maps are asymmetric due to
asymmetry in the instantaneous beam profiles and the nonuniform distribution
of scan angles over a year for sources away from the ecliptic poles.)
For Stokes Q and U, map measurement uncertainty is calculated either from
the 1$\sigma$ source peak uncertainty and base level uncertainty
from the beam fitting or as the Q or U flux times the fractional map
measurement uncertainty for I, whichever is largest.  The latter method is
the estimated uncertainty in Q or U due to beam asymmetry effects, assuming
the fractional uncertainty is the same in Q or U as it is in I.
The flux determination was tested on simulated sky maps containing
a population of point sources with no other signals and no noise, generated
using \wmap\ beam window functions as described in \citet{wright/etal:2009}.
Recovered flux densities were accurate to about 0.1\% or better, except in W
band where the recovered fluxes tended to be larger than the input fluxes
by up to 1\%.  This is allowed for by including an additional 1\%
uncertainty term in the quadrature sum for W band.

Fractional year-to-year variability for the selected sources has been obtained
using sky maps for individual years 1-7.  To remove confusion noise from the
CMB and Galactic foregrounds, we subtract the seven-year average map from 
each individual year map for each band.  A pixel-convolved beam plus flat
base level is fit to each difference map at each source position, giving
a flux difference $\Delta F_i$ for the $i$th year for each source in each band.
Uncertainty in $\Delta F_i$ is calculated from the quadrature sum of the source
peak uncertainty and the base level uncertainty.  The flux difference is divided
by the mean flux from the seven-year map to get the fractional flux variation
 $\Delta F_i/\langle F \rangle$.  For K, Ka, and Q bands, there
is a small ($\lesssim 0.2\%$) but significant year-to-year variation in the \wmap\ calibration,
which we have measured by correlating each yearly map against the seven-year map
(see Figure~1 of \citealt{jarosik/etal:prep}).  The measured fractional
flux variations in K-Q bands are corrected for these calibration variations $c_i$ using
$(\Delta F_i/\langle F \rangle)_{corrected} = (\Delta F_i/\langle F \rangle)_{measured}
 + (1 - c_i)$. 

\subsection{Results}
Source flux densities from the seven-year maps are presented in Table~\ref{tab:srcflux}.  
Fractional
uncertainties for the Stokes I fluxes are typically 1 to 3\%.  For some of the sources,
the maximum source extent given in Table~\ref{tab:calobj} is not entirely 
negligible relative to
the \wmap\ beam width in V or W band 
(FWHM 19.6\arcmin\ in V, FWHM 12.2\arcmin\ in W, \citealt{hinshaw/etal:2009}).  We
have estimated the possible error due to source extent for Tau~A.  The 1.4~GHz map
of the Tau~A region from the NVSS survey (beam size 45\arcsec\ FWHM, \citealt{condon/etal:1998})
was smoothed with the symmetrized \wmap\ V or W-band beam and converted to a resolution 9 
(Nside = 512) HEALPix map.
The flux determined by our method was found to underestimate the true flux by 1.5\% in 
V band and 3.7\% in W band. Spatial variations of the spectral index over Tau~A are 
very small
\citep{morsi/reich:1987, bietenholz/etal:1997, green/etal:2004} so the source extent 
at W band is probably
similar to that at 1.4~GHz.  Our V and W-band fluxes for Cas~A and 3C58 may also be
underestimated by similar amounts.  To allow for this, the uncertainty values listed for
these three sources in Table~\ref{tab:srcflux} include additional contributions of 
1.5\% in V and 3.7\% in W,
added in quadrature to the uncertainties calculated as described in 
Section~\ref{sec:calsrcmethods}.

\begin{figure}
  \begin{center}
    \includegraphics[width=6.5in]{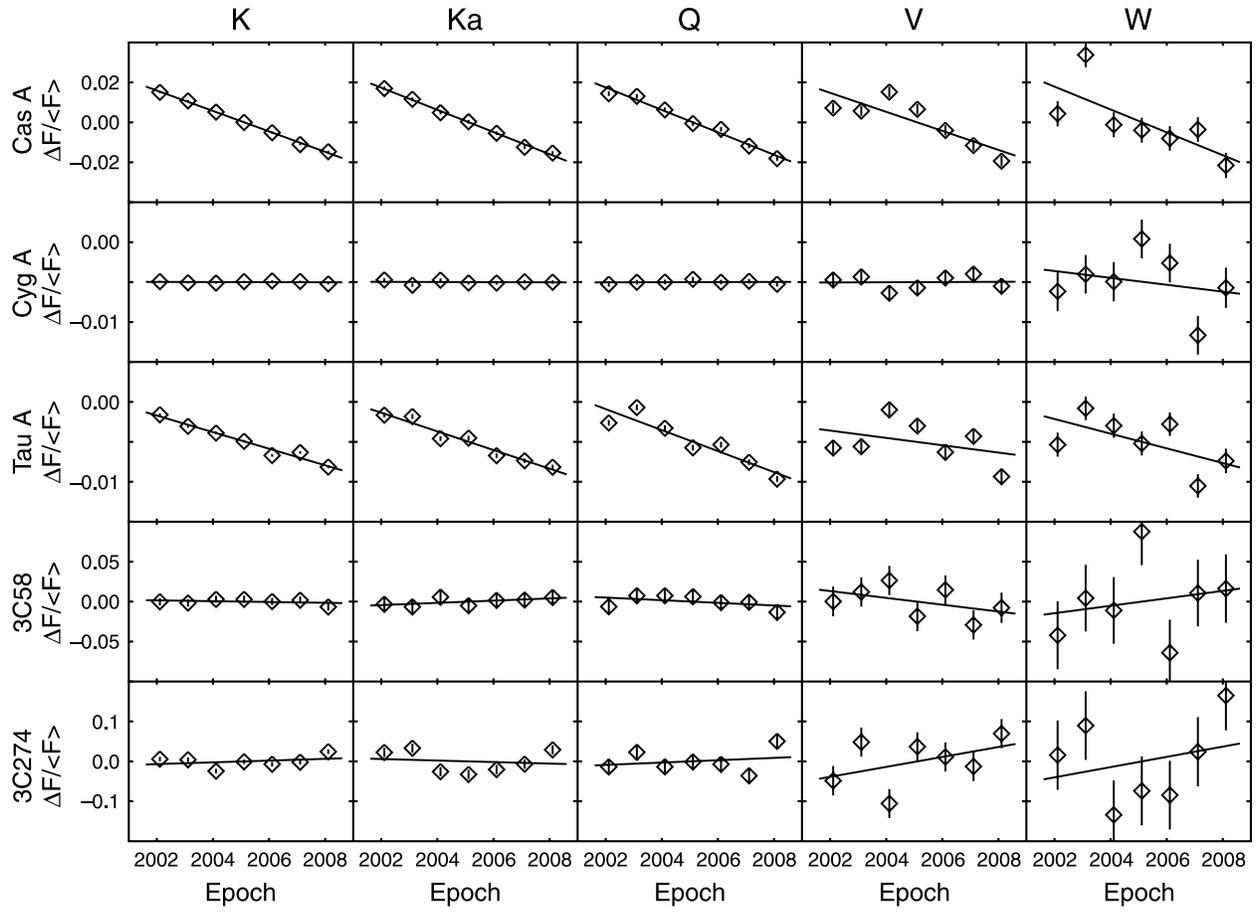}
    \caption{ Fractional yearly flux variation measured for the celestial calibrators over
            the \wmap\ mission.  Parameters of the linear fits are listed in 
	    Table~\ref{tab:calvar}.
      \label{fig:npovarplot}}
  \end{center}
\end{figure}

Fractional year-to-year variability results are presented in Figure~\ref{fig:npovarplot} and 
Table~\ref{tab:calvar}.
Significant secular decrease is seen for Cas~A and Tau~A.  The results are consistent
with a frequency independent decrease of about 0.53\% per year for Cas A and
0.22\% per year for Tau~A.  Our results for Cas~A fall between the $\sim0.6\%$ per year
decrease found by \citet{o'sullivan/green:1999} near 15~GHz from 1965 to 1995 and
the $0.394\% \pm 0.019\%$
decrease found by \citet{hafez/etal:2008} at 33~GHz from 2001 to 2004.
Dependence of the Cas~A decrease on frequency and epoch has recently been
discussed by \citet{reichart/stephens:2000} and \citet{hafez/etal:2008}.  For Tau~A, we
find a somewhat higher rate of decrease than that of $0.18\% \pm 0.01\%$ per year
found by \citet{vinyaikin/razin:1979a, vinyaikin/razin:1979b} at 927~MHz from 
1962 to 1977 and $0.167\% \pm 0.015\%$ per year
found by \citet{aller/reynolds:1985} at 8~GHz from 1968 to 1984. \citet{hafez/etal:2008}
found a decrease of $0.22\% \pm 0.07\%$ per year at 30~GHz from 2002 January to 2004 September.

No significant secular
trends are seen for the other sources, with $3\sigma$ upper limits in K band of 0.12\%
per year for Cyg~A, 0.21\% per year for 3C58, and 0.54\% per year for 3C274.  The limit
for Cyg~A is consistent with the limit of 0.10\% per year found by \citet{hafez/etal:2008}
at 33~GHz for the period 2001 March to 2004 May.  For 3C58, \citet{green:1987} reported
an increase of $0.32\% \pm 0.13\%$ per year at 408 MHz from 1967 to 1986.

Significant yearly flux variation is not seen for Cyg~A and 3C58; the rms year-to-year
variation is consistent with the uncertainties in each band.  The lowest rms variation
is in K band, and is 0.27\% for Cyg A and 0.33\% for 3C58.  \citet{carilli/barthel:1996}
give an upper limit of 10\% on Cyg A core variability from observations at 5, 15, and 90 GHz
over timescales from 10 months to 15 years.  The core contributes $\leq 10$\% of the total
flux at \wmap\ frequencies (e.g., \citealt{robson/etal:1998}).  For 3C274 there is evidence
for year-to-year variability of about 2\% in K, Ka, and Q bands.  The rms variation is 1.8 to 2.6
times the mean uncertainty in these bands, and the variations are correlated from band to band.
Previous observations of 3C274 have shown greater variability.  At 90~GHz, \citet{steppe/etal:1988}
reported a decrease of 1.4 Jy for the core between 1985 June and 1986 August, which corresponds
to $\sim$17\% of the total source flux.  At 43~GHz, VLBA monitoring showed an increase of
0.57~Jy for the core region between 2008 January and April \citep{wagner/etal:2009}, which
corresponds to 4\% of the total flux.  Smaller core flux variations, less than 1\%
of the total flux, have been observed at lower frequencies \citep{morabito/etal:1988, 
junor/biretta:1995, harris/etal:2009}.

\begin{figure}
  \begin{center}
    \includegraphics[width=6.5in]{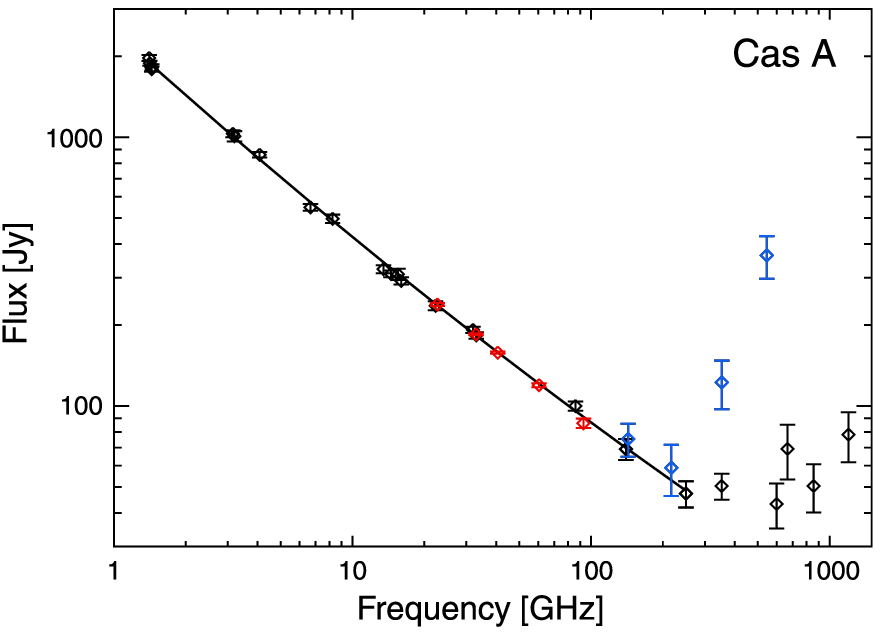}
    \caption{ Spectrum of Cas A for epoch 2000.  The red symbols show \wmap\ data.
      The black symbols show data from \citet{hafez/etal:2008}, 
      \citet{dunne/etal:2003} and \citet{sibthorpe/etal:2010}. The Archeops measurements of 
      \citet{desert/etal:2008}, shown in blue, are suspected of being affected by dust emission
      not associated with Cas~A. Black curve is a fit to the
      combined 1.4 - 250~GHz data of the form {log S(Jy) = a + b log $\nu$ + c log$^2\nu$};
      coefficients are provided in Table~\ref{tab:specfits}.
      \label{fig:casaspec}}
  \end{center}
\end{figure}

\begin{figure}
  \begin{center}
    \includegraphics[width=6.5in]{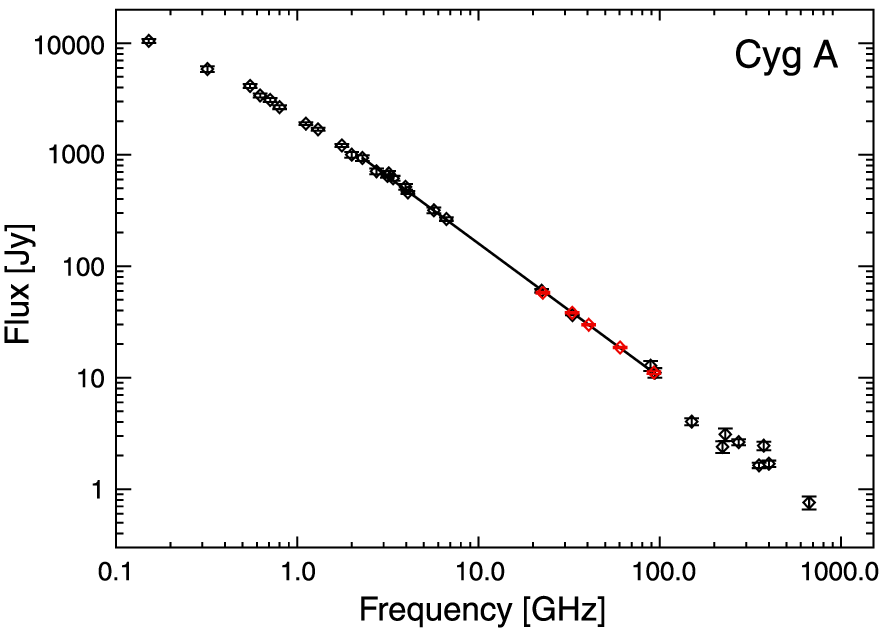}
    \caption{ Spectrum of Cyg A.  The red symbols show \wmap\ data.
     The black symbols show data from \citet{baars/etal:1977},
     \citet{wright/birkinshaw:1984}, \citet{salter/etal:1989b},
     \citet{eales/etal:1989}, \citet{wright/sault:1993}, \citet{robson/etal:1998},
     and \citet{hafez/etal:2008}.
     Black line is a power-law fit to the
      combined 2 - 94~GHz data;
      coefficients are provided in Table~\ref{tab:specfits}. 
      \label{fig:cygaspec}}
  \end{center}
\end{figure}

\begin{figure}
  \begin{center}
    \includegraphics[width=6.5in]{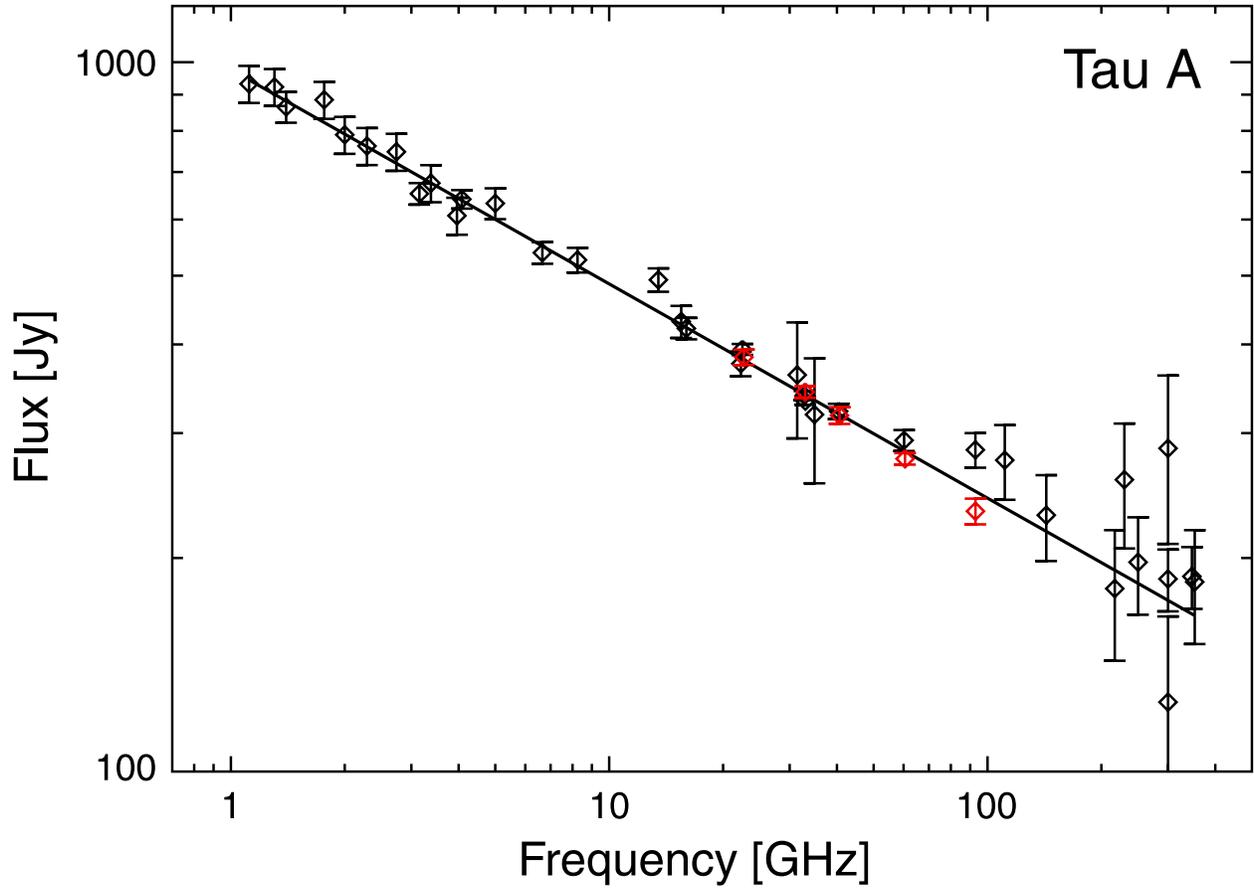}
    \caption{ Spectrum of Tau A for epoch 2005.  The red symbols show \wmap\ data.
      The black symbols show data from \citet{macias-perez/etal:2010} and \citet{hafez/etal:2008}.
     Black line is a power-law fit to the
      combined 1 - 353~GHz data;
      coefficients are provided in Table~\ref{tab:specfits}. 
      \label{fig:tauaspec}}
  \end{center}
\end{figure}

\begin{figure}
  \begin{center}
    \includegraphics[width=6.5in]{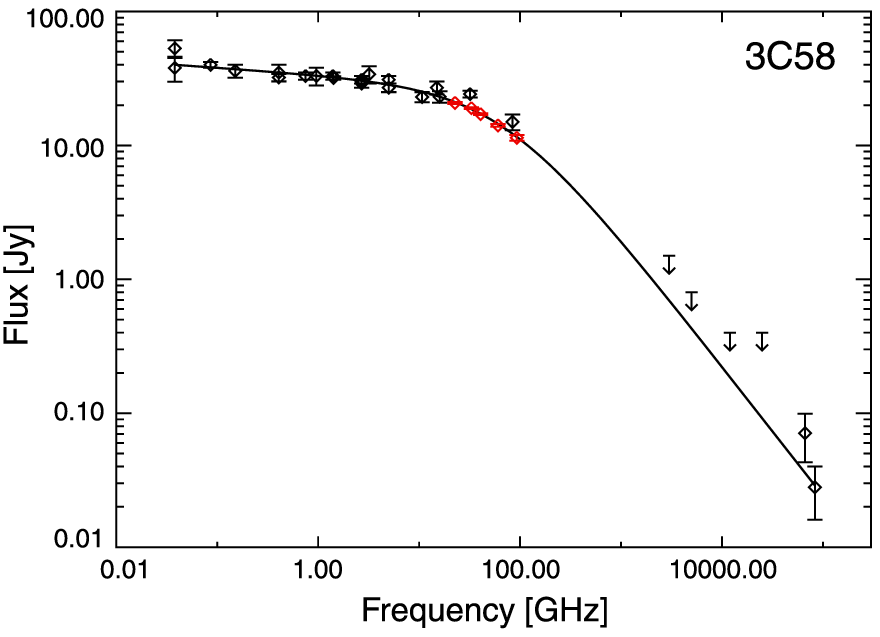}
    \caption{ Spectrum of 3C58.  The red symbols show \wmap\ data.  The black symbols
     show data from \citet{green:1986}, \citet{morsi/reich:1987}, \citet{salter/etal:1989}, 
     \citet{rees:1990}, 
       \citet{green/scheuer:1992}, \citet{kothes/etal:2006}, \citet{slane/etal:2008}, and
       \citet{hurley-walker/etal:2009}. Black curve is a fit to the
      combined 0.04 - 83000~GHz data of the form {S(Jy) = a $\nu^b$/ (1 + c $\nu^d$ ) };
      coefficients are provided in Table~\ref{tab:specfits}.
      \label{fig:spec3c58}}
  \end{center}
\end{figure}

\begin{figure}
  \begin{center}
    \includegraphics[width=6.5in]{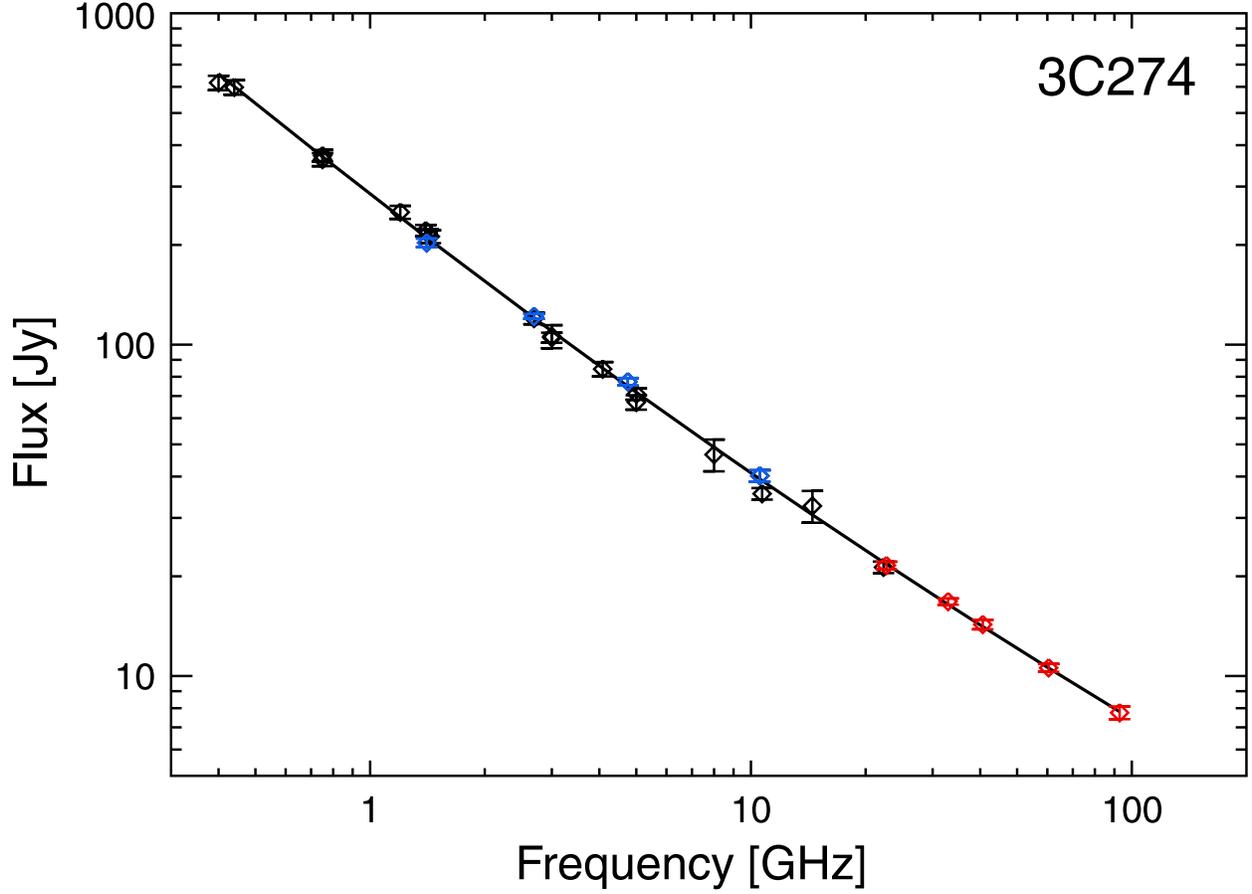}
    \caption{ Spectrum of 3C274.  The red symbols show \wmap\ data.  
    The black symbols show data from \citet{baars/etal:1977} and blue
    symbols show data from \citet{ott/etal:1994}. Black curve is a fit to the
      combined 0.4 - 93~GHz data of the form {log S(Jy) = a + b log $\nu$ + c log$^2\nu$};
      coefficients are provided in Table~\ref{tab:specfits}.
      \label{fig:spec3c274}}
  \end{center}
\end{figure}

Spectra of the seven-year \wmap\ fluxes together with previous measurements from the literature
are shown in Figures~\ref{fig:casaspec} - \ref{fig:spec3c274}.  
For Cas~A, we have scaled the \wmap\ fluxes, the Archeops fluxes of
\citet{desert/etal:2008}, the SCUBA fluxes of \citet{dunne/etal:2003}, and
the BLAST fluxes of \citet{sibthorpe/etal:2010} 
to epoch 2000 using a secular variation of $-0.53\%$ per year.
These are plotted with previous measurements that were scaled to epoch 2000 by 
\citet{hafez/etal:2008} using frequency-dependent scaling.
For Tau~A, we have scaled previous measurements from \citet{macias-perez/etal:2010} to the
epoch of the \wmap\ data using a secular variation of $-0.167\%$ per year at all frequencies.
Results are not significantly different if $-0.22\%$ per year is used.
Table~\ref{tab:specfits} presents parameters from fits to the spectra for \wmap\ data alone and for the
combined data.  The two fits are generally consistent within their uncertainties over the \wmap\
frequency range. Some notes on the individual sources follow.

\noindent{\textit{Cas~A}--} \hspace{0.15in}

A slightly curved spectrum gives a better fit to the combined data
than a power law (chi-squared per degree of freedom $\chi^2_{\nu} = 1.20$
compared to $\chi^2_{\nu} = 2.13$ for a power law).
The flattening of the spectrum with increasing frequency was previously
noted by \citet{hafez/etal:2008}.  This may be consistent with observations
of spatial variations of the spectrum in Cas~A (e.g., 
\citealt{wright.m/etal:1999},
\citealt{anderson/rudnick:1996}).  \citet{wright.m/etal:1999} presented spectra
of 26 brightness peaks from maps with $7\arcsec$ resolution at 1.5, 5, 28,
and 83~GHz.  The data were mostly consistent with power-law spectra,
with spectral indices ranging from -0.75 to -0.95. (For comparison, the overall spectral 
index from a power-law fit to our integrated spectrum from 1.4 to 93~GHz is -0.73).
Such a variation will lead to curvature in the integrated spectrum.
\citet{wright.m/etal:1999} also found curvature for some of the brightness peaks
with the spectra progressively flattening at the higher frequencies, and noted
that such curvature is expected from models of particle acceleration in cosmic
ray modified shocks \citep{reynolds/ellison:1992}.

Within the \wmap\ frequency range, all of the epoch 2000 scaled
fluxes are consistent within the uncertainties with the fit to the
combined data.  This includes the \wmap\ fluxes, the
absolutely calibrated fluxes from \citet{janssen/etal:1974} at 22.29~GHz
and \citet{Mason/etal:1999} at 32~GHz, the 33~GHz flux from
\citet{hafez/etal:2008}, which is calibrated using the five-year \wmap\
Jupiter temperature, and the 86~GHz flux from \citet{liszt/lucas:1999},
which is calibrated using the \citet{ulich:1981} Jupiter temperature.
 
Above 300~GHz, there is excess emission above that expected for synchrotron
emission, which has most recently been interpreted as emission from cool 
dust
by \citet{sibthorpe/etal:2010}.  The 353 and 545~GHz fluxes from Archeops
\citep{desert/etal:2008} are much higher than the 600~GHz flux from BLAST
\citep{sibthorpe/etal:2010} and the 353 and 666~GHz fluxes from SCUBA
\citep{dunne/etal:2003}.  The Archeops measurements were made with a larger
beam ($\sim12\arcmin$ compared to $\sim20\arcsec$ or better for SCUBA and BLAST)
and appear to be affected by dust emission that is not associated with
Cas~A.

\begin{figure}
  \begin{center}
    \includegraphics[width=6.5in]{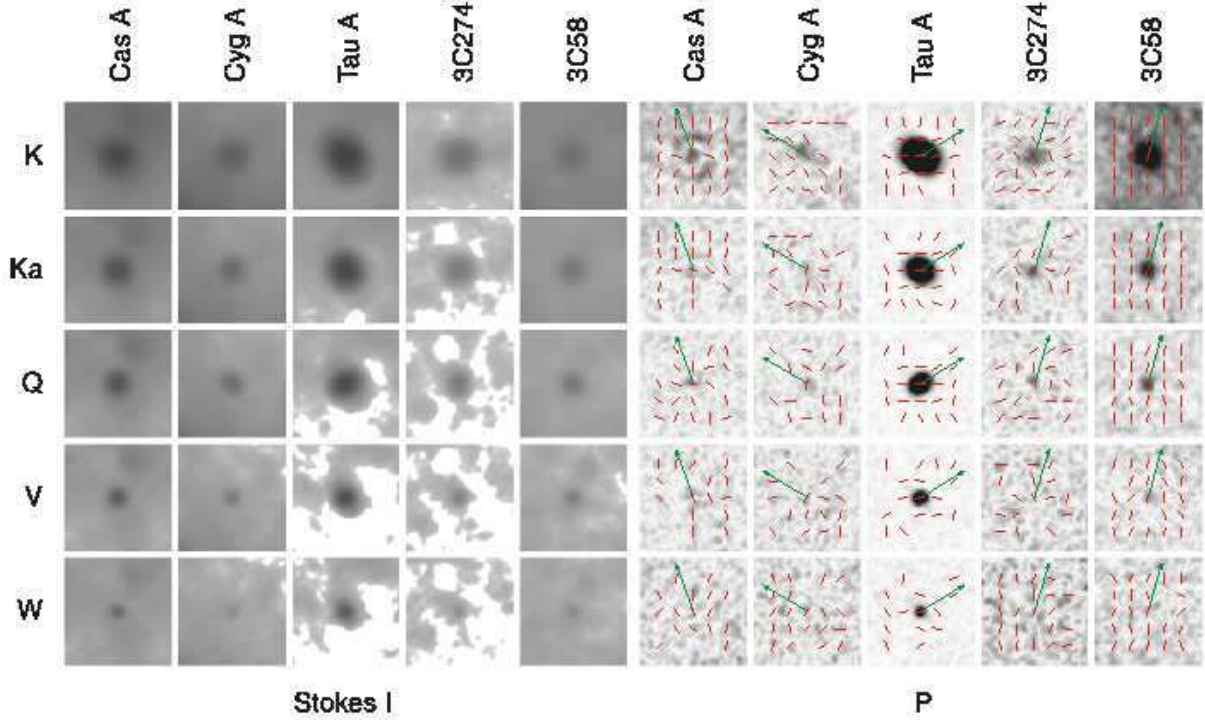}
    \caption{ Sky-fixed calibration sources in temperature and in polarization.
      Each picture is $4\fdg15$ in a side, and the vertical direction is
      aligned with Galactic north.  Sources are labeled along the top,
      and frequency bands are labeled along the left side.  Bright
      pixels are displayed as black. Left: Stokes $I$
      logarithmically scaled; peak pixel values in K band are 60.8,
      19.3, 95.8, 5.4, and 7.2 mK thermodynamic, respectively, for the five sources,
      at a native resolution of HEALPix $\texttt{nside} = 512$.
      Right:  polarization $P=(Q^2 + U^2)^{0.5}$, where $Q$ and
      $U$ are Stokes parameters, smoothed with a Gaussian of
      $\texttt{FWHM} = 13\farcm5$.  Scaling is linear between 0 and 0.25
      mK thermodynamic, except for Tau A, where scaling is linear between 
      0 and 1 mK thermodynamic.
      Green arrow:  north in the equatorial coordinate system.  Red
      bars:  polarization direction in a square pixel $39\arcmin$ wide,
      shown only where the S/N of $P$ is greater than $\sim$2.  The
      broken ring around Cas A in K band is spurious and results from
      beam ellipticity coupled with effective frequency differences between 
      the two orthogonally polarized K-band radiometers.
      \label{fig:ipforcelobj}}
  \end{center}
\end{figure}

\begin{figure}
  \begin{center}
    \includegraphics[width=6.5in]{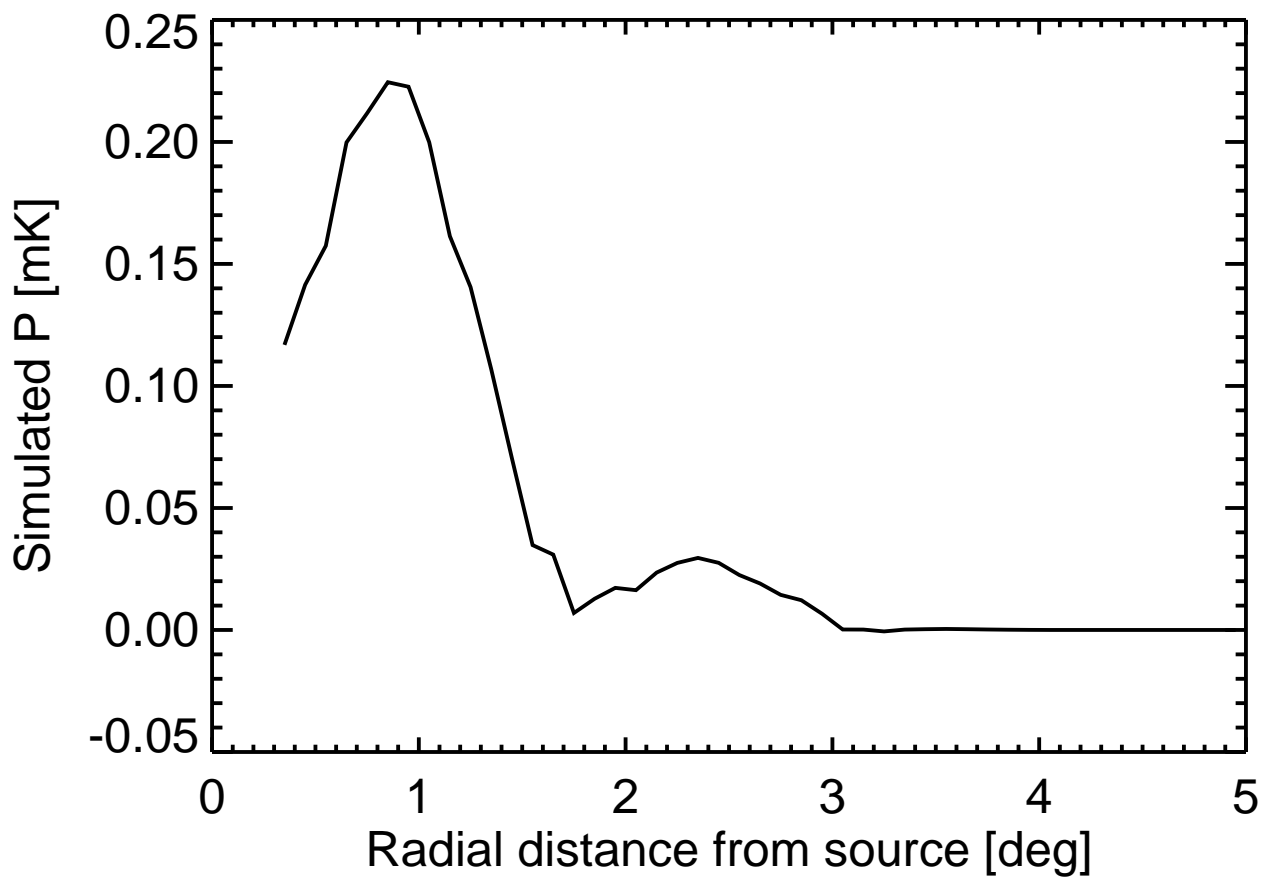}
    \caption{ Simplistic simulation of a false polarization signal which would
    resemble a ring encircling Cas~A at a radius of $\sim50\arcmin$.  The black line
    is the radial profile of an azimuthally symmetric polarization artifact centered 
    on the source, resulting from a combination of beam, detector and source properties.
    The simulation approximately reproduces the observed ring structure surrounding Cas~A in
    Figure~\ref{fig:ipforcelobj}, but is brighter by $\sim35\%$, and does not include details
    such as the asymmetric scan-angle distribution in the data.
    \label{fig:artifact}}
  \end{center}
\end{figure}

The \wmap\ 23~GHz (K band) polarization map for Cas~A exhibits unexpected structure.  
Figure~\ref{fig:ipforcelobj} shows 
seven-year mean intensity and polarization images centered on each source at each of the five \wmap\
frequency bands. These images are $4.15\arcdeg$ on a side, have not been background subtracted,
and are scaled such that brighter pixels are black.  The polarization (P) image for Cas~A shows
an irregular $\sim0.15$~mK ring at a radial distance roughly $40\arcmin-50\arcmin$ from
the source position.  The angular extent of Cas~A is $\sim5\arcmin$ (Table~\ref{tab:calobj}), leaving
the reality of the ring feature in question.
We have attempted to simulate this feature under the hypothesis that it is an artifact introduced by a 
combination of beam and  source spectrum characteristics.  Cas~A is a steep-spectrum source. 
As a result of effective frequency differences between the two K-band radiometers \citep{jarosik/etal:2003b},
the K11 radiometer (fed by the axial OMT port) has an FWHM roughly 3\% wider than that of K12 
(which is fed by the lateral OMT port), and
the peak observed signal in K11 is a few percent higher than that of K12.
Simulated beam maps for the two K-band radiometers were generated separately using Jupiter data as a template, 
and then individually scaled to peak 
values representative of Cas~A. The difference between the two beams as a function of azimuthal angle
and radial distance from beam center can be used to compute a rough estimate of induced Cas~A polarization signal,
under the assumption of complete scan-angle coverage.  Figure~\ref{fig:artifact} shows the results of
such a simulation, which produces a feature with an approximately correct peak position (near $50\arcmin$), but
slightly wider and 30\%-40\% brighter than that shown in the data image.  A more complete simulation would include
scan-angle coverage effects, which in this scenario are presumed responsible for the gaps in the
ring.  This is the only known instance of an apparent artifact in \wmap\ polarization data.

\noindent{\textit{Cyg~A}--} 

The fluxes from \citet{janssen/etal:1974} at 22.29~GHz, \citet{wright/birkinshaw:1984} at 89~GHz,
and \citet{wright/sault:1993} at 94~GHz are consistent with the \wmap\ 
results.  The flux from \citet{hafez/etal:2008} at 33~GHz is lower than \wmap\ Ka-band 
flux by $2.7\sigma$, taking both flux uncertainties into account.  We found that excluding the
\citet{hafez/etal:2008} flux from the power-law fit to the combined data improved
$\chi^2_{\nu}$ from 2.89 to 0.79, so it was excluded for the fit plotted in Figure~\ref{fig:cygaspec}
and the fit parameters given in Table~\ref{tab:specfits}.  Most of the measurements 
above 100~GHz in Figure~\ref{fig:cygaspec} are fluxes summed over the core and two hot spots in the 
radio lobes. These are probably valid measurements of the total flux, since the 
contribution of extended
emission from the steep-spectrum lobes appears to be small or negligible 
at these frequencies. At 230~GHz, \citet{salter/etal:1989b} found that the integrated flux 
over the entire source was only about 10\% greater than the summed flux of the hot spots and 
core, which did not amount to a significant detection of emission from the lobes.

\noindent{\textit{Tau~A}--}  

The \wmap\ flux agrees with that of \citet{janssen/etal:1974} at 
22.29~GHz and with that of
\citet{hafez/etal:2008} at 33~GHz.  Flux measurements from three-year \wmap\ sky maps
by \citet{macias-perez/etal:2010} are consistent with our results in K-V bands, but their W-band
measurement is 2.8$\sigma$ greater than ours.  The difference is probably 
due to difference in background subtraction error.

\begin{figure}
  \begin{center}
    \includegraphics[width=6.5in]{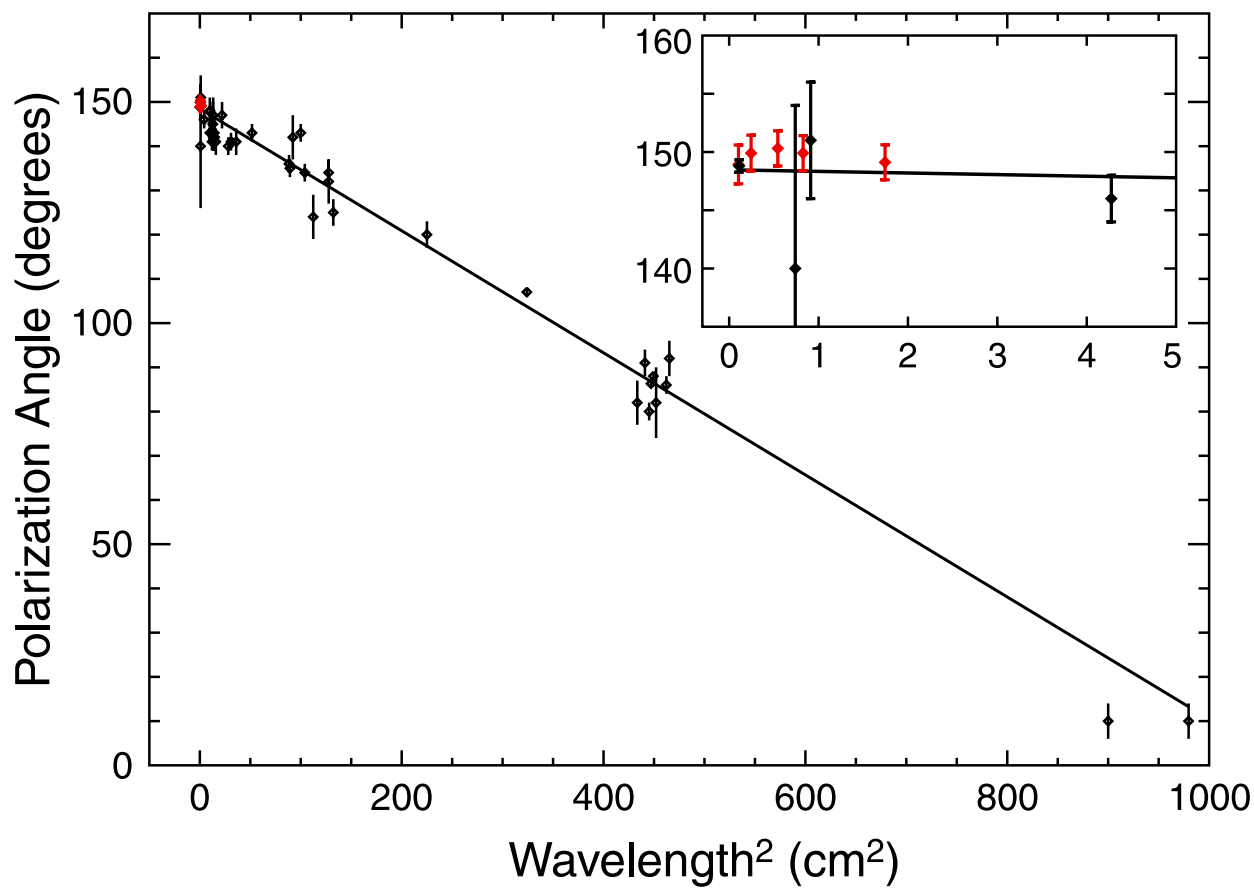}
    \caption{ Position angle of integrated polarization of Tau~A in equatorial coordinates
    as a function of wavelength squared.  The \wmap\ results from Table~\ref{tab:srcflux}
    are shown in red, with error bars that include a systematic uncertainty of $1.5\arcdeg$.
    Results shown in black are from references in \citet{wilson:1972} and from \citet{aumont/etal:2010}
    at 90 GHz for a 10 arcmin beam.  The linear fit gives an intrinsic polarization position angle
    of $148.5\arcdeg \pm 0.3\arcdeg$ and a rotation measure of $-24.1 \pm 0.2$ rad m$^{-2}$.
    \label{fig:polangles}}
  \end{center}
\end{figure}

Tau~A has been recommended as a polarization calibrator for CMB experiments 
by \citet{aumont/etal:2010},
who report a polarized flux density of $14.5 \pm 3.2$~Jy at 90~GHz.  The \wmap\ value at 93~GHz
is $16.6 \pm 0.7$~Jy.  For a $10\arcmin$ beam, \citet{aumont/etal:2010} predict a position angle
of $148.8\arcdeg \pm 0.2\arcdeg$ (statistical) $\pm 0.5\arcdeg$ (systematic) in equatorial
coordinates.  The \wmap\ value at 93~GHz is $148.9\arcdeg \pm 0.7\arcdeg$ (statistical) $\pm 1.5\arcdeg$ 
(systematic).  These measurements and previous position angle measurements integrated over the
source are shown as a function of wavelength squared in Figure~\ref{fig:polangles}.  A linear fit
to the data gives an intrinsic polarization position angle of $148.5\arcdeg \pm 0.3\arcdeg$ and
a rotation measure of $-24.1 \pm 0.2$ rad~m$^{-2}$.  This mean rotation measure appears to be
consistent with the rotation measure map obtained by \citet{bietenholz/kronberg:1991} from
VLA observations at $1.8\arcsec$ resolution.  They found a large-scale rotation measure of
$\sim-21$ rad~m$^{-2}$ that they attribute to the interstellar medium, and from the observed
depolarization they inferred absolute rotation measure values of a few hundred rad~m$^{-2}$
in unresolved filaments.

\citet{lopez-caniego/etal:2009} report polarized flux densities they measured from 5-year \wmap\ data 
that are smaller than the
values in Table~\ref{tab:srcflux} by 9\%, 15\%, 28\%, and 67\% at 23, 33, 41, and 61~GHz, respectively.
For Cas~A, Cyg~A, and 3C274, their polarized fluxes agree with those in 
Table~\ref{tab:srcflux} within the uncertainties.

\noindent{\textit{3C58}--}  

The \wmap\ data show the spectral steepening first suggested by the 84~GHz
measurement of \citet{salter/etal:1989} and confirmed by the upper limits of
\citet{green/scheuer:1992} in the infrared.
Our adopted form for the spectrum transitions between a low-frequency power law
and a high-frequency power law, and gives a good fit to the combined radio to
infrared data.  Our fit gives a change in spectral index of about 0.9.  
Additional steepening is needed to explain the observed X-ray spectrum, and 
\citet{slane/etal:2008} suggested there is a second spectral break
somewhere in the infrared.  They presented model calculations in which the break
at $\sim100$ GHz is due to a break in the electron spectrum injected into the nebula
and the higher frequency break is due to synchrotron losses.

\noindent{\textit{3C274}--}

The \wmap\ K-band flux agrees with that of \citet{janssen/etal:1974} at 22.29~GHz.  The combined
data are better fit by a quadratic ($\chi^2_{\nu} = 0.99$) than by a linear power law 
($\chi^2_{\nu} = 2.54$).
Curvature in the overall spectrum might be expected since 3C274 has a steep spectrum halo.  
Above 10~GHz,
the contribution of the halo to the total flux is $< 1\%$ \citep{baars/etal:1977}.
Over the \wmap\ frequency range, the quadratic fit and the \wmap\ only power-law fit are not significantly
different.  The data of \citet{ott/etal:1994} give a flatter 1.4-10.6~GHz spectral index than that of
\citet{baars/etal:1977}, and they suggested that this may be due to core activity that could
cause variability at the 5-10\% level. As noted above, variability of about 4\% and 17\% has
been observed at 43 GHz and 90 GHz, respectively.

\section{Conclusions\label{sec:summary}} 

\wmap\ data provide well-calibrated radiometry of the outer planets 
and bright sources.  The seven-year data provide the longest
\wmap\ baseline to date
from which to study temporal variability, plus an increase in
signal-to-noise which allows us to include fainter objects such as
Uranus and Neptune in the analysis.

Jupiter temperatures derived from the seven-year data are within
$1\sigma$ of the previously published five-year values of \citet{hill/etal:2009}.
The disk-integrated temperatures derived for Jupiter at \wmap\ frequencies 
have uncertainties of less than 1\%, with an apparent seasonal stability
of $\Delta T/T = 0.2\% \pm 0.4\%$. As a planet, Jupiter has additional
advantages in that it has a small 
apparent angular diameter and moves with respect to the fixed sky,
allowing for well-characterized background subtraction. 
Of the ten objects studied, \wmap\ uncertainties
for Jupiter are lowest, and we continue to recommend it as the best means
for transferring the \wmap\  dipole calibration to another microwave
instrument.

With the aid of models to predict geometrical and/or intrinsic variations,
the \wmap-derived temperatures for Mars and Saturn are predictable to $\le 3\%$. 
The recommendation for using the Mars values needs to be tempered by
the knowledge that these have been derived using mean seasonal properties only.  
The Saturn model is purely empirical, and can be improved with the addition
of the 2009 and 2010 \wmap\ observations when the rings are nearly
edge-on.

Uncertainties in the seven-year mean Uranus temperatures range from 3\% at 
W band to about 7\% in K and Ka band; only the Q, V and W measurements
are suitable for calibration at the $\lesssim5\%$ level.  
Some form of simple model including polar brightness effects would be required 
to extend the seven-year mean data here beyond the observing epoch, since 
long-term studies in the literature show clear seasonal temperature variations.
There is also the surprising hint of a dip in the microwave
spectrum near 30~GHz, where none is expected based on discussions in the
literature.

Seven-year mean temperature uncertainties for Neptune exceed 5\% at all 
\wmap\ frequencies and so are not recommended for serious calibration usage.

Flux densities in Stokes I for the five celestial calibrators
have typical uncertainties of 1\%-3\%,
and are in good agreement with previous measurements in the \wmap\ frequency range. 
For four of the sources, the \wmap\ data and other available data are 
consistent with a simple well-defined
spectrum over two decades in frequency or more.  The \wmap\ 
observations improve the accuracy
of the spectral fit parameters and in some cases they extend the 
frequency range where the object
may be used for calibration.  
The uncertainties in the spectral fits are typically
1\% or less in amplitude and 0.02 or less in spectral index.
We provide new estimates for the secular variation of Cas~A and Tau~A,
and provide limits and an estimate for year-to-year variability of the 
other sources.  We present
\wmap\ polarization data with uncertainties of a few percent for Tau~A.

The seven-year \wmap\ data products are available to the research community through
the Legacy Archive for Microwave Background Data Analysis (LAMBDA) at \\
\texttt{http://lambda.gsfc.nasa.gov}.

\acknowledgements

The \wmap\ mission is made possible by the support of the Science Mission Directorate
Office at NASA Headquarters.
This research has made use of NASA's Astrophysics Data System Bibliographic Services.
We acknowledge the use of the HEALPix package \citep{gorski/etal:2005}. 


%
\clearpage
\clearpage
\clearpage

\clearpage
\clearpage
\clearpage
\clearpage
\clearpage
\clearpage
\clearpage

\clearpage
\clearpage
\clearpage

\clearpage

\clearpage

\clearpage

\clearpage

\clearpage

\clearpage

\clearpage

\clearpage

%


\clearpage

\begin{deluxetable}{rclll}
  \tabletypesize{\scriptsize}
  \tablewidth{0pt}
  \tablecaption{ \wmap\ Planet Observing Seasons (2001-2008) \label{tab:planetseasons}}
  \tablehead{
    \colhead{Planet} &
    \colhead{Season\tablenotemark{a}} &
    \colhead{Begin}  &
    \colhead{End}    &
    \colhead{SP\tablenotemark{b}}}

\startdata
Jupiter &  1  & 2001 Oct  8  & 2001 Nov 22  & \\
 &   2  & 2002 Feb 11  & 2002 Mar 27  & G \\
 &   3  & 2002 Nov 10  & 2002 Dec 24  & \\
 &   4  & 2003 Mar 15  & 2003 Apr 29  & G \\
 &   5  & 2003 Dec 11  & 2004 Jan 23  & \\
 &   6  & 2004 Apr 15  & 2004 May 30  & \\
 &   7  & 2005 Jan  9  & 2005 Feb 21  & \\
 &   8  & 2005 May 16  & 2005 Jul  1  & \\
 &   9  & 2006 Feb  7  & 2006 Mar 24  & \\
 &  10  & 2006 Jun 16  & 2006 Aug  2  & \\
 &  11  & 2007 Mar 10  & 2007 Apr 24  & \\
 &  12  & 2007 Jul 19  & 2007 Sep  3  & \\
 &  13  & 2008 Apr 11  & 2008 May 27  & \\
\hline
Saturn &  1  & 2001 Sep  7  & 2001 Oct 22  & \\
 &   2  & 2002 Jan 14  & 2002 Feb 26  & \\
 &   3  & 2002 Sep 22  & 2002 Nov  5  & \\
 &   4  & 2003 Jan 28  & 2003 Mar 13  & G \\
 &   5  & 2003 Oct  7  & 2003 Nov 20  & \\
 &   6  & 2004 Feb 12  & 2004 Mar 26  & G \\
 &   7  & 2004 Oct 20  & 2004 Dec  3  & \\
 &   8  & 2005 Feb 25  & 2005 Apr 10  & \\
 &   9  & 2005 Nov  4  & 2005 Dec 17  & \\
 &  10  & 2006 Mar 11  & 2006 Apr 24  & \\
 &  11  & 2006 Nov 18  & 2006 Dec 31  & \\
 &  12  & 2007 Mar 25  & 2007 May  9  & \\
 &  13  & 2007 Dec  1  & 2008 Jan 13  & \\
 &  14  & 2008 Apr  7  & 2008 May 22  & \\
\hline
Mars &  1  & 2001 Aug 10  & 2001 Oct 14  & \\
 &   2  & 2003 Apr 19  & 2003 Jul 17  & U, N \\
 &   3  & 2003 Oct 10  & 2003 Dec 30  & U \\
 &   4  & 2005 Jul 13  & 2005 Oct  1  & \\
 &   5  & 2005 Dec 16  & 2006 Feb 19  & \\
 &   6  & 2007 Sep 18  & 2007 Nov 20  & \\
 &   7  & 2008 Jan 30  & 2008 Mar 30  & \\
\hline
\tablebreak
Uranus &  1  & 2001 Sep 30  & 2001 Nov 13  & \\
 &   2  & 2002 May 20  & 2002 Jul  5  & \\
 &   3  & 2002 Oct  4  & 2002 Nov 17  & \\
 &   4  & 2003 May 25  & 2003 Jul  9  & M \\
 &   5  & 2003 Oct  8  & 2003 Nov 21  & M \\
 &   6  & 2004 May 28  & 2004 Jul 13  & \\
 &   7  & 2004 Oct 12  & 2004 Nov 24  & \\
 &   8  & 2005 Jun  2  & 2005 Jul 17  & \\
 &   9  & 2005 Oct 16  & 2005 Nov 28  & \\
 &  10  & 2006 Jun  6  & 2006 Jul 22  & \\
 &  11  & 2006 Oct 20  & 2006 Dec  3  & \\
 &  12  & 2007 Jun 10  & 2007 Jul 26  & \\
 &  13  & 2007 Oct 24  & 2007 Dec  7  & \\
 &  14  & 2008 Jun 14  & 2008 Jul 29  & \\
\hline
Neptune &  1  & 2001 Sep 15  & 2001 Oct 29  & \\
 &    2  & 2002 May  2  & 2002 Jun 16  & \\
 &    3  & 2002 Sep 17  & 2002 Oct 31  & \\
 &    4  & 2003 May  4  & 2003 Jun 19  & M \\
 &    5  & 2003 Sep 20  & 2003 Nov  3  & \\
 &    6  & 2004 May  6  & 2004 Jun 21  & \\
 &    7  & 2004 Sep 21  & 2004 Nov  4  & \\
 &    8  & 2005 May  8  & 2005 Jun 23  & \\
 &    9  & 2005 Sep 23  & 2005 Nov  7  & \\
 &   10  & 2006 May 11  & 2006 Jun 26  & \\
 &   11  & 2006 Sep 26  & 2006 Nov  9  & \\
 &   12  & 2007 May 14  & 2007 Jun 28  & \\
 &   13  & 2007 Sep 28  & 2007 Nov 11  & \\
 &   14  & 2008 May 15  & 2008 Jun 30  & \\
\enddata
  \tablenotetext{a}{An observing season is defined as a contiguous time interval during which an object
               is in the \wmap\ viewing swath.}
  \tablenotetext{b}{sky coordinate proximity to any of the following: Galactic plane(G); Mars (M),
               Uranus (U), Neptune (N).}
\end{deluxetable}

\clearpage

\begin{deluxetable}{cccl}
\tablecolumns{4}
\tablewidth{0pt}
\tablecaption{Adopted Planetary Radii\tablenotemark{a} and fiducial Solid Angles\label{tab:radii}}
\tablehead{
    \colhead{Planet} &
    \colhead{$R_{\mathrm{eq}}$} &
    \colhead{$R_{\mathrm{pole}}$} &
    \colhead{$\Omega_\mathrm{ref}$\tablenotemark{b}} \\ &
    \colhead{(km)} &
    \colhead{(km)} &
    \colhead{(sr)}   }
\startdata
\hline
Mars    &   3396   &   3376  &  $7.153 \times 10^{-10}$ \\
Jupiter &  71492   &  66854  &  $2.481 \times 10^{-8}$  \\
Saturn  &  60268   &  54364  &  $5.096 \times 10^{-9}$ \\
Uranus  &  25559   &  24973  &  $2.482 \times 10^{-10}$ \\
Neptune &  24764   &  24341  &  $1.006 \times 10^{-10}$ \\
\hline
\enddata
  \tablenotetext{a}{From \citet{seidelmann/etal:2007}.}
  \tablenotetext{b}{Solid angle computed at fiducial distances of
               1.5, 5.2, 9.5, 19 and 29 AU respectively.
	       See Section~\ref{sec:methods} for further discussion.}
\end{deluxetable}

\clearpage

\begin{deluxetable}{ccccc}
  \tablewidth{0pt}
  \tablecolumns{5}
  \tabletypesize{\scriptsize}
  \tablecaption{Seven-year Mean Jupiter Temperatures \label{tab:juptemp}}
  \tablehead{
    \colhead{DA/band} &
    \colhead{$\nu_e^\mathrm{RJ}$ \tablenotemark{a}} &
    \colhead{$\lambda$ \tablenotemark{b}} &
    \colhead{$T$\tablenotemark{c}} &
    \colhead{$\sigma(T)$\tablenotemark{d}} \\ &
    \colhead{(GHz)}  &
    \colhead{(mm)}   &
    \colhead{(K)}    &
    \colhead{(K)}}
  \startdata
  \sidehead{Per DA}
  K1  & 22.85 & 13.1 & 136.2 & 0.85 \\
  Ka1 & 33.11 &  9.1 & 147.2 & 0.77 \\
  Q1  & 40.92 &  7.3 & 154.4 & 0.91 \\
  Q2  & 40.71 &  7.4 & 155.2 & 0.87 \\
  V1  & 60.41 &  5.0 & 165.0 & 0.78 \\
  V2  & 61.28 &  4.9 & 166.1 & 0.74 \\
  W1  & 93.25 &  3.2 & 173.0 & 0.84 \\
  W2  & 93.73 &  3.2 & 173.3 & 0.91 \\
  W3  & 93.72 &  3.2 & 173.7 & 0.94 \\
  W4  & 93.57 &  3.2 & 173.9 & 0.92 \\
  \sidehead{Per band}
   K  &   22.85	&  13.1	& 136.2 & 0.85 \\ 
   Ka &   33.11	&   9.1	& 147.2 & 0.77 \\	  
   Q  &   40.82	&   7.3	& 154.8 & 0.67 \\	  
   V  &   60.85	&   4.9	& 165.6 & 0.59 \\	  
   W  &   93.32	&   3.2	& 173.5 & 0.54 \\  
  \enddata
  \tablenotetext{a}{Seven-year values; see \citet{jarosik/etal:prep}.}
  \tablenotetext{b}{$\lambda = c/\nu_e^{\mathrm{RJ}}$.}
  \tablenotetext{c}{Brightness temperature
    calculated for a solid angle $\Omega_\mathrm{ref}
    = 2.481 \times 10^{-8}$ sr at a fiducial distance of 5.2 AU.
    Temperature is with respect to blank sky: absolute
    brightness temperature is obtained by adding 2.2, 2.0, 1.9, 1.5 and 1.1 K in bands
    K, Ka, Q, V and W respectively \citep{page/etal:2003b}. 
    Jupiter temperatures are uncorrected for a small synchrotron emission component
    (see Section~\ref{sec:jupsec}). 
    }
  \tablenotetext{d}{Computed from errors
    in $\Omega_B$ \citep{jarosik/etal:prep} summed in quadrature with absolute
    calibration error of $0.2\%$.}
\end{deluxetable}
\clearpage
\begin{deluxetable}{cccrcccc}
  \tablecolumns{8}
  \tablewidth{0pt}
  \tablecaption{Jupiter Temperature Changes by Season \label{tab:juptime}}
  \tablehead{
    \colhead{Season\tablenotemark{a}} & \colhead{Start} & \colhead{End} 
    & \multicolumn{2}{c}{$\Delta T/T$ (\%)} & \colhead{$\%$ Masked \tablenotemark{d}} 
    & \multicolumn{2}{c}{Mean Position \tablenotemark{e}}    \\
    & & & \colhead{Mean\tablenotemark{b} } & \colhead{Scatter\tablenotemark{c}} & 
    & \colhead{$l$} & \colhead{$b$}}
  \startdata
  1 & 2001 Oct 08 & 2001 Nov 22 & $ 0.11$   & 0.28 & 0   &  194 & 13 \\
  2 & 2002 Feb 11 & 2002 Mar 27 & \nodata  & \nodata & 95  &  189 & 5  \\
  3 & 2002 Nov 10 & 2002 Dec 24 & $-0.13$  & 0.28 & 0   &  214 & 40 \\
  4 & 2003 Mar 15 & 2003 Apr 29 & $-0.28$  & 0.48 & 60  &  206 & 32 \\
  5 & 2003 Dec 11 & 2004 Jan 23 & $ 0.01$  & 0.22 & 0   &  254 & 59 \\
  6 & 2004 Apr 15 & 2004 May 30 & $-0.01$  & 0.32 & 0   &  237 & 55 \\
  7 & 2005 Jan 09 & 2005 Feb 21 & $ 0.00$  & 0.27 & 12  &  311 & 56 \\
  8 & 2005 May 16 & 2005 Jul 01 & $-0.08$  & 0.39 & 0   &  294 & 60 \\
  9 & 2006 Feb 07 & 2006 Mar 24 & $ 0.13$  & 0.29 & 18  &  344 & 35 \\
  10 & 2006 Jun 16 & 2006 Aug 02 & $ 0.03$  & 0.45 & 0   &  335 & 43 \\
  11 & 2007 Mar 10 & 2007 Apr 24 & $ 0.41$  & 0.31 & 0   &  2  & 9 \\
  12 & 2007 Jul 19 & 2007 Sep 03 & $-0.11$  & 0.49 & 42  &  356 & 17 \\
  13 & 2008 Apr 11 & 2008 May 27 & $-0.20$  & 0.37 & 0   &  17 & -19 \\
  \enddata
  \tablenotetext{a}{Season 2 omitted from analysis because Jupiter is aligned with the Galactic
    plane.}
  \tablenotetext{b}{Mean of the percentage temperature change among the
    DAs for each season, relative to the seven-year mean.}
  \tablenotetext{c}{$1\sigma$ scatter in the percentage temperature change among the
    DAs for each season.}
  \tablenotetext{d}{Typical percentage of available observations within $0.5\arcdeg$ of planet center 
    which are excluded from analysis because of Galactic plane masking.}
  \tablenotetext{e}{Approximate seasonal Jupiter position in Galactic coordinates at mean time of unmasked
  observations.}

\end{deluxetable}
\clearpage

\begin{deluxetable}{cccccccc}
  \tablewidth{0pt}
  \tablecaption{Jupiter Linear Polarization \label{tab:juppoln}}
  \tablehead{
    \colhead{DA}   &
    \colhead{$Q/I$}  &
    \colhead{$\sigma(Q/I)$\tablenotemark{a}} &
    \colhead{$U/I$}  &
    \colhead{$\sigma(U/I)$\tablenotemark{b}} &
    \colhead{$p_{\mathrm{lin}}$\tablenotemark{c}}     &
    \colhead{$\sigma(p_{\mathrm{lin}})$\tablenotemark{d}} &
    \colhead{$ S/N $\tablenotemark{e}}  \\ &
    \colhead{$(\%)$} &
    \colhead{$(\%)$} &
    \colhead{$(\%)$} &
    \colhead{$(\%)$} &
    \colhead{$(\%)$} &
    \colhead{$(\%)$} &
    }
  \startdata
  K1  & -0.306  &   0.601  &   0.658  &   0.428  &   0.726  &   0.464  &   1.62 \\
  Ka1 & -0.181  &   0.408  &  -0.191  &   0.293  &   0.263  &   0.352  &   0.79 \\
  Q1  & -0.168  &   0.430  &  -0.244  &   0.304  &   0.296  &   0.349  &   0.90 \\
  Q2  & -0.151  &   0.428  &  -0.062  &   0.297  &   0.163  &   0.411  &   0.41 \\
  V1  & -0.149  &   0.358  &   0.462  &   0.252  &   0.486  &   0.263  &   1.88 \\
  V2  &  0.143  &   0.305  &   0.067  &   0.223  &   0.158  &   0.292  &   0.56 \\
  W1  &  0.040  &   0.319  &   0.393  &   0.233  &   0.395  &   0.234  &   1.69 \\
  W2  &  0.156  &   0.380  &   0.188  &   0.264  &   0.245  &   0.317  &   0.82 \\
  W3  &  0.006  &   0.397  &  -0.377  &   0.277  &   0.377  &   0.277  &   1.36 \\
  W4  & -0.058  &   0.371  &   0.311  &   0.273  &   0.316  &   0.277  &   1.15 \\
  \enddata
  \tablenotetext{a}{Propagated error from $\sigma_I$ and $\sigma_Q$.}
  \tablenotetext{b}{Propagated error from $\sigma_I$ and $\sigma_U$.}
  \tablenotetext{c}{$p_{\mathrm{lin}} = \sqrt{(Q/I)^2 + (U/I)^2}$. }
  \tablenotetext{d}{Propagated error.}
  \tablenotetext{e}{$S/N = p_{\mathrm{lin}}/\sigma(p_{\mathrm{lin}})$.}
\end{deluxetable}
\clearpage

\begin{deluxetable}{ccccccc}
\tablecolumns{7}
\tablewidth{0pt}
\tablecaption{Derived Mars Temperatures per Observing Season per Frequency
\label{tab:mars}}
\tablehead{
    \colhead{Season} &
    \colhead{RJD\tablenotemark{a}} &
    \multicolumn{5}{c}{$T_b$(K)\tablenotemark{b}} \\ 
    \cline{3-7} \\ & &  
    \colhead{K} &
    \colhead{Ka} &
    \colhead{Q} &
    \colhead{V} &
    \colhead{W}  }
\startdata
\hline
1 &  $ 2182.62 $ & $ 178 \pm   4 $ & $ 182 \pm   3 $ & $ 186 \pm   4 $ & $ 191 \pm   3 $ & $ 189 \pm   2 $ \\
2 &  $ 2776.39 $ & $ 183 \pm   4 $ & $ 187 \pm   3 $ & $ 191 \pm   4 $ & $ 197 \pm   3 $ & $ 204 \pm   2 $ \\
3 &  $ 2983.75 $ & $ 191 \pm   4 $ & $ 195 \pm   3 $ & $ 185 \pm   4 $ & $ 193 \pm   3 $ & $ 195 \pm   2 $ \\
4 &  $ 3586.17 $ & $ 200 \pm   3 $ & $ 199 \pm   2 $ & $ 203 \pm   3 $ & $ 209 \pm   2 $ & $ 213 \pm   1 $ \\
5 &  $ 3758.26 $ & $ 191 \pm   5 $ & $ 184 \pm   3 $ & $ 189 \pm   4 $ & $ 186 \pm   3 $ & $ 185 \pm   2 $ \\
6 &  $ 4389.29 $ & $ 186 \pm   4 $ & $ 187 \pm   3 $ & $ 196 \pm   4 $ & $ 197 \pm   3 $ & $ 198 \pm   2 $ \\
7 &  $ 4530.49 $ & $ 174 \pm   6 $ & $ 177 \pm   4 $ & $ 176 \pm   5 $ & $ 181 \pm   4 $ & $ 182 \pm   2 $ \\
\hline
\enddata
  \tablenotetext{a}{Approximate mean time of observations in 
               each season, Julian Day $-2450000$. Masking of data in proximity to Uranus and Neptune
	       skews the mean data time of the second season toward the beginning of the observing season.}
  \tablenotetext{b}{Brightness temperature calculated for a solid angle $\Omega_{\mathrm{ref}}= 7.153 \times 10^{-10}$ sr at a
               fiducial distance of 1.5 AU.  Temperature is with respect to blank sky: absolute
	       brightness temperature is obtained by adding 2.2, 2.0, 1.9, 1.5 and 1.1 K in bands
	       K, Ka, Q, V and W respectively \citep{page/etal:2003b}.}
\end{deluxetable}

\clearpage

\begin{deluxetable}{crcc}
\tablecolumns{4}
\tablewidth{0pt}
\tablecaption{COBE/DIRBE Mars fluxes\tablenotemark{a}\label{tab:dirbemars}}
\tablehead{
    \colhead{$\lambda$} &
    \colhead{Flux} &
    \colhead{Abscal Err\tablenotemark{b}} & 
    \colhead{$F/F_{\mathrm{model}}$}\tablenotemark{c} \\
    \colhead{($\micron$)} &
    \colhead{(Jy)} &
    \colhead{(\%)} & }
\startdata
\hline
 $ 12  $ &  148594  &	3.1  & 0.89  \\
 $ 25  $ &  186136  &  14.6  & 0.92  \\
 $ 60  $ &   50606  &  10.0  & 0.66  \\
 $ 100 $ &   24477  &  12.8  & 0.73  \\
 $ 140 $ &   18991  &  10.1  & 1.02  \\
 $ 240 $ &    7217  &  10.1  & 1.05  \\
\hline
\enddata
  \tablenotetext{a}{Fluxes have been averaged over the 146-day observing period, 
               after correction to a fiducial distance of 1.5 AU and 
	       application of color corrections (see the text).}
  \tablenotetext{b}{Uncertainty in absolute calibration, as documented in the DIRBE 
               Explanatory Supplement \citep{hauser/etal:1998}.}
  \tablenotetext{c}{Ratio of observed DIRBE flux to Wright model prediction. The model is processed in
               the same manner as the data. Model uncertainty is quoted as roughly $\pm 5\%$.}
\end{deluxetable}

\clearpage

\begin{deluxetable}{cccc}
\tablecolumns{4}
\tablewidth{0pt}
\tablecaption{Mars Model Scaling Factors for \wmap\ Frequencies
\label{tab:marsmodelresid}}
\tablehead{
    \colhead{Freq} &
    \colhead{$f_{\mathrm{scl}}(3.2$~mm)\tablenotemark{a}} &
    \colhead{$f_{\mathrm{scl}}(350 \micron)$\tablenotemark{b}} &
    \colhead{$\sigma_{f_{\mathrm{scl}}}(\%)$\tablenotemark{c}} }
\startdata
\hline
 K  &  $ 0.908 $ &  $ 0.897 $ &$ 2 $  \\
 Ka &  $ 0.914 $ &  $ 0.903 $ &$ 1 $  \\
 Q  &  $ 0.923 $ &  $ 0.912 $ &$ 1 $  \\
 V  &  $ 0.943 $ &  $ 0.932 $ &$ 0.7 $  \\
 W  &  $ 0.953 $ &  $ 0.941 $ &$ 0.5 $  \\
\hline
\enddata
  \tablenotetext{a}{Scaling factor applied to Wright model evaluated at 3.2~mm which
               will reproduce \wmap\ observational means at the specified frequencies.}
  \tablenotetext{b}{Scaling factor applied to Wright model evaluated at 350~\micron\ which
               will reproduce \wmap\ observational means at the specified frequencies.}
  \tablenotetext{c}{Approximate $1\sigma$ error in both scaling factors.}

\end{deluxetable}

\clearpage

\begin{deluxetable}{ccccccccccccc}
\tablecolumns{13}
\tablewidth{0pt}
\tabletypesize{\scriptsize}

\tablecaption{Derived Saturn Temperatures per Observing Season per DA
\label{tab:saturn}}

\tablehead{
    \colhead{Season\tablenotemark{a}} &
    \colhead{RJD\tablenotemark{b}} &
    \colhead{$B$\tablenotemark{c}} &
    \multicolumn{10}{c}{$T_b$ (K)\tablenotemark{d}} \\ 
    \cline{4-13} \\ & & &    
    \colhead{K}  &
    \colhead{Ka} &
    \colhead{Q1} &
    \colhead{Q2} &
    \colhead{V1} &
    \colhead{V2} &
    \colhead{W1} & 
    \colhead{W2} & 
    \colhead{W3} & 
    \colhead{W4}  
   }
\startdata
  1& 2175.58 &  -26& $ 133 \pm  1 $ & $ 141 \pm  1 $ & $ 146 \pm  1 $ & $ 148 \pm  1 $ & $ 157 \pm  1 $ & $ 156 \pm  1 $ & $ 164 \pm  1 $ & $ 164 \pm  1 $ & $ 166 \pm  1 $ & $ 167 \pm  1 $ \\
  2& 2305.74 &  -25& $ 133 \pm  2 $ & $ 143 \pm  1 $ & $ 146 \pm  1 $ & $ 148 \pm  1 $ & $ 155 \pm  1 $ & $ 156 \pm  1 $ & $ 161 \pm  1 $ & $ 165 \pm  1 $ & $ 164 \pm  1 $ & $ 165 \pm  1 $ \\
  3& 2554.16 &  -26& $ 131 \pm  2 $ & $ 141 \pm  1 $ & $ 149 \pm  1 $ & $ 150 \pm  1 $ & $ 158 \pm  1 $ & $ 157 \pm  1 $ & $ 166 \pm  1 $ & $ 166 \pm  1 $ & $ 164 \pm  1 $ & $ 166 \pm  1 $ \\
  5& 2931.78 &  -24& $ 131 \pm  1 $ & $ 138 \pm  1 $ & $ 144 \pm  1 $ & $ 146 \pm  1 $ & $ 153 \pm  1 $ & $ 153 \pm  1 $ & $ 161 \pm  1 $ & $ 161 \pm  1 $ & $ 160 \pm  1 $ & $ 161 \pm  1 $ \\
  7& 3310.06 &  -21& $ 126 \pm  1 $ & $ 135 \pm  1 $ & $ 140 \pm  1 $ & $ 140 \pm  1 $ & $ 147 \pm  1 $ & $ 148 \pm  1 $ & $ 154 \pm  1 $ & $ 154 \pm  1 $ & $ 154 \pm  1 $ & $ 154 \pm  1 $ \\
  8& 3441.94 &  -23& $ 130 \pm  2 $ & $ 138 \pm  1 $ & $ 142 \pm  1 $ & $ 141 \pm  1 $ & $ 149 \pm  1 $ & $ 150 \pm  1 $ & $ 155 \pm  1 $ & $ 159 \pm  1 $ & $ 159 \pm  1 $ & $ 158 \pm  1 $ \\
  9& 3690.17 &  -17& $ 122 \pm  1 $ & $ 130 \pm  1 $ & $ 135 \pm  1 $ & $ 134 \pm  1 $ & $ 141 \pm  1 $ & $ 141 \pm  1 $ & $ 146 \pm  1 $ & $ 146 \pm  1 $ & $ 147 \pm  1 $ & $ 147 \pm  1 $ \\
 10& 3813.21 &  -20& $ 125 \pm  2 $ & $ 131 \pm  1 $ & $ 132 \pm  3 $ & $ 134 \pm  3 $ & $ 143 \pm  2 $ & $ 142 \pm  1 $ & $ 150 \pm  1 $ & $ 149 \pm  2 $ & $ 148 \pm  2 $ & $ 152 \pm  2 $ \\
 11& 4068.06 &  -12& $ 123 \pm  1 $ & $ 130 \pm  1 $ & $ 132 \pm  1 $ & $ 137 \pm  1 $ & $ 139 \pm  1 $ & $ 140 \pm  1 $ & $ 142 \pm  1 $ & $ 144 \pm  1 $ & $ 143 \pm  1 $ & $ 144 \pm  1 $ \\
 12& 4196.39 &  -15& $ 121 \pm  2 $ & $ 131 \pm  2 $ & $ 132 \pm  1 $ & $ 136 \pm  1 $ & $ 139 \pm  1 $ & $ 141 \pm  1 $ & $ 143 \pm  1 $ & $ 143 \pm  1 $ & $ 143 \pm  1 $ & $ 144 \pm  2 $ \\
 13& 4443.26 &   -6& $ 128 \pm  2 $ & $ 131 \pm  1 $ & $ 135 \pm  1 $ & $ 138 \pm  1 $ & $ 140 \pm  1 $ & $ 139 \pm  1 $ & $ 143 \pm  1 $ & $ 146 \pm  1 $ & $ 142 \pm  1 $ & $ 146 \pm  1 $ \\
 14& 4577.96 &   -9& $ 123 \pm  2 $ & $ 130 \pm  1 $ & $ 132 \pm  1 $ & $ 133 \pm  1 $ & $ 140 \pm  1 $ & $ 141 \pm  1 $ & $ 140 \pm  1 $ & $ 141 \pm  1 $ & $ 141 \pm  1 $ & $ 141 \pm  1 $ \\
\enddata
  \tablenotetext{a}{Seasons 4 and 6 omitted from analysis because Saturn is aligned with the Galactic plane.}

  \tablenotetext{b}{Approximate mean time of observations in 
                    each season, Julian Day $-2450000$.}
  \tablenotetext{c}{Approximate mean ring opening angle for 
                    each season, degrees.}
  \tablenotetext{d}{Brightness temperature calculated for a solid angle $\Omega_{\mathrm{ref}}= 5.096 \times 10^{-9}$ sr at a
               fiducial distance of 9.5 AU.  A correction for planetary disk oblateness has not been applied,
	       as that is accounted for in modeling.
	       Temperature is with respect to blank sky: absolute
	       brightness temperature is obtained by adding 2.2, 2.0, 1.9, 1.5 and 1.1 K in bands
	       K, Ka, Q, V and W respectively \citep{page/etal:2003b}.}

\end{deluxetable}

\clearpage

\begin{deluxetable}{cccc}
\tablecolumns{4}
\tablewidth{0pt}
\tablecaption{Saturn Model Fixed Parameters\tablenotemark{a} \label{tab:satmodel1}}
\tablehead{
    \colhead{Ring Sector\tablenotemark{b}} &
    \colhead{$R_{\mathrm{inner}}$} &
    \colhead{$R_{\mathrm{outer}}$} &
    \colhead{$\tau_0$\tablenotemark{c}}  \\   &
    \colhead{$[R_{\mathrm{Saturn}}\tablenotemark{d}]$} &
    \colhead{$[R_{\mathrm{Saturn}}\tablenotemark{d}]$} & \\ }
\startdata
\hline
 A		   & $ 2.025 $ & $ 2.27  $ & $ 0.7  $ \\
 Cassini division  & $ 1.95  $ & $ 2.025 $ & $ 0.1  $ \\
 Outer B	   & $ 1.64  $ & $ 1.95  $ & $ 2.0  $ \\
 Inner B	   & $ 1.525 $ & $ 1.64  $ & $ 1.0  $ \\
 Outer C	   & $ 1.43  $ & $ 1.525 $ & $ 0.1  $ \\
 Middle C	   & $ 1.29  $ & $ 1.43  $ & $ 0.15 $ \\
 Inner C	   & $ 1.24  $ & $ 1.29  $ & $ 0.08 $ \\
\hline
\enddata
  \tablenotetext{a}{Disk radii are specified in Table~\ref{tab:radii}. }
  \tablenotetext{b}{Following Table IV of \citet{dunn/molnar/fix:2002}.}
  \tablenotetext{c}{Ring-normal optical depth.}
  \tablenotetext{d}{$R_{\mathrm{Saturn}}$ refers to the equatorial radius.}
\end{deluxetable}

\begin{deluxetable}{ccccccc}
\tablecolumns{7}
\tablewidth{0pt}
\tablecaption{Saturn Model Fit Parameters\label{tab:satmodel2}}
\tablehead{
    \colhead{Freq} &
    \multicolumn{3}{c}{Disk} &
    \multicolumn{3}{c}{Rings} \\ &
    \colhead{$T_{\mathrm{disk}}$} &
    \colhead{$\sigma_{\mathrm{fit}}$} &
    \colhead{$\sigma_{\mathrm{adopted}}$} &
    \colhead{$T_{\mathrm{ring}}$} &
    \colhead{$\sigma_{\mathrm{fit}}$} &
    \colhead{$\sigma_{\mathrm{adopted}}$} \\ &
    \colhead{(K)} &
    \colhead{(K)} &
    \colhead{(K)} &
    \colhead{(K)} &
    \colhead{(K)} &
    \colhead{(K)} }
\startdata
\hline
 K   &  133.8 & 1.5 & 4.0 &   6.7   &    1.1  &  2.1 \\
 Ka  &  138.8 & 1.2 & 3.7 &   9.6   &    0.8  &  1.8 \\
 Q   &  142.1 & 0.9 & 3.4 &  11.3   &    0.6  &  1.6 \\
 V   &  146.1 & 0.7 & 3.2 &  14.5   &    0.5  &  1.5 \\
 W   &  146.0 & 0.6 & 3.1 &  19.8   &    0.4  &  1.4 \\
\hline
\enddata
\end{deluxetable}

\clearpage

\begin{deluxetable}{cccccccc}
\tablecolumns{8}
\tablewidth{0pt}
\tablecaption{Derived Uranus Temperatures per Observing Season per Frequency
\label{tab:uran}}
\tablehead{
    \colhead{Season} &
    \colhead{RJD\tablenotemark{a}} &
    \colhead{$D_\mathrm{W}$\tablenotemark{b}} &
    \multicolumn{5}{c}{$T_b$(K)\tablenotemark{c}} \\ 
    \cline{4-8} \\ & & & 
    \colhead{K} &
    \colhead{Ka} &
    \colhead{Q} &
    \colhead{V} &
    \colhead{W}  }
\startdata
\hline
1 & 2204 & -26.4 & $ 163 \pm  45 $ & $  89 \pm  34 $ & $ 133 \pm  25 $ & $ 107 \pm  18 $ & $ 144 \pm  15 $ \\
2 & 2437 & -19.0 & $ 192 \pm  37 $ & $ 139 \pm  28 $ & $ 135 \pm  24 $ & $ 162 \pm  17 $ & $ 130 \pm  15 $ \\
3 & 2573 & -22.4 & $ 172 \pm  41 $ & $ 150 \pm  31 $ & $ 193 \pm  22 $ & $ 117 \pm  18 $ & $ 110 \pm  15 $ \\
4 & 2807 & -15.2 & $ 148 \pm  57 $ & $  68 \pm  44 $ & $ 193 \pm  34 $ & $ 132 \pm  31 $ & $  66 \pm  25 $ \\
5 & 2942 & -18.5 & $ 189 \pm  58 $ & $  94 \pm  43 $ & $ 147 \pm  31 $ & $ 143 \pm  25 $ & $ 106 \pm  21 $ \\
6 & 3176 & -11.2 & $ 192 \pm  37 $ & $  97 \pm  28 $ & $ 144 \pm  22 $ & $ 177 \pm  19 $ & $ 141 \pm  16 $ \\
7 & 3312 & -14.5 & $ 146 \pm  38 $ & $ 148 \pm  29 $ & $ 122 \pm  23 $ & $ 160 \pm  18 $ & $ 117 \pm  15 $ \\
8 & 3546 &  -7.3 & $ 121 \pm  37 $ & $ 116 \pm  28 $ & $ 167 \pm  23 $ & $ 109 \pm  18 $ & $ 108 \pm  14 $ \\
9 & 3681 & -10.7 & $ 175 \pm  38 $ & $ 117 \pm  29 $ & $ 127 \pm  22 $ & $ 134 \pm  18 $ & $ 112 \pm  15 $ \\
10 & 3915 &  -3.4 & $ 167 \pm  37 $ & $  87 \pm  28 $ & $ 117 \pm  22 $ & $ 157 \pm  18 $ & $ 137 \pm  15 $ \\
11 & 4050 &  -6.7 & $ 139 \pm  38 $ & $ 116 \pm  29 $ & $ 148 \pm  23 $ & $ 124 \pm  18 $ & $ 138 \pm  15 $ \\
12 & 4284 &   0.5 & $ 148 \pm  37 $ & $ 152 \pm  28 $ & $ 161 \pm  21 $ & $ 143 \pm  18 $ & $ 114 \pm  15 $ \\
13 & 4419 &  -2.8 & $ 246 \pm  40 $ & $ 167 \pm  30 $ & $ 203 \pm  23 $ & $ 138 \pm  19 $ & $ 120 \pm  16 $ \\
14 & 4654 &   4.4 & $ 135 \pm  37 $ & $ 173 \pm  29 $ & $ 167 \pm  22 $ & $ 129 \pm  18 $ & $  97 \pm  15 $ \\
MEAN & \nodata  &  \nodata  & $ 166 \pm  11 $ & $ 126 \pm   9 $ & $ 153 \pm   6 $ & $ 138 \pm   5 $ & $ 120 \pm   4 $ \\
\hline
\enddata
  \tablenotetext{a}{Approximate mean time of observations in 
                    each season, Julian Day $-2450000$.}
  \tablenotetext{b}{Sub-\wmap\ latitude, degrees.}
  \tablenotetext{c}{Brightness temperature calculated for a solid angle $\Omega_{\mathrm{ref}}= 2.482 \times 10^{-10}$ sr at a
               fiducial distance of 19.0 AU.  Temperature is with respect to blank sky: absolute
	       brightness temperature is obtained by adding 2.2, 2.0, 1.9, 1.5 and 1.1 K in bands
	       K, Ka, Q, V and W respectively \citep{page/etal:2003b}.}
\end{deluxetable}

\begin{deluxetable}{cccccccc}
\tablecolumns{8}
\tablewidth{0pt}
\tablecaption{Derived Neptune Temperatures per Observing Season per Frequency
\label{tab:nept}}
\tablehead{
    \colhead{Season} &
    \colhead{RJD\tablenotemark{a}} &
    \colhead{$D_\mathrm{W}$\tablenotemark{b}} &
    \multicolumn{5}{c}{$T_b$(K)\tablenotemark{c}} \\ 
    \cline{4-8} \\ & & & 
    \colhead{K} &
    \colhead{Ka} &
    \colhead{Q} &
    \colhead{V} &
    \colhead{W}  }
\startdata
\hline
1 & 2185 &  -27.4 & $ 279 \pm 106 $ & $ 126 \pm  81 $ & $  13 \pm  65 $ & $  78 \pm  49 $ & $ 199 \pm  42 $ \\
2 & 2420 &  -27.9 & $ 303 \pm  91 $ & $  43 \pm  70 $ & $ 191 \pm  54 $ & $ 134 \pm  41 $ & $ 197 \pm  36 $ \\
3 & 2555 &  -27.6 & $ 205 \pm 106 $ & $  27 \pm  82 $ & $  75 \pm  60 $ & $  72 \pm  45 $ & $ 132 \pm  38 $ \\
4 & 2799 &  -27.9 & $ 137 \pm 119 $ & $ 108 \pm  90 $ & $ 120 \pm  68 $ & $ 164 \pm  54 $ & $ 152 \pm  46 $ \\
5 & 2923 &  -27.6 & $ 167 \pm 102 $ & $ 146 \pm  77 $ & $  76 \pm  66 $ & $ 178 \pm  48 $ & $ 133 \pm  39 $ \\
6 & 3155 &  -28.0 & $  24 \pm  90 $ & $  21 \pm  70 $ & $ 188 \pm  52 $ & $ 310 \pm  43 $ & $ 151 \pm  37 $ \\
7 & 3290 &  -27.7 & $ 216 \pm  91 $ & $ 121 \pm  69 $ & $ -14 \pm  59 $ & $ 194 \pm  51 $ & $ 196 \pm  41 $ \\
8 & 3522 &  -28.0 & $ 246 \pm  93 $ & $ 180 \pm  73 $ & $ 137 \pm  52 $ & $  53 \pm  43 $ & $ 128 \pm  36 $ \\
9 & 3657 &  -27.9 & $ 184 \pm  96 $ & $ 263 \pm  74 $ & $ 140 \pm  55 $ & $  18 \pm  51 $ & $ 154 \pm  41 $ \\
10 & 3890 &  -27.9 & $ 177 \pm  96 $ & $ 150 \pm  74 $ & $ 235 \pm  55 $ & $ 131 \pm  40 $ & $  76 \pm  36 $ \\
11 & 4025 &  -27.9 & $  89 \pm 107 $ & $ 195 \pm  83 $ & $ 151 \pm  54 $ & $ 181 \pm  48 $ & $ 162 \pm  42 $ \\
12 & 4257 &  -27.7 & $ 104 \pm  93 $ & $ 232 \pm  72 $ & $  52 \pm  54 $ & $ 161 \pm  42 $ & $  82 \pm  38 $ \\
13 & 4391 &  -27.6 & $ -86 \pm 116 $ & $ 193 \pm  88 $ & $ 125 \pm  57 $ & $ 161 \pm  49 $ & $ 132 \pm  42 $ \\
14 & 4625 &  -27.6 & $ 142 \pm  90 $ & $ 242 \pm  68 $ & $ 180 \pm  53 $ & $  82 \pm  46 $ & $ 116 \pm  40 $ \\
MEAN & \nodata   &  \nodata   & $ 161 \pm  27 $ & $ 147 \pm  21 $ & $ 126 \pm  15 $ & $ 138 \pm  12 $ & $ 142 \pm  11 $ \\
\hline
\enddata
  \tablenotetext{a}{Approximate mean time of observations in 
                    each season, Julian Day $-2450000$.}
  \tablenotetext{b}{Sub-\wmap\ latitude, degrees.}
  \tablenotetext{c}{Brightness temperature calculated for a solid angle $\Omega_{\mathrm{ref}}= 1.006 \times 10^{-10}$ sr at a
               fiducial distance of 29.0 AU.  Temperature is with respect to blank sky: absolute
	       brightness temperature is obtained by adding 2.2, 2.0, 1.9, 1.5 and 1.1 K in bands
	       K, Ka, Q, V and W respectively \citep{page/etal:2003b}.}
\end{deluxetable}

\begin{deluxetable}{lrrrrlcl}
\tablecolumns{8}
\tabletypesize{\scriptsize}
\tablewidth{0pt}
\tablecaption{Selected Celestial Calibration Sources
\label{tab:calobj}}
\tablehead{
    \colhead{Source Name} &
    \colhead{R.A.$_{\mathrm{J2000}}$ \tablenotemark{a}} &
    \colhead{Decl.$_{\mathrm{J2000}}$ } &
    \colhead{$l$} &
    \colhead{$b$} &
    \colhead{Type} &
    \colhead{Angular Extent} &
    \colhead{Reference} \\ &
    \colhead{(hms)} &
    \colhead{(dms)} &
    \colhead{(deg)} &
    \colhead{(deg)} & &
    \colhead{(arcmin)}  
    }
\startdata
Cas A          & 23  23  24  &  +58  48  54  & 111.74 &  -2.13 & shell type SNR    &        $5$       & \citet{green:2009}    \\ 
Cyg A          & 19  59  28  &  +40  44  02  &  76.19 &   5.76 & radio galaxy      & $2.8 \times 0.8$ & \citet{ott/etal:1994} \\ 
Tau A (Crab)   & 05  34  32  &  +22  00  52  & 184.56 &  -5.78 & filled-center SNR &   $7 \times 5$   & \citet{green:2009}    \\ 
3C58           & 02  05  38  &  +64  49  42  & 130.72 &   3.08 & filled-center SNR &  $9 \times 5$    & \citet{green:2009}    \\ 
3C274 (M87, Vir A) & 12  30  49  &  +12  23  28  & 283.78 &  74.49 & radio galaxy  &  $ 2.5$          & \citet{ott/etal:1994} \\ 
\hline
\enddata
 \tablenotetext{a}{Positions for the galaxies are from NED (\texttt{http://nedwww.ipac.caltech.edu/}) and positions
              for the SNRs are from  SIMBAD (\texttt{http://simbad.u-strasbg.fr/simbad/}).}
\end{deluxetable}

\clearpage
\begin{deluxetable}{lccccccccc}
\tablecolumns{9}
\tabletypesize{\scriptsize}
\tablewidth{0pt}
\tablecaption{Flux Densities of Celestial Calibration Sources
\label{tab:srcflux}} 						
\tablehead{
\colhead{Source} & \colhead{$\nu_{\mathrm{eff}}$\tablenotemark{a}}  &
\colhead{$\Gamma$\tablenotemark{a}} & \colhead{I Flux} & \colhead{Q Flux} & \colhead{U Flux} & \colhead{P Flux\tablenotemark{b}} & \colhead{P/I\tablenotemark{b}} & \colhead{Pol Angle\tablenotemark{c}} \\
\colhead{} & \colhead{(GHz)} & \colhead{(mK Jy$^{-1}$)} &
\colhead{(Jy)}  & \colhead{(Jy)} & \colhead{(Jy)} & \colhead{(Jy)}  & \colhead{(\%)} & \colhead{(deg)} 
}
\startdata 
\hline
 Cas A &      22.68 &      0.2497 &     232.1$\pm$ 2.9 &    -0.92$\pm$0.09 &   -0.15$\pm$0.09 &  0.93 $\pm$ 0.09 &  0.41$\pm$0.04 &     85.5$\pm$ 2.7 (66.0) & \\
 Cas A &      32.94 &      0.2041 &     179.8$\pm$ 1.5 &    -0.56$\pm$0.08 &   -0.19$\pm$0.09 &  0.59 $\pm$ 0.08 &  0.33$\pm$0.05 &     80.6$\pm$ 4.2 (61.1) & \\
 Cas A &      40.62 &      0.2162 &     153.7$\pm$ 1.5 &    -0.50$\pm$0.09 &   -0.17$\pm$0.10 &  0.53 $\pm$ 0.09 &  0.35$\pm$0.06 &     80.6$\pm$ 5.3 (61.1) & \\
 Cas A &      60.48 &      0.2084 &     116.3$\pm$ 2.0$^d$ &    -0.25$\pm$0.16$^d$ &    0.12$\pm$0.16$^d$ &  0.28 $\pm$ 0.16 &  0.24$\pm$0.14 &    -77.6$\pm$16.8 (82.9) & \\
 Cas A &      92.90 &      0.1800 &      84.0$\pm$ 3.4$^e$ &    -0.05$\pm$0.28$^e$ &    0.12$\pm$0.28$^e$ &  0.13 $\pm$ 0.28 &  0.16$\pm$0.34 &    -56.1$\pm$59.9 (104.4) & \\
 \phn &   \phn     &    \phn     &      \phn          &         \phn      &    \phn          &    \phn          &     \phn          & \\
 Cyg A &      22.65 &      0.2493 &      57.7$\pm$ 1.3 &     0.02$\pm$0.05 &   -0.48$\pm$0.06 &  0.48 $\pm$ 0.06 &  0.84$\pm$0.11 &     43.6$\pm$ 2.9 (164.8) & \\
 Cyg A &      32.91 &      0.2039 &      38.2$\pm$ 0.6 &     0.40$\pm$0.08 &    0.20$\pm$0.10 &  0.45 $\pm$ 0.08 &  1.18$\pm$0.22 &    -13.5$\pm$ 6.0 (107.7) & \\
 Cyg A &      40.58 &      0.2159 &      29.8$\pm$ 0.4 &     0.58$\pm$0.08 &    0.18$\pm$0.10 &  0.61 $\pm$ 0.09 &  2.04$\pm$0.29 &     -8.4$\pm$ 4.8 (112.8) & \\
 Cyg A &      60.42 &      0.2083 &      18.6$\pm$ 0.2 &     0.38$\pm$0.15 &    0.26$\pm$0.20 &  0.46 $\pm$ 0.17 &  2.48$\pm$0.89 &    -16.9$\pm$11.4 (104.3) & \\
 Cyg A &      92.83 &      0.1802 &      11.1$\pm$ 0.3 &     0.26$\pm$0.26 &    0.11$\pm$0.34 &  0.28 $\pm$ 0.27 &  2.55$\pm$2.45 &    -11.5$\pm$33.2 (109.7) & \\
 \phn &   \phn     &    \phn     &      \phn          &         \phn      &    \phn          &    \phn          &     \phn          & \\ 
 Tau A &      22.70 &      0.2501 &     383.8$\pm$ 9.6 &   -27.13$\pm$0.68 &    1.40$\pm$0.08 &  27.17 $\pm$ 0.68 &  7.08$\pm$0.25 &    -88.5$\pm$ 0.1 (149.1) & \\
 Tau A &      32.96 &      0.2043 &     342.8$\pm$ 6.4 &   -23.72$\pm$0.45 &    1.88$\pm$0.12 &  23.80 $\pm$ 0.44 &  6.94$\pm$0.18 &    -87.7$\pm$ 0.1 (149.9) & \\
 Tau A &      40.64 &      0.2163 &     317.7$\pm$ 8.6 &   -22.03$\pm$0.60 &    2.06$\pm$0.14 &  22.12 $\pm$ 0.60 &  6.97$\pm$0.27 &    -87.3$\pm$ 0.2 (150.3) & \\
 Tau A &      60.53 &      0.2085 &     276.0$\pm$ 5.2$^d$ &   -19.25$\pm$0.36$^d$ &    1.52$\pm$0.24$^d$ &  19.31 $\pm$ 0.36 &  7.00$\pm$0.19 &    -87.7$\pm$ 0.4 (149.9) & \\
 Tau A &      92.95 &      0.1798 &     232.8$\pm$ 9.7$^e$ &   -16.58$\pm$0.73$^e$ &    0.75$\pm$0.42$^e$ &  16.60 $\pm$ 0.73 &  7.13$\pm$0.43 &    -88.7$\pm$ 0.7 (148.9) & \\
 \phn &   \phn     &    \phn     &      \phn          &         \phn      &    \phn          &    \phn          &     \phn          & \\
 3C58 &      22.70 &      0.2500 &      20.8$\pm$ 0.4 &     1.11$\pm$0.04 &    0.45$\pm$0.06 &  1.20 $\pm$ 0.05 &  5.77$\pm$0.25 &    -11.0$\pm$ 1.4 (5.4) & \\
 3C58 &      32.96 &      0.2043 &      19.0$\pm$ 0.3 &     1.20$\pm$0.07 &    0.25$\pm$0.10 &  1.23 $\pm$ 0.07 &  6.46$\pm$0.39 &     -5.8$\pm$ 2.3 (10.7) & \\
 3C58 &      40.64 &      0.2163 &      17.1$\pm$ 0.3 &     0.99$\pm$0.08 &    0.46$\pm$0.09 &  1.09 $\pm$ 0.08 &  6.37$\pm$0.50 &    -12.6$\pm$ 2.3 (3.9) & \\
 3C58 &      60.52 &      0.2085 &      14.1$\pm$ 0.3$^d$ &     0.77$\pm$0.12$^d$ &    0.07$\pm$0.17$^d$ &  0.77 $\pm$ 0.12 &  5.48$\pm$0.87 &     -2.7$\pm$ 6.3 (13.7) & \\
 3C58 &      92.94 &      0.1798 &      11.4$\pm$ 0.6$^e$ &     0.52$\pm$0.23$^e$ &    0.42$\pm$0.30$^e$ &  0.67 $\pm$ 0.26 &  5.87$\pm$2.28 &    -19.3$\pm$11.7 (-2.9) & \\
 \phn &   \phn     &    \phn     &      \phn          &         \phn      &    \phn          &    \phn          &     \phn          & \\
3C274 &      22.68 &      0.2497 &      21.6$\pm$ 0.6 &     0.49$\pm$0.08 &   -0.73$\pm$0.06 &  0.88 $\pm$ 0.06 &  4.06$\pm$0.31 &     28.1$\pm$ 2.3 (45.5) & \\
3C274 &      32.94 &      0.2041 &      16.8$\pm$ 0.4 &     0.43$\pm$0.12 &   -0.69$\pm$0.09 &  0.81 $\pm$ 0.10 &  4.84$\pm$0.61 &     28.9$\pm$ 4.0 (46.3) & \\
3C274 &      40.61 &      0.2161 &      14.3$\pm$ 0.5 &     0.40$\pm$0.14 &   -0.55$\pm$0.10 &  0.68 $\pm$ 0.12 &  4.72$\pm$0.84 &     26.9$\pm$ 5.5 (44.3) & \\
3C274 &      60.48 &      0.2084 &      10.6$\pm$ 0.3 &     0.16$\pm$0.25 &   -0.11$\pm$0.18 &  0.20 $\pm$ 0.23 &  1.85$\pm$2.15 &     17.2$\pm$29.9 (34.6) & \\
3C274 &      92.89 &      0.1800 &       7.8$\pm$ 0.3 &     0.41$\pm$0.43 &   -0.43$\pm$0.31 &  0.59 $\pm$ 0.37 &  7.66$\pm$4.83 &     23.0$\pm$18.2 (40.4) & \\
\hline
\enddata
\tablenotetext{a}{The second and third columns list effective band center frequency and peak antenna temperature 
             to flux conversion factor $\Gamma$, calculated for spectral index values 
             $\alpha$ of -0.72, -1.17, -0.36, -0.42, and -0.73
             for Cas A, Cyg A, Tau A, 3C58, and 3C274 respectively.}

\tablenotetext{b}{Polarized flux P = (Q$^2$ + U$^2$)$^{1/2}$ with no correction for noise bias.}

\tablenotetext{c}{Polarization position angle is calculated as ${1\over 2}\tan^{-1}(-U/Q)$. It is positive for Galactic 
north through east.  The value for equatorial coordinates is given in parentheses.  The errors listed are statistical
and do not include the potential systematic error of $1.5\arcdeg$ \citep{page/etal:2003b,page/etal:2007}.}

\tablenotetext{d}{The uncertainty includes an uncertainty of 0.015 times the flux, to allow for error due 
               to source extent.}

\tablenotetext{e}{The uncertainty includes an uncertainty of 0.037 times the flux, to allow for error due to 
              source extent.}

\end{deluxetable}

\clearpage

\begin{deluxetable}{crrrrr}
  \tablewidth{0pt}
  \tablecolumns{6}
  \tablecaption{Variability Fit Results for Celestial Calibration Sources
    \label{tab:calvar}}
  \tablehead{
    \colhead{} & 
    \colhead{K Band\tablenotemark{a}} &
    \colhead{Ka Band\tablenotemark{a}} &
    \colhead{Q Band\tablenotemark{a}} &
    \colhead{V Band\tablenotemark{a}} &
    \colhead{W Band\tablenotemark{a}} \\
    \colhead{Source} & 
    \colhead{(\%/yr)} &
    \colhead{(\%/yr)} &
    \colhead{(\%/yr)} &
    \colhead{(\%/yr)} &
    \colhead{(\%/yr)}}
  \startdata
 Cas A  & $-0.51 \pm 0.01$ & $-0.55 \pm 0.02$ & $-0.56 \pm 0.02$ & $-0.48 \pm 0.05$ & $-0.57 \pm 0.12$ \\
 Cyg A  & $-0.02 \pm 0.03$ & $-0.02 \pm 0.08$ & $ 0.02 \pm 0.11$ & $ 0.03 \pm 0.32$ & $-0.86 \pm 0.93$ \\
 Tau A  & $-0.21 \pm 0.01$ & $-0.23 \pm 0.01$ & $-0.26 \pm 0.02$ & $-0.09 \pm 0.03$ & $-0.18 \pm 0.06$ \\
 3C58   & $-0.05 \pm 0.09$ & $ 0.14 \pm 0.15$ & $-0.17 \pm 0.18$ & $-0.43 \pm 0.35$ & $ 0.47 \pm 0.80$ \\
 3C274  & $ 0.22 \pm 0.11$ & $-0.19 \pm 0.22$ & $ 0.29 \pm 0.29$ & $ 1.24 \pm 0.69$ & $ 1.28 \pm 1.64$ \\
  \enddata
  \tablenotetext{a}{Trends over the seven-year \wmap\ mission, from the linear fits to the
  fractional variability data shown in Figure~\ref{fig:npovarplot}.}
\end{deluxetable}

\clearpage
\begin{deluxetable}{lccccccccc}
\tablecolumns{8}
\tabletypesize{\scriptsize}
\tablewidth{0pt}
\tablecaption{Spectral Fits for Celestial Calibration Sources
  \label{tab:specfits}}                                          
\tablehead{
\colhead{Source} & \colhead{Data}  &
\colhead{Frequency Range} & \colhead{Fit Type\tablenotemark{a}} & \colhead{a} & \colhead{b} & \colhead{c} & \colhead{d} \\
\colhead{} & \colhead{} & \colhead{(GHz)} & \colhead{} &
\colhead{}  & \colhead{} & \colhead{} & \colhead{}
}
\startdata 
\hline
 Cas A &      WMAP only &     22 - 93 &     Power law &    2.204$\pm$0.002 &   -0.712$\pm$0.018 &    \nodata &    \nodata \\
 Cas A &      Combined$^b$ & 1.4 - 250 &    Curved    &    2.204$\pm$0.002 &   -0.682$\pm$0.011 & 0.038 $\pm$ 0.008 &    \nodata\\
 \phn  &      \phn      &     \phn    &     \phn      &    \phn            &   \phn             &    \phn    &    \phn \\
 Cyg A &      WMAP only &     22 - 93 &     Power law   &  1.480$\pm$0.003 &   -1.172$\pm$0.020 &  \nodata &    \nodata \\
 Cyg A &      Combined &       2 - 94 &     Power law &    1.482$\pm$0.003 &   -1.200$\pm$0.006 &    \nodata &    \nodata \\
 \phn  &      \phn      &     \phn    &     \phn      &    \phn            &   \phn             &    \phn    &    \phn \\
 Tau A &      WMAP only &     22 - 93 &     Power law &    2.502$\pm$0.005 &   -0.350$\pm$0.026 &    \nodata &    \nodata \\
 Tau A &      Combined$^c$ &   1 - 353 &    Power law &    2.506$\pm$0.003 &   -0.302$\pm$0.005 &    \nodata &    \nodata \\
 \phn  &      \phn      &     \phn    &     \phn      &    \phn            &   \phn             &    \phn    &    \phn \\
 3C58 &      Combined &    0.04 - 83000 &     Rolloff   &     33.7$\pm$0.7   &   -0.053$\pm$0.013 &  0.021$\pm$0.006 & 0.913$\pm$0.045 & \\
 \phn  &      \phn      &     \phn    &     \phn      &    \phn            &   \phn             &    \phn    &    \phn \\
3C274 &      WMAP only &      22 - 93 &     Power law &    1.159$\pm$0.005 &   -0.731$\pm$0.030 &    \nodata &    \nodata \\
3C274 &      Combined &      0.4 - 93 &     Curved    &    1.154$\pm$0.005 &   -0.733$\pm$0.013 & 0.050 $\pm$ 0.008 &    \nodata \\

\hline
\enddata
\tablenotetext{a}{Power law: log S(Jy) = a + b log ($\nu$/40 GHz), Curved: log S(Jy) = a + b log ($\nu$/40 GHz) +c log$^2$ ($\nu$/40 GHz),
 Rolloff: S(Jy) = a ($\nu$/1 GHz)$^b$/ (1 + c ($\nu$/1 GHz)$^d$ ). }
\tablenotetext{b}{The combined data for Cas A were scaled to epoch 2000.}
\tablenotetext{c}{The non-WMAP data for Tau A were scaled to epoch 2005.}
\end{deluxetable}

\end{document}